\documentclass[11pt,a4paper]{article}

\usepackage[utf8]{inputenc}
\usepackage{amsmath,latexsym}
\usepackage{mathtools}
\usepackage[dvipsnames,table]{xcolor} 
\usepackage{graphicx}
\usepackage{slashed}

\usepackage{tikz}
\usepackage[compat=1.0.0]{tikz-feynman}
\usepackage[nomessages]{fp}
\usepackage{jheppub}
\usepackage{graphicx}
\usepackage{soul}

\usepackage{changepage}
\usepackage{braket}
\usepackage{relsize}
\usepackage{lscape} 
\usepackage{bbm} 
 
\newcommand{\matrixx}[1]{\begin{matrix} #1 \end{matrix}} 
\usepackage{pdfpages}
\newcommand{\minus}{\scalebox{0.75}[1.0]{$-$}}
\usepackage{nicematrix}
\usepackage{array,xparse,l3keys2e,expl3}
\usepackage{float}
\usepackage{multicol}
\usepackage{scalerel}

\def\beq{\begin{equation}}   
\def\eeq{\end{equation}}
\def\bea{\begin{eqnarray}}  
\def\eea{\end{eqnarray}} 
\def\nn{\nonumber}
\def\r{\right} 
\def\l{\left} 

\def\eps{\varepsilon}
\def\O{\mathcal{O}}
\def\L{\mathcal{L}}
\def\Z{\mathcal{Z}}
\def\E{\mathcal{E}}
\def\p{\phi}
\def\d{\partial}
\def\T{\mathcal{T}}
\def\spacetimeInt{\int\!\!d^{\scaleto{D}{4.5pt}}\!x\ }

\definecolor{darkgreen}{rgb}{0,0.5,0}

\definecolor{what}{rgb}{0,0.5,0.5}

\title{Renormalization and non-renormalization of scalar EFTs at higher orders}

\author[a,b]{Weiguang Cao,}
\author[c]{Franz Herzog,}
\author[a]{Tom Melia,}
\author[d]{and Jasper Roosmale Nepveu}

\affiliation[a]{Kavli Institute for the Physics and Mathematics of the Universe (WPI),\\ UTIAS, The University of Tokyo, Kashiwa, Chiba 277-8583, Japan}
\affiliation[b]{Department of Physics, Graduate School of Science, The University of Tokyo, Tokyo 113-0033, Japan}
\affiliation[c]{Higgs Centre for Theoretical Physics, School of Physics and Astronomy, The University of Edinburgh, Edinburgh EH9 3FD, Scotland, UK}
\affiliation[d]{Institut f\"ur Physik, Humboldt-Universit\"at zu Berlin,\\  
D-12489 Berlin, Germany}

\emailAdd{weiguang.cao@ipmu.jp, fherzog@ed.ac.uk, tom.melia@ipmu.jp, jasper.roosmalenepveu@physik.hu-berlin.de}

\abstract{
We renormalize massless scalar effective field theories (EFTs) to higher loop orders and higher orders in the EFT expansion. To facilitate EFT calculations with the R* renormalization method, we construct suitable operator bases using Hilbert series and related ideas in commutative algebra and conformal representation theory, including their novel application to off-shell correlation functions. We obtain new results ranging from full one loop at mass dimension twelve  to five loops at mass dimension six. We explore the structure of the anomalous dimension matrix with an emphasis on its zeros, and investigate the effects of conformal and orthonormal operators. For the real scalar, the zeros can be explained by a `non-renormalization' rule recently derived by Bern et al.
For the complex scalar we find two new selection rules for mixing $n$- and $(n-2)$-field operators, with $n$ the maximal number of fields at a fixed mass dimension. 
The first appears only when the $(n-2)$-field operator is conformal primary, and is valid at one loop. The second appears in more generic bases, and is valid at three loops. 
Finally, we comment on how the Hilbert series we construct may be used to provide a systematic enumeration of a class of evanescent operators that appear at a particular mass dimension in the scalar EFT.
}

\begin{document}
\begin{flushright}
HU-EP-21/14-RTG
\end{flushright}

\vspace{-9.15mm}


\maketitle
\newpage

 \section{Introduction}\label{sec:intro}

Motivated by the need to match experimental accuracy, quantum field theory calculations have had to evolve to tackle problems of increasing complexity. For example, the Large Hadron Collider (LHC) is a powerful driver for advancing the boundaries of perturbative QCD and electroweak theory calculations, where $S$-matrix elements are needed at higher loop order to match the precision measurements, and may involve many `legs' (number of external particles) to describe the large multiplicity processes offered by the high collision energy. For instance, recent years have witnessed state of the art QCD amplitude calculations being pushed to three-loops for up to four external legs \cite{Caola:2020dfu}, and to two-loops for up to five external legs, see for instance \cite{Badger:2021nhg,Chawdhry:2021mkw,Abreu:2021fuk}. Similarly, extremely precise measurements of the fine structure constant from atom interferometry~\cite{Parker191,Morel:2020dww} and the magnetic dipole moment of the electron~\cite{Hanneke_2008} have pushed calculations up to five loops in QED~\cite{Aoyama:2012wj,Aoyama:2014sxa,Aoyama:2019ryr}. 

In addition to such experimental drivers, there is also strong motivation to advance perturbative calculational frontiers from purely theoretical considerations, in an endeavour to better understand underlying structures ({\it e.g.}~\cite{Arkani_Hamed_2014}) in gauge and gravity theories, {and to make comparisons with non-perturbative predictions of critical phenomena.} For instance, maximally supersymmetric gravity has recently been confirmed to be finite up to five loops~\cite{Bern:2018jmv}; planar maximally supersymmetric Yang-Mills theory has now been explored at seven loops~\cite{Caron_Huot_2019}; and, {in scalar $\phi^4$ theory, renormalization group functions have been published at six loops for use in the $\epsilon$-expansion approach to the 3D Ising model~\cite{Chetyrkin:1981jq,Kleinert:1991rg,Gorishnii:1983gp,Kompaniets:2017yct,Batkovich:2016jus,Schnetz:2016fhy}. 
} 

Effective field theories
(EFTs) are valid up to some cutoff scale, with a hierarchy under which operators in
an action are organized, namely the EFT expansion. 
There are many cases where one is interested in higher order terms in the EFT expansion to capture the relevant physics. For example, the Standard Model (SM) EFT, which has an expansion organized in operator mass dimension, is able to parameterize the indirect effects of new physics at the LHC. Higher-order (and higher leg) calculations in this EFT are recently being studied, following the full one-loop renormalization in Refs.~\cite{Jenkins2013,Jenkins:2013wua,Alonso:2013hga}. These studies are in part motivated by the need to understand the role of loops in the eventual reach of the LHC in constraining the EFT parameter space, see {\it e.g.}~\cite{passarino2016standard},
and in part to understand {underlying structures in EFT and in calculational approaches to them more generally {\it e.g.}~\cite{Alonso:2014rga,Cheung:2015aba,Caron-Huot:2016cwu,Craig:2019wmo,Bern2020,Baratella:2020dvw,EliasMiro:2020tdv,Baratella:2020lzz,Bern:2020ikv,Jiang:2021tqo}.}

Going to higher orders in an EFT expansion compounds the complexities faced at large loops and legs: it comes with an (exponential) increase in Feynman vertices and a corresponding diagrammatic explosion, and implies higher rank tensors appear in loop integrands. 
In addition, one has to keep track of the organization of an independent set of operators which span all physical observables; this is further complicated beyond tree-level by operator mixing under renormalization group flow, and by the fact that at intermediate stages of the renormalization procedure, one needs to include unphysical operators to absorb divergences.

In this paper we study the renormalization of (the parity even sector of) EFTs of real and complex massless scalar fields. Throughout, we work exclusively at linear order in the higher dimension operators.  The results we obtain are thus also relevant to the study of $\phi^4$ theory at the Wilson-Fisher fixed point~\cite{Wilson:1971dc}.
Our focus is two-fold: i) to explore new boundaries in mass dimension and loops for EFTs of scalar fields, and the structures that appear there and ii) the development of new general techniques to enable the renormalization of EFTs at higher order. Regarding i), we summarize the results we obtain in Fig~\ref{fig1} (structures in Fig.~\ref{fig:adm_pic} in section~\ref{sec:results} below). As for ii), a main result is the elucidation of the role of Hilbert series, conformal representation theory, and polynomial ring (or amplitude\footnote{More specifically, the polynomial pieces, or contact terms.}) techniques in organizing an EFT basis relevant for off-shell Green's functions, and how this dovetails with the R* renormalization technique. 

 \begin{figure}
 \centering
 \includegraphics[scale=0.6]
 {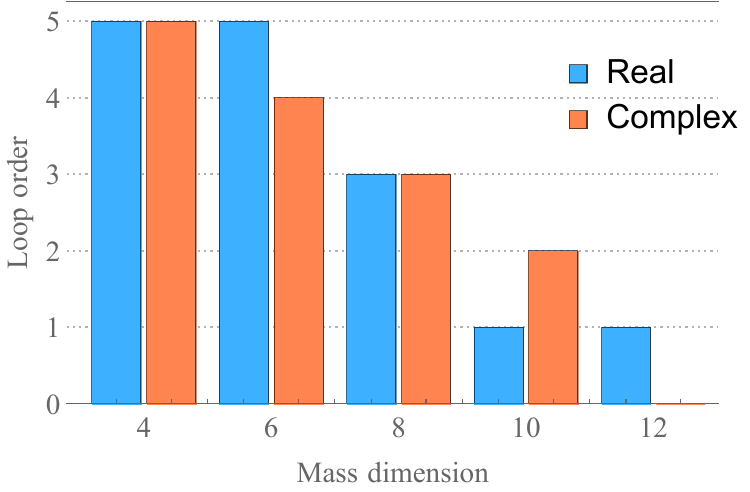}
 \caption{Summary of the operator anomalous dimensions we obtain in the real and complex $\phi^4$ effective theories in this work.
 }\label{fig1} 
 \end{figure}

In $\phi^4$ theory, or more generally $O(n)$ models (of which $O(2)$ is equivalent to the $C$-even sector of the complex scalar), around four spacetime dimensions, the renormalization group functions have been published up to seven loops~\cite{Chetyrkin:1981jq,Kleinert:1991rg,Gorishnii:1983gp,Batkovich:2016jus,Kompaniets:2017yct,Schnetz:2016fhy} (results up to eight loops have been obtained with graphical functions \cite{Borinsky:2021gkd,Borinsky:2021jdb}); operators of higher mass dimension have been studied in general at leading perturbative order, and  subsets at two-loop~\cite{Kehrein:1992fn,Kehrein:1994ff,Kehrein:1995ia,Rychkov:2015naa,Sen:2015doa,Hogervorst:2015akt}.
 In QCD and flavour physics, higher loop order anomalous dimensions (ADs) have been calculated for low mass dimension operators \cite{Altarelli:1980fi,Morozov:1985ef,Buras:1989xd,Buras:1991jm,Ciuchini:1992tj,Ciuchini:1993vr,Misiak:1994zw,Chetyrkin:1996vx,Chetyrkin:1997fm,Ciuchini:1998ix,Buras:2000if,Gracey:2000am,Gracey:2002he,Gambino:2003zm,Gorbahn:2004my,Gorbahn:2005sa,Degrassi:2005zd, Czakon:2006ss,deVries:2019nsu}, going up to four loops in the last reference.
The two loop structure of the SM EFT AD matrix has been explored recently \cite{Bern:2020ikv} using an on-shell approach \cite{Caron-Huot:2016cwu}.
Regarding calculations that go to higher EFT dimensions, QCD gluonic operators up to mass dimensions 16 were recently computed using generalized unitarity at two loops \cite{Jin:2020pwh}. Another related problem is the renormalization of spin-$N$ light-cone operators. Here the four-loop ADs of fermionic light-cone operators were computed up to mass dimension 23 by direct calculation of two-point correlators~\cite{Moch:2017uml}. It should be noted that in the latter two calculations there are restrictions on the minimal number of fields appearing in the operators, making direct calculational approaches feasible.

Beyond unitarity or direct approaches, many of the above calculations were done using a technique called infrared rearrangement, originally pioneered by Vladimirov \cite{Vladimirov:1979zm}. It is based on the observation that ADs in minimal subtraction (MS) schemes are independent of the external kinematics or the internal masses. 
A powerful extension of this method, on which we focus in this work, is the R* method \cite{Chetyrkin:1982nn,Chetyrkin:1984xa,Chetyrkin:2017ppe}; see also \cite{Batkovich:2014rka}. It allows for the subtraction of the IR-divergences created in the process of differentiation and nullification of the external momenta. The R* method thus allows for arbitrary IR rearrangements and thereby enables to maximally simplify the loop integrals. Its power is summarized in a theorem which states that the ADs of any $L$-loop Feynman diagram can be extracted from products of massless self-energies of at most $L-1$ loops \cite{Chetyrkin:1982nn}. The required master integrals are now available up to 5-loops \cite{Smirnov:2010hd,Baikov:2010hf,Lee:2011jt,Georgoudis:2021onj}; paving the way for 6-loop R* calculations. Although its mathematical structure is involved, the method has by now been developed into a very general tool. A global method has been developed and used in calculations up to five loops; see e.g. \cite{Baikov:2016tgj, Chetyrkin:2018dce}. The local R* method was further extended to generic numerator Feynman graphs \cite{Herzog:2017bjx}. This enabled a number of state-of-the-art five-loop calculations in QCD/gauge theory \cite{Herzog:2017ohr,Herzog:2017dtz,Herzog:2018kwj}. A further extension was also used to compute the two- and three-loop ADs of the dimension-six CP-odd gluonic operator in the SM EFT \cite{deVries:2019nsu}. On the more formal level the Hopf-algebraic structure of the R* operation has been unraveled \cite{Beekveldt:2020kzk, Brown:2015fyf}, further enriching our understanding and control of its intricate combinatorics.

Separately to the complication of multi-loop integrals, the EFT expansion adds its own unique problems. At higher order in this expansion, the organization of what is termed the operator basis of the EFT---a set of operators that lead to independent\footnote{Independent under operator redundancies owing to field redefinitions (often referred to as equations of motion redundancies in the literature), integration by parts identities, and other spacetime and internal symmetry group redundancies.} physical effects---is a significant complication in its own right, both from the technical point of view, and in the mapping out of the connection to experimental observations. It requires the development of techniques that are orthogonal to refinements of the ideas of (on-shell or off-shell) amplitude recursion techniques, master integral reduction, R*, IR subtractions \textit{etc.}~that render loops and legs tractable. 

Hilbert series (more generally, the mathematical structure of polynomial rings) have recently been utilized to organize and ameliorate the difficulties surrounding the construction of EFT operator bases and to study the structure of EFTs~\cite{Henning:2015daa,Henning:2015alf,Henning:2017fpj,Henning:2019enq,Henning:2019mcv,Graf:2020yxt,Melia:2020pzd} (see also the developments~\cite{Christensen:2018zcq,Shadmi:2018xan,Ma:2019gtx,Durieux:2019eor,Aoude:2019tzn,Christensen:2019mch,Durieux:2019siw,Durieux:2020gip,Li:2020gnx,Jiang:2021tqo,Dong:2021yak}). 
The scalar EFTs we consider fall into the class where their operator bases are controlled by an underlying conformal representation theory, which can be directly used in the construction of a Hilbert series. In~\cite{Henning:2015alf} it was shown that the EFT operator basis can be identified as the set of conformal primary operators (in~\cite{Henning:2019enq, Henning:2019mcv} these primaries were further identified as harmonics of the manifold of phase space).%
\footnote{This is one aspect of a beautiful picture in integer dimensions $d=3,4$, utilizing spinors of $SL(2,R)$ and $SL(2,C)$ and the oscillator representation of the conformal algebra. The current work, calculating at higher loop order in perturbation theory, of course relies heavily on dimensional regularisation \textit{i.e.}~working in $D=4-2\epsilon$. It would be fascinating to understand whether/how aspects of the picture laid out in~\cite{Henning:2019mcv} may survive/be deformed to non-integer spacetime dimensions, perhaps in connection with the ideas laid out in~\cite{Binder:2019zqc}. } 
This represents a mathematically singled out (up to rotations in the space of primaries) basis for the $S$-matrix {\it i.e.} the number of {\it on-shell} physical measurements one can make in a theory. We explore how this picture generalizes for {\it off-shell} correlation functions and the calculation of quantum corrections; here a larger basis is required at an intermediate stage in the EFT renormalization procedure. We will see that many of the techniques established in~\cite{Henning:2015daa,Henning:2015alf,Henning:2017fpj} can be leveraged to organize this larger operator basis. 

We emphasize that both the Hilbert series and R* techniques we develop are general, and can apply beyond the scalar EFTs we consider here. In particular, for Hilbert series, applications to spin~\cite{Henning:2015alf,Henning:2017fpj,Henning:2019enq,Henning:2019mcv}, non-linearly realized internal symmetries~\cite{Henning:2017fpj,Graf:2020yxt}, gravity~\cite{Ruhdorfer:2019qmk} and non-relativistic EFTs~\cite{Kobach:2017xkw,Kobach:2018pie}, have all been developed. The R* method has already found application in gauge theories and the SM EFT, as already mentioned above.

Before proceeding, we comment on two theoretical matters of interest regarding the structure of perturbative EFT: zeros in the anomalous dimension matrix, and evanescent operators. Despite the fact that we are working with an EFT of a single scalar field, the story is quite interesting on both counts. 

Regarding the anomalous dimension matrix (ADM), a number of recent papers \cite{Cheung:2015aba,Bern2020,Elias-Miro:2014eia,Jiang:2021tqo} have observed and proven the existence of a priori unexpected zeros at various loop orders. 
Operator basis choice is clearly important for the observation of such structures---given the freedom in basis choice one may think it should always be possible to \textit{e.g.}~rotate to a basis where the ADM can even be diagonal, the question more being `is there anything special/identifiable about such a basis?’ (or indeed, the inverse: `how unnatural does a basis have to be to diagonalize the ADM?'). For example, the observed one-loop holomorphy of the SM EFT is apparent in an `on-shell' basis where one favourably eliminates derivatives in operators.\footnote{Such a basis could also be termed an `amplitude' basis: a choice where the fields that appear in all operators indicate the leading on-shell amplitude to which the operator contributes.} We note that beyond one-loop, this question also becomes a scheme dependent one, see for example the discussion in~\cite{Bern2020}, such that one can only discuss structure within a given scheme (we will use the minimal subtraction scheme in this paper). One obviously identifiable feature at hand for the scalar EFT we consider is the singled out basis of primary operators. At leading order in perturbation theory, it is known~\cite{Craigie:1983fb} that, in a basis of primary operators, the ADM of operators in $\phi^4$ theory takes a block diagonal form. We detail the breakdown of this structure in moving beyond the leading-order perturbations: even at one loop\footnote{The counting of perturbative order is not aligned with the loop order.} we find a non-primary basis of operators is necessary to block-diagonalize the ADM. We are unable to identify any further principle that dictates the particular choice of non-primary basis that achieves this. 
At higher loop order, we find zeros in the ADM in concordance with the theorem in \cite{Bern2020}. We do however find two particular zeros {that are stronger than what is minimally implied} by this theorem, which we detail in Sec.~\ref{sec:results}.

Evanescent operators play an important role in higher loop calculations in dimensional regularization, and have been particularly well-studied in the context of four fermion interactions {\it e.g.}~\cite{DUGAN1991239,Gracey:2008mf,BOS1988177,Buras:1998raa,Gracey:2016mio} (see also {\it e.g.}~\cite{Boughezal:2019xpp} for a recent discussion in the SMEFT). Such operators are not typically associated with theories of scalars, but at high enough mass dimension a class of such operators does in fact exist, see~\cite{Hogervorst:2015akt,Hogervorst:2014rta}. This is due to the fact that only $D$ vectors can be linearly independent in $D$ spacetime dimensions. We discuss these evanescent operators in the Hilbert series and polynomial ring analysis of the scalar EFTs in Sec.~\ref{sec:operatorbasis}, and include there a method to systematically enumerate them. The mass dimension at which they appear---16---is beyond that at which we perform the renormalization of the EFT (Fig.~\ref{fig1}), so we will not deal with them in the explicit computations. Their treatment at higher loop order in dimensional regularisation is potentially interesting; we postpone a study of this to a future publication.

The paper is organized as follows. We begin with background on the renormalization of (scalar) EFT in Sec.~\ref{sec:background}, which serves to set out our conventions. In Sec.~\ref{sec:operatorbasis} we discuss details of EFT operator basis, and introduce the technical components---Hilbert series, primary operator bases---that we will utilize. In Sec.~\ref{sec:rstar} we describe the R* method and its application to the EFT. In Sec.~\ref{sec:results} we give our results. We conclude with a discussion in Sec.~\ref{sec:conclusions}.

\section{Background}\label{sec:background}
   In the following, we review the massless $Z_2$-symmetric 
 real scalar EFT, focusing in particular on general properties of operator bases and renormalization. The discussion for the massless complex scalar EFT (and indeed more general theories) follows analogously. Throughout this paper, we restrict ourselves to the spacetime parity even sector.

\subsection{Dimension four Lagrangian} \label{s:dim4background}
In dimensional regularisation ($D=4 \minus 2\eps$ dimensions),
the bare fields and coupling can be expressed in terms of the renormalized fields and coupling, 
\beq
\phi^b=\sqrt{Z_2}\phi\,,\qquad g^b={(4\pi)^2} \, {Z_g \, g(\mu) \, \mu^{2\eps}}.
\eeq
This leads to the following expression of the renormalized dimension 4 Lagrangian:
\beq \label{Lagr4}
\L^{(4)}(\phi,\d_\mu\p)=\O_2^{b(4)}-g^b \, \O_4^{b(4)}\,,
\eeq
where we define the bare operators in the 
real scalar theory to be
\beq
\O_2^{b(4)}=\frac{1}{2} Z_2(\d_\mu\p)(\d^\mu\p),\qquad \O_4^{b(4)}=\frac{1}{4!} Z_2^2 \, \p^4 \, .
\eeq

\subsection{EFT Lagrangian and renormalization}
\label{section:EFTLandR}
The effective Lagrangian is defined by
\beq
\L=\L^{(4)}(\phi,\d_\mu\p)+\sum_{n>4}
 \frac{1}{\Lambda^{n\text{\,-\,}4}}
\L^{(n)}(\phi,\d_\mu\p,\d_{\mu}\d_\nu\p,...)\,,
\eeq
with $\Lambda$ some presumably large scale below which the effective Lagrangian is valid and with the bare higher order interaction terms given by
\beq \label{eq:ibplagr}
\L^{(n)}(\phi,\d_\mu\p,\d_{\mu}\d_\nu\p,...)=\sum_i \tilde c^{\, b(n)}_i \, \tilde \O_i^{b(n)}\,.
\eeq
Here, $\tilde c_i^{\, b}$ are the bare coupling constants (or Wilson coefficients), and we define the bare operators as follows:
\beq \label{bareoperators}
\tilde \O_i^{b(n)}=
\big(g^b\big)^{(l(\tilde \O_i)-2)/2} \, Z_2^{\, l(\tilde \O_i)/2}
\,\tilde \O_i^{(n)}\,,
\eeq
where we write the operators and couplings in this Lagrangian with tildes ($\sim$) to contrast them with the operators in a physical basis, to be defined below.
 The renormalized operators $\tilde \O^{(n)}_{i}$ are in general a function of renormalized fields $\phi$ and derivatives acting on them. We will conventionally include fractional constants in the definition of operators, such that they give rise to a Feynman rule with prefactor +1.
By
 $l(\tilde{\O}_i)$ we denote the length of an operator $\tilde{\O}_i$, \textit{i.e.}~its number of fields.
The included factor of $g^b$ is convenient as it carries the correct powers of the renormalization scale $\mu$ to render the renormalized couplings $\tilde c^{\, (n)}_i$ 
 dimensionless in $D=4\minus 2\eps$ dimensions, while it also generates the conventional factors of $4\pi$ to avoid them in loop integrals in dimensional regularisation.

We assume that the sum in \eqref{eq:ibplagr} runs over a complete set of operators $\tilde \O_i$ that are independent under integration by parts (IBP). That is, these operators are independent at the level of the classical action, 
\beq
I[\phi]=
\spacetimeInt
\L(\phi(x),\d_\mu\p(x),\d_{\mu}\d_\nu\p(x),...)\,,
\eeq
which in essence means that they cannot be related via IBP identities,
\beq
\spacetimeInt \partial_\mu( \bullet )=0\,.
\eeq
Such a set of operators will be called an \textit{off-shell basis}, as it spans the possible counterterms for off-shell correlation functions.\footnote{In this section, we assume that we work in an infinite dimensional spacetime, or with operators of low enough mass dimension, such that Gram conditions can be ignored. This basis is thus consistent with the usual assumptions of dimensional regularisation. See section \ref{sec:operatorbasis} for more details on this point.}
This also explains why the off-shell basis is relevant: 
the full set of IBP-independent operators is necessary to absorb the divergences of Green's functions when renormalizing the theory.

The couplings $\tilde c_i^{\, (n)}$ mix under renormalization, \textit{i.e.}~the bare renormalization constants $\tilde c_i^{\, b(n)}$ are generally power series in the renormalized couplings constants $\{ g, \tilde c_i^{(n)} \}$,
 \begin{equation} \label{eq:Ztildemixing}
 {\tilde c^{\, b(n)}_{i}} = 	
 			\sum_{r\ge1}\;	
 			\sum_{\substack{ n_1>4,..,n_r>4 \\n_1+..+n_r=m(n,r)}} 
 			\sum_{\substack{j_1,..,j_r}}
 			\tilde Z^{(n_1...n_r)}_{{i}j_1...j_r}(g)\prod_{a=1}^{r} \tilde c_{j_a}^{\,(n_a)} \,,
 \end{equation}
where 
$$
m(n,r)=n-4(1-r).
$$
In the sum each partition $n_1,...,n_r$ of mass dimensions determines a set of indices $j_1,...,j_r$ 
which label different couplings (or operators) at the same mass dimension. The corresponding renormalization tensors $\tilde Z_{ij_1...j_r}^{(n)}(g)$, appearing in the sum, are symmetric under interchanging $j_a$ and $j_b$ whenever $n_a{=}n_b$ for some $a$ and $b$. Equation \eqref{eq:Ztildemixing} covers all the possible mixing in a massless EFT. In a massive EFT, one would obtain series in $m^2/\Lambda^2$ for each entry, thereby allowing for mixing down in mass dimension (the sum over partitions should then be replaced by $n_1+..+n_r\geq m(n,r)$). In this paper we only focus on terms of linear order in the EFT expansion, that is, we only consider $r=1$ in the above.

Although the off-shell basis of operators forms an independent basis when inserted into off-shell Green's functions, it is well known that they are not independent when inserted into S-matrix elements. Operators can be related by field redefinitions (FR) of the form
\beq \label{FR}
\phi
\, \stackrel{\textsc{fr\hspace{0.5mm}}}{\longrightarrow} \, \phi+F(\Lambda,\p,\d_\mu\p,\d_\mu\d_\nu\p,...)\,,
\eeq
with $F\sim O(1/\Lambda)$ a polynomial (local) function of the fields and their derivatives multiplied by appropriate powers of $1/\Lambda$ such that the mass dimension of $F$ is equal to that of $\p$ in $D=4$ spacetime dimensions. See for instance \cite{Criado} for a detailed discussion on field redefinitions in EFT and \cite{Balduf2020} for a recent mathematical perspective. To ensure that $F$ carries the correct factors that preserve normalization of bare operators in \eqref{bareoperators} it will be convenient to also include the power of $g^b \, Z_2$ that gives $F$ the mass dimension of $\p$ in $D=4 \minus 2\eps$. We will discuss examples of field redefinitions in section \ref{s:excalc}.

The FR can be used to reduce the off-shell basis,
\beq \label{eq:physical lagr}
\L
\, \stackrel{\textsc{fr\hspace{0.5mm}}}{\longrightarrow} \,
\L_{\text{phys}}= \L^{(4)}(\phi,\d_\mu\p)
\, + \sum_{n>4}
\frac{1}{\Lambda^{n\text{\,-\,}4}} \,
\sum_i 
c_i^{b(n)} \, \O_i^{b(n)} \, ,
\eeq
where the second sum runs over a subset of IBP-independent operators at each mass dimension, which we call the physical basis (no tildes). The FR leave the S-matrix invariant, although not the Green's functions. 
 The physical couplings $c_i$ and their renormalization constants $Z_i$ are linear combinations of the non-physical quantities $\tilde c_i$ and $\tilde Z_i$. 

Similarly to eq.\ (\ref{eq:Ztildemixing}), the bare couplings $c_i^b$ can be expanded as power series in the renormalized physical couplings, (where we now restrict to linear order)
\beq
\label{eq:Zmixing}
c_i^{b(n)}=\sum_j Z_{ij}^{(n)} c_j^{(n)}
\,.
\eeq
The anomalous dimension matrix (ADM) $\gamma^{(n)}_{ij}$ of the couplings $c_i^{(n)}$ are defined as: 
\beq \label{eq:gammadef}
\mu\frac{d }{d\mu}c_{i}^{(n)}=\sum_j \gamma^{(n)}_{ij}c_{j}^{(n)}
\eeq
 It can be derived from the renormalization constants using
\beq
\label{eq:bareRGE}
\mu\frac{d}{d\mu}c_i^{b(n)}=0\,,
\eeq
which gives 
\begin{align}\label{iterative}
Z_{ij}^{(n)}\ \mu\frac{d}{d\mu}c_j^{(n)}
&=-
\mu\frac{d Z_{ij}^{(n)}}{d \mu} c_j^{(n)}
 \, .
\end{align}
The ADM that encodes the mixing between operators of the same mass dimension is then determined to be
\beq \label{ADlowesdim}
\gamma_{ij}^{(n)}=-\beta(g,\eps)\big(Z_{ij}^{(n)}\big)^{\text{-} \scaleto{1}{5pt}} 
\, \frac{d Z_{jk}^{(n)}}{d g} \,,
\eeq
where 
\beq
\beta(g,\eps) \equiv \mu \frac{dg}{d \mu} \, ,
\eeq
is the $\beta$-function.

\subsection{Field Redefinitions, EoM operators and the ADM}
\label{sec:EOMandFR}
The full effect of a field redefinition \eqref{FR} on the level of the action is
\beq
  I[\p] 
  \, \stackrel{\textsc{fr\hspace{0.5mm}}}{\longrightarrow} \,
  I[\p] + \,
  \spacetimeInt
  F(\Lambda,\p,\d_\mu\p,\d_\mu\d_\nu\p,...) \frac{\delta I[\p]}{\delta \p} + O(F^2) \, .
\eeq
This generally has a complicated form because all terms up to the relevant order in $1/\Lambda$ need to be considered, which may include terms of order $O(F^2)$. However, when one restricts to analysing the mixing matrix of the couplings within one mass dimension, 
given by $\gamma^{(n)}_{ij}$ in \eqref{ADlowesdim}, 
the relevant change in the action is
\beq \label{eq:lowFR}
  I[\p] 
  \, \stackrel{\textsc{fr\hspace{0.5mm}}}{\longrightarrow} \,
  I[\p] + \,
\spacetimeInt
   \frac{1}{\Lambda^{n\text{\,-\,}4}}
  F^{(n)}(\p,\d_\mu\p,\d_\mu\d_\nu\p,...) \frac{\delta I^{(4)}[\p]}{\delta \p} + O(1/\Lambda^{n\text{\,-\,}2},F) \, ,
\eeq
for a field redefinition 
\beq \label{FR-oneDim}
\phi\to\phi+
 \frac{1}{\Lambda^{n\text{\,-\,}4}}
F^{(n)}(\p,\d_\mu\p,\d_\mu\d_\nu\p,..)\,,
\eeq
where $F^{(n)}$ has mass dimension $n{-}3$ (in $D=4$). 
We observe the appearance of the classical equations of motion at mass dimension 4, 
\beq \label{eq:ClassicalEoM}
E^{(4)} \coloneqq \frac{\delta I^{(4)}[\p]}{\delta\phi} \, .
\eeq
This implies that the addition of any operator 
  $$\E^{b(n)} = 
   \frac{1}{\Lambda^{n\text{\,-\,}4}}
  E^{(4)} F^{(n)}(\p,\d_\mu\p,\d_\mu\d_\nu\p,..)$$
to the Lagrangian leaves $\gamma^{(n)}_{ij}$ unchanged, even when $F^{(n)}$ is parametrised by the couplings $c^{(n)}_i\,$.%
\footnote{Note that the addition of $\E$ does change the theory if the EFT is defined up to $1/\Lambda^m$ for $m>n$, since it corresponds to only part of the full field redefinition.} 
Such operators $\E$ will be called EoM operators.
Field redefinitions \eqref{FR-oneDim} can therefore be accounted for at lowest order by splitting the off-shell basis into the physical basis and an \emph{unphysical} part spanned by the EoM operators,
\beq
\L^{(n)}=\L^{(n)}_{\text{phys}}+\L^{(n)}_{\text{EoM}}
\eeq
with $\L^{(n)}_{\text{phys}}$ defined in \eqref{eq:physical lagr}, and
\beq
 \L^{(n)}_{\text{EoM}}=\sum_{i}\hat c_i^b \E_i^{b(n)}\,.
\eeq
We denote by $\hat c_i^b$ the bare couplings corresponding to EoM operators. 
The coefficients of EoM operators can then freely be changed to rewrite the Lagrangian without affecting the part of the anomalous dimensions of interest.
As is well known~\cite{PhysRevD.12.3159,Joglekar:1975nu}, the EoM operators do not mix into the physical operators, thus allowing them to be dropped from the calculation (see below).

Organizing the Lagrangian in this way is convenient because, as noted before, only the full $\mathcal{L}$ contains all independent operators when inserted in Green's functions (the off-shell basis). 
On the other hand, only $\L_{\text{phys}}$ contains the physical operators and couplings of interest for the S-matrix. When the non-physical part of the Lagrangian (which only produces counterterm vertices) is spanned by the EoM operators, the necessary field redefinition to reduce the off-shell basis as in \eqref{eq:physical lagr} becomes trivial. 

In summary, one typically calculates the divergences of off-shell Green's functions using generic vertices of a physical operator basis and counterterm vertices of an off-shell basis. 
There exist two equivalent ways to compute the \emph{physical} anomalous dimension matrix $\gamma_{ij}^{(n)}$:
\begin{enumerate}
 \item[(i)] Project the counterterms onto an arbitrary off-shell basis and then perform a field redefinition to obtain the information in the desired physical basis.
 \item[(ii)] Project the counterterms onto the basis which contains the physical basis as well as a maximal set of independent EoM operators $\E_i$. The EoM operators can then simply be dropped. 
\end{enumerate}
We will exemplify both of these methods in some detail below in section \ref{s:excalc}.  Note that beyond linear order in the EFT operators, only the approach (i) above is valid, as higher order effects of the FR must be included.

\subsection{Changing between physical operator bases} \label{s:alpha}

Let us now observe the effect of changing the physical basis of operators on the anomalous dimension matrices, \textit{i.e.}~the basis dependence of $\gamma^{(n)}_{ij}$.
Since we are only concerned with the mixing between operators of the same mass dimension, it is convenient to split 
the Lagrangian into physical and non-physical parts; the latter being spanned by the EoM operators. A change of basis can be achieved by a general linear transformation of the coupling constants. Let us denote the original coupling constants by the vector:
\beq
\vec{C}=(c_1,...,c_{k},\hat c_1,..,\hat c_{m}) 
\eeq
where we suppress for the moment the dimension-superscript ($n$). The transformed coupling vector $\vec{C}'$ is then given by 
\beq
\vec{C}'=A^{-1}\vec{C}\,,
\eeq
with $A$ a general real valued invertible $(k+m)\times (k+m)$ matrix. 
However, we know that the unphysical couplings, as well as their renormalization constants, are irrelevant to the physical part, so we are less interested in that part of the matrix. The relevant physical information of the matrix $A$ is instead captured by writing it in block form as
\beq
\begin{pmatrix}
\vec{c} \;\\
\vec{\hat c}\;  
\end{pmatrix}
=
\begin{pmatrix}
B & 0\\
R & I
\end{pmatrix}
\begin{pmatrix}
\vec{c}\,' \\
\vec{\hat c}\, '  
\end{pmatrix}
\eeq
Now $B$ is a $k\times k$ matrix while $R$ is $k\times m$. Let us now study how such a transformation acts on the Lagrangian. 
While the unphysical part is left invariant, due to the identity matrix $I$, the physical part changes as follows:
\beq
\L_{\text{phys}}'=\sum_{i,j}\O_i^b B_{ij} (c')_{j}^{b}  +\sum_{i,j}\E_i^b R_{ij} (c')_j^b 
\eeq
which we can rewrite as
\beq
\L_{\text{phys}}'=\sum_{i}(c')_{i}^b \O_i'\,,\qquad \O_i'=\sum_j  \O_j^b B_{ji} +\sum_{j}\E_j^b R_{ji} \,.
\eeq
where we have identified the new basis of physical operators $\O_i'$. The matrix $B$ therefore allows us to transform among the physical operators.
In addition, we can add or subtract EoM operators via the matrix $R$, to obtain operators that were not present in the initial physical basis. Note that only the matrix $B$ enters the transformed ADM $\gamma_{ij}'$ of the new basis. If
\beq
c_i^b=B_{ij}{c_j^{b}}'\,,
\eeq
then we also have that the renormalized couplings are related by the same transformation (this follows from the finiteness of $B_{ij}$)
\beq
c_i=B_{ij}c_j'\,,\qquad Z_{ij}=B_{ik}Z_{kl}'B^{-1}_{lj}\,.
\eeq
Due to the conventional rescaling of operators by $g^b$ (see eq.\ \eqref{bareoperators}), $B$ does not carry any scale dependence through factors of $g(\mu)$. Therefore, the anomalous dimension matrix transforms as
\beq
\gamma_{ij}=B_{ik}\gamma_{kl}'B^{-1}_{lj}
\eeq
and is independent of $R$. In other words, it does not ``know'' that the redefined operators $\O'_i$ have received contributions from the EoM operators. The anomalous dimensions only change when a new basis is generated by taking linear combinations of the original couplings; the application of field redefinitions and IBP themselves do not affect the anomalous dimensions.

From this argument, it also follows directly that a change in the operator content of the form
$\partial^2\phi\, \O_{\scaleto{N}{4.5pt} \text{-}\scaleto{1}{4.5pt}}$ (after IBP) does not affect the part of the mixing matrix that describes the mixing of $N$-point operators at low loop order.%
\footnote{For example, consider the change of basis
$$
c\,(\O_{\scaleto{N}{4.5pt}}+\partial^2\!\phi\,\O_{\scaleto{N}{4.5pt}\text{-}\scaleto{1}{4.5pt}}') + ...
\rightarrow 
c\,\O_{\scaleto{N}{4.5pt}} + ...\,,
$$ which leaves the coupling $c$ and therefore its anomalous dimension unchanged}
Instead, since such operators can be related to an ($N{+}2$)-point operator using the EoM, such a change of basis is effectively achieved by a rotation with the physical couplings of $N+2$ number of fields. At low loop order, the operators with $N{+}2$ fields do not mix into operators with $N$ fields, see the results below or ref.\ \cite{Bern2020}, so this rotation does not affect the submatrix that describes mixing between $N$-point operators.

\section{Operator bases}\label{sec:operatorbasis}
  
\begin{figure}
  \centering
  \includegraphics[width=11cm]{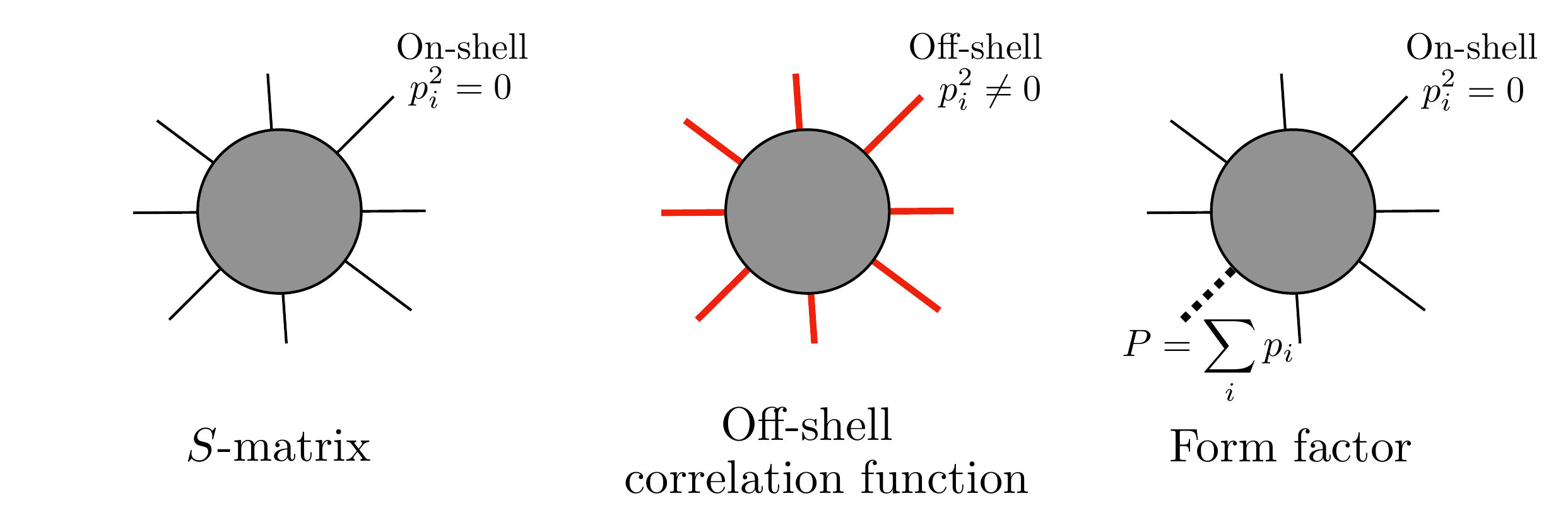}
  \caption{The three types of objects in QFT calculations for which we discuss kinematic bases for: $S$-matrices, off-shell correlation functions, and form factors.   \label{fig:qftobjects}}
\end{figure}

We now turn to discussing operator bases for various objects of interest in quantum field theory, namely $S$-matrices, off-shell correlation functions, and form factors, see Fig.~\ref{fig:qftobjects}. In Sec.~\ref{sec:rings} we discuss the construction of the kinematic bases for these three types of objects under the different kinematic constraints they require, and how the kinematic bases can be used to identify operator bases. We then define a {\it multigraph operator basis}, which is a very simply obtained basis that we show is non-redundant. The multigraph operator basis can be used for both the calculation of off-shell correlation functions and form factors.\footnote{That is, until evanescent operators become important, see Sec.~\ref{sec:hs}. However, in the following we do not obtain results at high enough mass dimension to probe this issue.} In Sec.~\ref{sec:hs} we detail the Hilbert series that can be constructed for each type of object, we discuss the effect of evanescent operators at high mass dimension, and explain how Hilbert series can be used to count such operators. In Sec.~\ref{sec:prim} we detail the construction of primary operators in the theories we consider, for use in the study of the anomalous dimension matrix in the below.

\subsection{$S$-matrices, correlation functions, and form factors}
\label{sec:rings}

We begin by considering the three objects of interest---$N$-particle $S$-matrices, off-shell correlation functions, and form factors---in momentum space. In what follows we only focus on the piece that is polynomial in $N$ momenta, $p_i^\mu$. Since all objects are Lorentz covariant, the spacetime indices of the momenta must be contracted; in restricting to the spacetime parity even sector, we will deal with polynomials in invariants $s_{ii}\equiv p_i^2$ and $s_{ij}\equiv p_i\cdot p_j$.\footnote{Note our convention for $s_{ij}$ does not include a factor of two, in contrast to the typical the definition of Mandelstam variables.} Further, since we are dealing with indistinguishable particles, the form of the polynomials must be invariant under the relevant permutation group. 
The three objects have different constraints which must be satisfied:
\begin{eqnarray}
  S\text{-matrix:}&~&\text{on-shell, momenta sum to zero}\,, \nonumber \\
  \text{Off-shell correlation function (C.F.):}&~&\text{off-shell, momenta sum to zero} \,, \nonumber \\
  \text{Form factor (F.F.):}&~&\text{on-shell, momenta sum to }P\,, \nonumber
\end{eqnarray}
where $P$ is some inflow momentum to the form factor. 

We use notation from commutative algebra, following \cite{Henning:2017fpj}. The polynomial rings of interest are
\begin{eqnarray}
  M_{\text{real}}^N &=& \left[\mathbb{C}[ \{s_{ii},s_{ij}\}] / I \right]^{S_N} \,, \\
  M_{\text{complex}}^N &=& \left[\mathbb{C}[ \{s_{ii},s_{ij}\}] / I \right]^{S_{N/2}\times S_{N/2}} \,, 
\end{eqnarray}
where the superscript means invariance under $S_N$ ($S_{N/2}\times S_{N/2}$) and is imposed to account for the indistinguishability. For the complex scalar, we use the fact that there must be an equal number of external $\phi$ and $\phi^\dagger$ fields to conserve charge; invariance under $S_{N/2}\times S_{N/2}$ means invariance under permutations that exchange the $N/2$ momenta corresponding to $\phi$ within themselves, and also for the $N/2$ momenta corresponding to $\phi^\dagger$. In the above, $I$ denotes an ideal that implements the relevant constraints:
\begin{eqnarray}
  I_{S\text{-matrix}} &=& \langle \{s_{ii}\}, \{X_i\} , \{ G\} \rangle \,, \\
  I_{\text{C.F.}} &=& \langle \{X_i\}, \{G\} \rangle \,, \\
  I_{\text{F.F.}} &=& \langle \{s_{ii}\} , \{G\} \rangle \,,
\end{eqnarray}
where we define $X_i=p_{i\,\mu}\cdot\sum_jp_j^\mu=\sum_{j}s_{ij}$, and its appearance in the ideal implements momentum conservation in the ring, $X_i=0$. Similarly $s_{ii}=0$ implements on-shell conditions. Note that momentum conservation is not imposed for the form factor, since in this case the momenta sum to the inflow momentum $P=\sum_i p_i$, and we wish to keep track of instances of this total momentum. The conditions $\{G\}$ that appear in all of the above ideals impose constraints that arise from the finite dimensionality of spacetime---the fact that only $D$ vectors can be linearly independent in $D$ dimensions. These `Gram', or rank, conditions are specified by requiring that all $(D{+}1)\times (D{+}1)$ minors of the Gram matrix,
\begin{equation}
  \begin{pmatrix} 
  s_{11} & s_{12} & \dots & s_{1N}\\
  s_{21} & s_{22} & \dots & s_{2N}\\
  \vdots &\vdots & \ddots & \vdots \\
  s_{N1} & s_{N2} &\dots & s_{NN} 
  \end{pmatrix} \,,
  \label{eq:gram}
\end{equation}
 vanish. The set of all such constraints is what is denoted $\{ G\}$.
 
 If $D$ is taken to be large enough, no such relations between the $s_{ij}$ exist; the fact that in lower spacetime dimensions certain polynomials in $s_{ij}$ are redundant is tied to the existence of a class of evanescent operators. We will return to discuss these further in the following subsection. 

The above polynomial rings capture the possible allowed polynomial forms of the three types of objects in Fig.~\ref{fig:qftobjects}. A detailed study of the $S$-matrix case was given in \cite{Henning:2017fpj}, and we will not focus on it further in the present work. Instead, we turn to the rings relevant for off-shell correlation functions---the objects we will work with in the next section---and form factors. 

We analyse the rings assuming $D$ is large enough, such that we can safely ignore Gram conditions (consistent with our use of dimensional regularisation). In this case, there exists a ring isomorphism between the two rings. That is, with $\Sigma = S_N \ (S_{N/2} \times S_{N/2})$ for the real (complex) scalar,
\begin{equation}
  \left[\mathbb{C}[ \{s_{ii},s_{ij}\}] / \langle \{X_i\} \rangle \right]^{\Sigma} \simeq \left[\mathbb{C}[ \{s_{ii},s_{ij}\}] / \langle \{s_{ii}\} \rangle \right]^{\Sigma} = \left[\mathbb{C}[ \{s_{ij}\}]\right]^{\Sigma} \,.
  \label{eq:multring}
\end{equation}
In the case of the form factor one removes $s_{ii}$ from the ring by setting all instances to zero, using its appearance in the ideal; for the off-shell correlation function one may use $X_i=s_{ii} + \sum_{j\ne i} s_{ij}=0$ to eliminate $s_{ii}$ in favour of $s_{ij}$ with $j\ne i$. This motivates that the two constraints result in isomorphic rings, and we provide a rigorous treatment in Appendix~\ref{s:isoapp}.

There is a useful graphical way to represent elements of the ring in Eq.~\eqref{eq:multring}. One considers a graph with $N$ vertices, and edges connecting vertices which correspond to a factor of $s_{ij}$. Note the absence of $s_{ii}$ excludes self-loops in the graph. More than one edge can connect the same two vertices; for example a graph that has two edges connecting the same two vertices corresponds to a factor of $s_{ij}^2$. Permutation symmetry is applied to the monomial in the $s_{ij}$ one obtains from a particular graph. For the complex scalar, two colors of vertices are needed so as to keep track of momenta associated to $\phi$ and $\phi^\dagger$ separately. The graphs described above are called (colored) multigraphs.\footnote{We take the definition of multigraph to mean no self-loops.} All non-isomorphic multigraphs constitute a basis for the ring Eq.~\eqref{eq:multring}. 

For example, for the real scalar, the non-isomorphic graphs with four vertices and two edges, and their corresponding ring elements are: 
\begin{align}
\nonumber
%
%
%
\begin{gathered}
\begin{tikzpicture}
\begin{feynman}[small, baseline=g1]
%
%
\tikzfeynmanset{every vertex={dot,black,minimum size=1mm}}
\vertex  (g1);
\vertex [right =0.3cm of g1] (g2) ;
\vertex [below =0.3cm of g1] (g3) ;
\vertex [right =0.3cm of g3] (g4) ;
\tikzfeynmanset{every vertex={dot,black,minimum size=0mm}}
\vertex [above =0.15cm of g1];
\diagram* {
	(g1) -- [out=30,in=150] (g2) -- [out=-150,in=-30] (g1),
};
\end{feynman}
\end{tikzpicture}
\end{gathered}
=4\, \mathcal{P}_4\left(s_{12}^2\right) \,,~~~
%
%
\begin{gathered}
\begin{tikzpicture}
\begin{feynman}[small, baseline=g1]
%
%
\tikzfeynmanset{every vertex={dot,black,minimum size=1mm}}
\vertex  (g1);
\vertex [right =0.3cm of g1] (g2) ;
\vertex [below =0.3cm of g1] (g3) ;
\vertex [right =0.3cm of g3] (g4) ;
\tikzfeynmanset{every vertex={dot,black,minimum size=0mm}}
\vertex [above =0.15cm of g1];
\diagram* {
	(g1) -- [] (g2),
	(g1) -- [] (g3)
};
\end{feynman}
\end{tikzpicture}
\end{gathered}
=2\, \mathcal{P}_4\left(s_{12}s_{13}\right) \,,~~~  
%
%
%
\begin{gathered}
\begin{tikzpicture}
\begin{feynman}[small, baseline=g1]
\tikzfeynmanset{every vertex={dot,black,minimum size=1mm}}
\vertex  (g1);
\vertex [right =0.3cm of g1] (g2) ;
\vertex [below =0.3cm of g1] (g3) ;
\vertex [right =0.3cm of g3] (g4) ;
\tikzfeynmanset{every vertex={dot,black,minimum size=0mm}}
\vertex [above =0.15cm of g1];
\diagram* {
	(g1) -- [] (g2),
	(g3) -- [] (g4)
};
\end{feynman}
\end{tikzpicture}
\end{gathered}
=8\, \mathcal{P}_4\left(s_{12}s_{34}\right) \, .
\end{align}	
where we define the permutation such that it runs over non-identical contributions (\textit{i.e.}~any overall numerical factors are removed), \textit{e.g.}
	\begin{align*}
		\mathcal{P}_4(s_{12}^2) &= s_{12}^2+s_{13}^2+s_{14}^2+s_{23}^2+s_{24}^2+s_{34}^2 \, .
	\end{align*}
	
\noindent
As another example, for the complex scalar, the non-isomorphic graphs with four vertices and two edges, and their corresponding ring elements are: 
\begin{alignat*}{4}
%
			&\begin{gathered}
			\begin{tikzpicture}	
			\begin{feynman}[small, baseline=g1]
					\tikzfeynmanset{every vertex={dot,black,minimum size=1mm}}
				\vertex  (l1);
				\vertex [below =0.3cm of l1] (l2);
					\tikzfeynmanset{every vertex={empty dot,black,minimum size=1mm}}
				\vertex [right =0.3cm of l1] (r1) ;
				\vertex [below =0.3cm of r1] (r2);
\tikzfeynmanset{every vertex={dot,black,minimum size=0mm}}
\vertex [above =0.15cm of g1];
				\diagram* {
					(l1) -- [out=30,in=150] (r1) -- [out=-150,in=-30] (l1)
				};	
			\end{feynman}
			\end{tikzpicture}
			\end{gathered} =\mathcal{P}_{2\bar{2}}\left(s_{1\bar{1}}^2\right)\,, 
%
%
			&&\begin{gathered}
			\begin{tikzpicture}	
			\begin{feynman}[small, baseline=g1]
					\tikzfeynmanset{every vertex={dot,black,minimum size=1mm}}
				\vertex  (l1);
				\vertex [below =0.3cm of l1] (l2);
					\tikzfeynmanset{every vertex={empty dot,black,minimum size=1mm}}
				\vertex [right =0.3cm of l1] (r1) ;
				\vertex [below =0.3cm of r1] (r2);
\tikzfeynmanset{every vertex={dot,black,minimum size=0mm}}
\vertex [above =0.15cm of l1];
				\diagram* {
					(l1) -- [out=-60,in=60] (l2) -- [out=120,in=-120] (l1)
				};	
			\end{feynman}
			\end{tikzpicture}
			\end{gathered} =4\, \mathcal{P}_{2\bar{2}}\left(s_{12}^2\right)\,, 
%
%
%
			&&\begin{gathered}\begin{tikzpicture} \begin{feynman}[small, baseline=g1]
					\tikzfeynmanset{every vertex={dot,black,minimum size=1mm}}
				\vertex  (l1);
				\vertex [below =0.3cm of l1] (l2);
					\tikzfeynmanset{every vertex={empty dot,black,minimum size=1mm}}
				\vertex [right =0.3cm of l1] (r1) ;
				\vertex [below =0.3cm of r1] (r2);
\tikzfeynmanset{every vertex={dot,black,minimum size=0mm}}
\vertex [above =0.15cm of l1];
				\diagram* {
					(r1) -- [out=-60,in=60] (r2) -- [out=120,in=-120] (r1)
				};	
			\end{feynman}\end{tikzpicture}\end{gathered}=4\, \mathcal{P}_{2\bar{2}}\left(s_{\bar{1}\bar{2}}^2\right)\,, \\
%
%
%
			&\begin{gathered}
			\begin{tikzpicture}	
			\begin{feynman}[small, baseline=g1]
					\tikzfeynmanset{every vertex={dot,black,minimum size=1mm}}
				\vertex  (l1);
				\vertex [below =0.3cm of l1] (l2);
					\tikzfeynmanset{every vertex={empty dot,black,minimum size=1mm}}
				\vertex [right =0.3cm of l1] (r1) ;
				\vertex [below =0.3cm of r1] (r2);
\tikzfeynmanset{every vertex={dot,black,minimum size=0mm}}
\vertex [above =0.15cm of l1];
				\diagram* {
					(l1) -- [] (r1),
					(l1) -- [] (l2)
				};	
			\end{feynman}
			\end{tikzpicture}
			\end{gathered}= \mathcal{P}_{2\bar{2}}\left(s_{12}s_{1\bar{1}}\right)\,,  
%
%
%
%
%
			&&\begin{gathered}\begin{tikzpicture} \begin{feynman}[small, baseline=g1]
					\tikzfeynmanset{every vertex={dot,black,minimum size=1mm}}
				\vertex  (l1);
				\vertex [below =0.3cm of l1] (l2);
					\tikzfeynmanset{every vertex={empty dot,black,minimum size=1mm}}
				\vertex [right =0.3cm of l1] (r1) ;
				\vertex [below =0.3cm of r1] (r2);
\tikzfeynmanset{every vertex={dot,black,minimum size=0mm}}
\vertex [above =0.15cm of l1];
				\diagram* {
					(l1) -- [] (r1),
					(r1) -- [] (r2)
				};	
			\end{feynman}\end{tikzpicture}\end{gathered}= \mathcal{P}_{2\bar{2}}\left(s_{1\bar{1}}s_{\bar{1}\bar{2}}\right)\,, 
%
%
			&&\begin{gathered}\begin{tikzpicture} \begin{feynman}[small, baseline=g1]
					\tikzfeynmanset{every vertex={dot,black,minimum size=1mm}}
				\vertex  (l1);
				\vertex [below =0.3cm of l1] (l2);
					\tikzfeynmanset{every vertex={empty dot,black,minimum size=1mm}}
				\vertex [right =0.3cm of l1] (r1) ;
				\vertex [below =0.3cm of r1] (r2);
\tikzfeynmanset{every vertex={dot,black,minimum size=0mm}}
\vertex [above =0.15cm of l1];
				\diagram* {
					(l1) -- [] (r1),
					(l1) -- [] (r2),
				};	
			\end{feynman}\end{tikzpicture}\end{gathered}=2\, \mathcal{P}_{2\bar{2}}\left(s_{1\bar{1}}s_{1\bar{2}}\right)\,, 
%
\\ 
			&\begin{gathered}\begin{tikzpicture} \begin{feynman}[small, baseline=g1]
					\tikzfeynmanset{every vertex={dot,black,minimum size=1mm}}
				\vertex  (l1);
				\vertex [below =0.3cm of l1] (l2);
					\tikzfeynmanset{every vertex={empty dot,black,minimum size=1mm}}
				\vertex [right =0.3cm of l1] (r1) ;
				\vertex [below =0.3cm of r1] (r2);
\tikzfeynmanset{every vertex={dot,black,minimum size=0mm}}
\vertex [above =0.15cm of l1];
				\diagram* {
					(r1) -- [] (l1),
					(r1) -- [] (l2),
				};	
			\end{feynman}\end{tikzpicture}\end{gathered}=2\, \mathcal{P}_{2\bar{2}}\left(s_{1\bar{1}}s_{2\bar{1}}\right)\,,\quad 
%
%
			&&\begin{gathered}\begin{tikzpicture} \begin{feynman}[small, baseline=g1]
					\tikzfeynmanset{every vertex={dot,black,minimum size=1mm}}
				\vertex  (l1);
				\vertex [below =0.3cm of l1] (l2);
					\tikzfeynmanset{every vertex={empty dot,black,minimum size=1mm}}
				\vertex [right =0.3cm of l1] (r1) ;
				\vertex [below =0.3cm of r1] (r2);
\tikzfeynmanset{every vertex={dot,black,minimum size=0mm}}
\vertex [above =0.15cm of l1];
				\diagram* {
					(l1) -- [] (l2),
					(r1) -- [] (r2),
				};	
			\end{feynman}\end{tikzpicture}\end{gathered}=4\, \mathcal{P}_{2\bar{2}}\left(s_{12}s_{\bar{1}\bar{2}}\right)\,,\quad 
%
%
			&&\begin{gathered}\begin{tikzpicture} \begin{feynman}[small, baseline=g1]
					\tikzfeynmanset{every vertex={dot,black,minimum size=1mm}}
				\vertex  (l1);
				\vertex [below =0.3cm of l1] (l2);
					\tikzfeynmanset{every vertex={empty dot,black,minimum size=1mm}}
				\vertex [right =0.3cm of l1] (r1) ;
				\vertex [below =0.3cm of r1] (r2);
\tikzfeynmanset{every vertex={dot,black,minimum size=0mm}}
\vertex [above =0.15cm of l1];
				\diagram* {
					(l1) -- [] (r1),
					(l2) -- [] (r2),
				};	
			\end{feynman}\end{tikzpicture}\end{gathered}=2\, \mathcal{P}_{2\bar{2}}\left(s_{1\bar{1}}s_{2\bar{2}}\right)\,. 
			\nonumber
%
\end{alignat*}
In this case we similarly define the permutation over non-identical contributions, while taking into account that the momenta of $\phi$ ($1,2,...$) and $\phi^\dagger$ ($\bar{1},\bar{2},...$) have to be permuted separately, as they arise from distinguishable fields. For example,
 $$\mathcal{P}_{2\bar{2}}\left(s_{1\bar{1}}^2\right) = s_{1\bar{1}}^2 + s_{1\bar{2}}^2+s_{2\bar{1}}^2+s_{2\bar{2}}^2 \ .$$

From this point, it is straightforward to identify a basis of operators for $N$-point correlation functions of form factors. The logic is the following. The above ring captures all possible polynomial contributions to the objects. We can consider all such polynomial contributions as coming from contact terms 
\textit{ i.e.}~from operators involving $N$ fields. Each element of the ring Eq.~\eqref{eq:multring} can be interpreted as the Feynman rule for a particular $N$-point operator; the operator is unambiguously re-constructed from this. In the real scalar example above, the operators are reconstructed as 
\begin{align}
\nonumber
%
%
%
\begin{gathered}
\begin{tikzpicture}
\begin{feynman}[small, baseline=g1]
%
%
\tikzfeynmanset{every vertex={dot,black,minimum size=1mm}}
\vertex  (g1);
\vertex [right =0.3cm of g1] (g2) ;
\vertex [below =0.3cm of g1] (g3) ;
\vertex [right =0.3cm of g3] (g4) ;
\tikzfeynmanset{every vertex={dot,black,minimum size=0mm}}
\vertex [above =0.15cm of g1];
\diagram* {
	(g1) -- [out=30,in=150] (g2) -- [out=-150,in=-30] (g1),
};
\end{feynman}
\end{tikzpicture}
\end{gathered}
=    \phi^2 \, \partial_\mu \partial_\nu \phi \, \partial^\mu \partial^\nu  \phi \,,~~~
%
%
\begin{gathered}
\begin{tikzpicture}
\begin{feynman}[small, baseline=g1]
%
%
\tikzfeynmanset{every vertex={dot,black,minimum size=1mm}}
\vertex  (g1);
\vertex [right =0.3cm of g1] (g2) ;
\vertex [below =0.3cm of g1] (g3) ;
\vertex [right =0.3cm of g3] (g4) ;
\tikzfeynmanset{every vertex={dot,black,minimum size=0mm}}
\vertex [above =0.15cm of g1];
\diagram* {
	(g1) -- [] (g2),
	(g1) -- [] (g3)
};
\end{feynman}
\end{tikzpicture}
\end{gathered}
=\phi \,  \partial_\mu \phi \, \partial_\nu \phi \, \partial^\mu \partial^\nu  \phi \,,~~~  
%
%
%
\begin{gathered}
\begin{tikzpicture}
\begin{feynman}[small, baseline=g1]
\tikzfeynmanset{every vertex={dot,black,minimum size=1mm}}
\vertex  (g1);
\vertex [right =0.3cm of g1] (g2) ;
\vertex [below =0.3cm of g1] (g3) ;
\vertex [right =0.3cm of g3] (g4) ;
\tikzfeynmanset{every vertex={dot,black,minimum size=0mm}}
\vertex [above =0.15cm of g1];
\diagram* {
	(g1) -- [] (g2),
	(g3) -- [] (g4)
};
\end{feynman}
\end{tikzpicture}
\end{gathered}
=\partial_\mu \phi \,\partial^\mu  \phi \,\partial_\nu \phi\, \partial^\nu  \phi \, .
\end{align}	
As can be seen, the graphical notation allows for a simple determination of the operators. Every vertex then corresponds to a field and the edges represent pairs of contracted derivatives, which act on the vertices (fields) they are attached to.
In the complex scalar example above, the operators are 
\begin{alignat*}{4}
%
			&\begin{gathered}
			\begin{tikzpicture}	
			\begin{feynman}[small, baseline=g1]
					\tikzfeynmanset{every vertex={dot,black,minimum size=1mm}}
				\vertex  (l1);
				\vertex [below =0.3cm of l1] (l2);
					\tikzfeynmanset{every vertex={empty dot,black,minimum size=1mm}}
				\vertex [right =0.3cm of l1] (r1) ;
				\vertex [below =0.3cm of r1] (r2);
\tikzfeynmanset{every vertex={dot,black,minimum size=0mm}}
\vertex [above =0.15cm of g1];
				\diagram* {
					(l1) -- [out=30,in=150] (r1) -- [out=-150,in=-30] (l1)
				};	
			\end{feynman}
			\end{tikzpicture}
			\end{gathered} =
			\phi^\dagger
			\phi \, 
			\partial_\mu \partial_\nu \phi^\dagger
			\partial^\mu \partial^\nu \phi
			\,, \quad
%
%
			&&\begin{gathered}
			\begin{tikzpicture}	
			\begin{feynman}[small, baseline=g1]
					\tikzfeynmanset{every vertex={dot,black,minimum size=1mm}}
				\vertex  (l1);
				\vertex [below =0.3cm of l1] (l2);
					\tikzfeynmanset{every vertex={empty dot,black,minimum size=1mm}}
				\vertex [right =0.3cm of l1] (r1) ;
				\vertex [below =0.3cm of r1] (r2);
\tikzfeynmanset{every vertex={dot,black,minimum size=0mm}}
\vertex [above =0.15cm of l1];
				\diagram* {
					(l1) -- [out=-60,in=60] (l2) -- [out=120,in=-120] (l1)
				};	
			\end{feynman}
			\end{tikzpicture}
			\end{gathered} =
				\phi^{\dagger2}
						\partial_\mu \partial_\nu \phi \,
						\partial^\mu \partial^\nu \phi
			\,, \quad
%
%
%
			&&\begin{gathered}\begin{tikzpicture} \begin{feynman}[small, baseline=g1]
					\tikzfeynmanset{every vertex={dot,black,minimum size=1mm}}
				\vertex  (l1);
				\vertex [below =0.3cm of l1] (l2);
					\tikzfeynmanset{every vertex={empty dot,black,minimum size=1mm}}
				\vertex [right =0.3cm of l1] (r1) ;
				\vertex [below =0.3cm of r1] (r2);
\tikzfeynmanset{every vertex={dot,black,minimum size=0mm}}
\vertex [above =0.15cm of l1];
				\diagram* {
					(r1) -- [out=-60,in=60] (r2) -- [out=120,in=-120] (r1)
				};	
			\end{feynman}\end{tikzpicture}\end{gathered}=
			\phi^2
				\partial_\mu \partial_\nu \phi^\dagger 
				\partial^\mu \partial^\nu \phi^\dagger
			\,, \\
%
%
%
			&\begin{gathered}
			\begin{tikzpicture}	
			\begin{feynman}[small, baseline=g1]
					\tikzfeynmanset{every vertex={dot,black,minimum size=1mm}}
				\vertex  (l1);
				\vertex [below =0.3cm of l1] (l2);
					\tikzfeynmanset{every vertex={empty dot,black,minimum size=1mm}}
				\vertex [right =0.3cm of l1] (r1) ;
				\vertex [below =0.3cm of r1] (r2);
\tikzfeynmanset{every vertex={dot,black,minimum size=0mm}}
\vertex [above =0.15cm of l1];
				\diagram* {
					(l1) -- [] (r1),
					(l1) -- [] (l2)
				};	
			\end{feynman}
			\end{tikzpicture}
			\end{gathered}=
			\phi^\dagger
			\partial_\mu \phi\,
			\partial_\nu \phi^\dagger 
			\partial^\mu\partial^{\nu}\phi
			\,,
%
%
%
%
%
			&&\begin{gathered}\begin{tikzpicture} \begin{feynman}[small, baseline=g1]
					\tikzfeynmanset{every vertex={dot,black,minimum size=1mm}}
				\vertex  (l1);
				\vertex [below =0.3cm of l1] (l2);
					\tikzfeynmanset{every vertex={empty dot,black,minimum size=1mm}}
				\vertex [right =0.3cm of l1] (r1) ;
				\vertex [below =0.3cm of r1] (r2);
\tikzfeynmanset{every vertex={dot,black,minimum size=0mm}}
\vertex [above =0.15cm of l1];
				\diagram* {
					(l1) -- [] (r1),
					(r1) -- [] (r2)
				};	
			\end{feynman}\end{tikzpicture}\end{gathered}= 
			\phi \,
			\partial_\mu \phi^\dagger 
			\partial_\nu \phi\,
			\partial^\mu\partial^{\nu}\phi^\dagger
			\,, \quad
%
%
			&&\begin{gathered}\begin{tikzpicture} \begin{feynman}[small, baseline=g1]
					\tikzfeynmanset{every vertex={dot,black,minimum size=1mm}}
				\vertex  (l1);
				\vertex [below =0.3cm of l1] (l2);
					\tikzfeynmanset{every vertex={empty dot,black,minimum size=1mm}}
				\vertex [right =0.3cm of l1] (r1) ;
				\vertex [below =0.3cm of r1] (r2);
\tikzfeynmanset{every vertex={dot,black,minimum size=0mm}}
\vertex [above =0.15cm of l1];
				\diagram* {
					(l1) -- [] (r1),
					(l1) -- [] (r2),
				};	
			\end{feynman}\end{tikzpicture}\end{gathered}=
			\phi \,
			\partial_\mu \phi^\dagger 
			\partial_\nu \phi^\dagger
			\partial^\mu\partial^{\nu}\phi
			\,, 
%
\\ 
			&\begin{gathered}\begin{tikzpicture} \begin{feynman}[small, baseline=g1]
					\tikzfeynmanset{every vertex={dot,black,minimum size=1mm}}
				\vertex  (l1);
				\vertex [below =0.3cm of l1] (l2);
					\tikzfeynmanset{every vertex={empty dot,black,minimum size=1mm}}
				\vertex [right =0.3cm of l1] (r1) ;
				\vertex [below =0.3cm of r1] (r2);
\tikzfeynmanset{every vertex={dot,black,minimum size=0mm}}
\vertex [above =0.15cm of l1];
				\diagram* {
					(r1) -- [] (l1),
					(r1) -- [] (l2),
				};	
			\end{feynman}\end{tikzpicture}\end{gathered}=
						\phi^\dagger 
						\partial_\mu \phi \,
						\partial_\nu \phi \,
						\partial^\mu\partial^{\nu}\phi^\dagger
			\,, 
%
%
			&&\begin{gathered}\begin{tikzpicture} \begin{feynman}[small, baseline=g1]
					\tikzfeynmanset{every vertex={dot,black,minimum size=1mm}}
				\vertex  (l1);
				\vertex [below =0.3cm of l1] (l2);
					\tikzfeynmanset{every vertex={empty dot,black,minimum size=1mm}}
				\vertex [right =0.3cm of l1] (r1) ;
				\vertex [below =0.3cm of r1] (r2);
\tikzfeynmanset{every vertex={dot,black,minimum size=0mm}}
\vertex [above =0.15cm of l1];
				\diagram* {
					(l1) -- [] (l2),
					(r1) -- [] (r2),
				};	
			\end{feynman}\end{tikzpicture}\end{gathered}=
			\partial_\mu \phi \,
			\partial^\mu \phi \,
			\partial_\nu \phi^\dagger 
			\partial^\nu \phi^\dagger
			\,, 
%
%
			&&\begin{gathered}\begin{tikzpicture} \begin{feynman}[small, baseline=g1]
					\tikzfeynmanset{every vertex={dot,black,minimum size=1mm}}
				\vertex  (l1);
				\vertex [below =0.3cm of l1] (l2);
					\tikzfeynmanset{every vertex={empty dot,black,minimum size=1mm}}
				\vertex [right =0.3cm of l1] (r1) ;
				\vertex [below =0.3cm of r1] (r2);
\tikzfeynmanset{every vertex={dot,black,minimum size=0mm}}
\vertex [above =0.15cm of l1];
				\diagram* {
					(l1) -- [] (r1),
					(l2) -- [] (r2),
				};	
			\end{feynman}\end{tikzpicture}\end{gathered}=
						\partial_\mu \phi^\dagger
						\partial^\mu \phi \,
						\partial_\nu \phi^\dagger 
						\partial^\nu \phi
			\,, 
			\nonumber
\end{alignat*}
where the different coloured vertices correspond to the different fields (in the below we will typically combine Hermitian conjugates $\mathcal{O}$ and $\mathcal{O}^\dagger$ into the combinations $\frac{1}{2}\left(\mathcal{O}+\mathcal{O}^\dagger\right)$ and $\frac{i}{2}\left(\mathcal{O}-\mathcal{O}^\dagger\right)$ so as to consider purely real Wilson coefficients).

The above logic---{\it i.e.} starting with the kinematic polynomial ring---proves the result that one can use a multigraph basis as a non-redundant basis for either off-shell correlation functions or form factors, so long as we work at low enough mass dimension (or high enough spacetime dimension) to avoid Gram constraints.

\subsubsection*{Example} It is instructive to work through a full example at fixed mass dimension; we consider the case of the real scalar EFT at mass dimension 6, and work out an operator basis for use in the calculation of an off-shell correlation function. For heuristic purposes, we follow the more familiar `route', that is first discussing the operators and the relations between them, and then work our way to the polynomials in kinematic variables we have been using above.  

Let us begin by simply writing down all possible non-total derivative operators we can think of:
\begin{align}
    \phi^6\,,
    ~~~\partial^2 \phi \, \phi^3\,,
    ~~~\partial_\mu \phi \, \partial^\mu \phi \, \phi^2\,,
    ~~~\partial^4  \phi \, \phi \,,
    ~~~\partial^2 \phi \, \partial^2 \phi \,,
    ~~~ \partial^\mu \partial^2  \phi \,\partial_\mu \phi\,,
    ~~~\partial^\mu\partial^\nu \phi \, \partial_\mu\partial_\nu \phi \,.
\end{align}
Note that we have not (yet) removed operators containing $\partial^2 \phi$ using a field redefinition since we wish to find a basis for calculating off-shell correlation functions, and we might wonder if such operators should then be kept (we will see they can indeed be removed).
The graphs corresponding to the above operators are:
\begin{equation*} 
\begin{gathered}
	\begin{tikzpicture}
	\begin{feynman}[small, baseline=g4]
	\tikzfeynmanset{every vertex={dot,black,minimum size=1mm}}
	\vertex  (g1);
	\vertex [right =0.3cm of g1] (g2) ;
	\vertex [below =0.3cm of g1] (g3) ;
	\vertex [right =0.3cm of g3] (g4) ;
	\vertex [below =0.3cm of g3] (g5) ;
	\vertex [right =0.3cm of g5] (g6) ;
	\tikzfeynmanset{every vertex={dot,black,minimum size=0mm}}
	\vertex [above =0.3cm of g1] ;
	\end{feynman}
	\end{tikzpicture}
\end{gathered}\ , \hspace{3mm}
\begin{gathered}
	\begin{tikzpicture}
	\begin{feynman}[small, baseline=g4]
	\tikzfeynmanset{every vertex={dot,black,minimum size=1mm}}
	\vertex  (g1);
	\vertex [right =0.3cm of g1] (g2) ;
	\vertex [below =0.3cm of g1] (g3) ;
	\vertex [right =0.3cm of g3] (g4) ;
	\tikzfeynmanset{every vertex={dot,black,minimum size=0mm}}
	\vertex [above =0.3cm of g1] ;	
	\tikzfeynmanset{every vertex={dot,black,minimum size=0mm}}
	\vertex [above =2.5mm of g1] (loop1);
	\diagram* {
		(g1) -- [out=45,in=0] (loop1) -- [out=180,in=135] (g1)
	};
	\end{feynman}
	\end{tikzpicture}
\end{gathered}\ , \hspace{3mm}
\begin{gathered}
	\begin{tikzpicture}
	\begin{feynman}[small, baseline=g4]
	\tikzfeynmanset{every vertex={dot,black,minimum size=1mm}}
	\vertex  (g1);
	\vertex [right =0.3cm of g1] (g2) ;
	\vertex [below =0.3cm of g1] (g3) ;
	\vertex [right =0.3cm of g3] (g4) ;
	\tikzfeynmanset{every vertex={dot,black,minimum size=0mm}}
	\vertex [above =0.3cm of g1] ;
	\tikzfeynmanset{every vertex={dot,black,minimum size=0mm}}
	\vertex [above =2.5mm of g1] (loop1);
	\diagram* {
		(g1) -- (g2) 
	};
	\end{feynman}
	\end{tikzpicture}
\end{gathered}\ , \hspace{3mm}
\begin{gathered}
	\begin{tikzpicture}
	\begin{feynman}[small, baseline=g4]
	\tikzfeynmanset{every vertex={dot,black,minimum size=1mm}}
	\vertex  (g1);
	\vertex [right =0.3cm of g1] (g2) ;
	\tikzfeynmanset{every vertex={dot,black,minimum size=0mm}}
	\vertex [above =2.5mm of g1] (loop1);
	\vertex [below =2.5mm of g1] (loop2);	\tikzfeynmanset{every vertex={dot,black,minimum size=0mm}}
	\vertex [above =0.5cm of g1] ;
	\diagram* {
		(g1) -- [out=45,in=0] (loop1) -- [out=180,in=135] (g1),
		(g1) -- [out=-45,in=0] (loop2) -- [out=-180,in=-135] (g1)
	};
	\end{feynman}
	\end{tikzpicture}
\end{gathered}\ , \hspace{3mm}
\begin{gathered}
	\begin{tikzpicture}
	\begin{feynman}[small, baseline=g4]
	\tikzfeynmanset{every vertex={dot,black,minimum size=1mm}}
	\vertex  (g1);
	\vertex [right =0.3cm of g1] (g2) ;
	\tikzfeynmanset{every vertex={dot,black,minimum size=0mm}}
	\vertex [above =2.5mm of g1] (loop1);
	\vertex [above =2.5mm of g2] (loop2);
	\diagram* {
		(g1) -- [out=45,in=0] (loop1) -- [out=180,in=135] (g1),
	    (g2) -- [out=45,in=0] (loop2) -- [out=180,in=135] (g2)
	};
	\end{feynman}
	\end{tikzpicture}
\end{gathered}\ , \hspace{3mm}
\begin{gathered}
	\begin{tikzpicture}
	\begin{feynman}[small, baseline=g4]
	\tikzfeynmanset{every vertex={dot,black,minimum size=1mm}}
	\vertex  (g1);
	\vertex [right =0.3cm of g1] (g2) ;
	\tikzfeynmanset{every vertex={dot,black,minimum size=0mm}}
	\vertex [above =2.5mm of g1] (loop1);
	\diagram* {
	    (g1) -- [] (g2),
		(g1) -- [out=45,in=0] (loop1) -- [out=180,in=135] (g1)
	};
	\end{feynman}
	\end{tikzpicture}
\end{gathered}\ , \hspace{3mm}
\begin{gathered}
	\begin{tikzpicture}
	\begin{feynman}[small, baseline=g4]
	\tikzfeynmanset{every vertex={dot,black,minimum size=1mm}}
	\vertex  (g1);
	\vertex [right =0.3cm of g1] (g2) ;
	\tikzfeynmanset{every vertex={dot,black,minimum size=0mm}}
	\diagram* {
         (g1) -- [out=30,in=150] (g2) -- [out=-150,in=-30] (g1)
	};
	\end{feynman}
	\end{tikzpicture}
\end{gathered}\ , \hspace{3mm}
\end{equation*}

 Now we examine the relations between these operators coming from total derivatives. Let us look first at the operators with  two $\phi$ fields
\begin{align}\label{eq:2pts2edges}
    &0=\partial^\mu\left( \partial_\mu\partial^2 \phi \phi\right)= \partial^4  \phi \phi + \partial^\mu \partial^2  \phi \partial_\mu \phi \,,\\
    &0=\partial^\mu\left( \partial^2 \phi \partial_\mu\phi\right)= \partial^2  \phi \partial^2 \phi + \partial^\mu \partial^2  \phi \partial_\mu \phi \,,\\
    &0=\partial^\mu\left( \partial_\mu\partial^\nu \phi \partial_\nu \phi\right)=\partial^\mu \partial^2  \phi \partial_\mu \phi +\partial^\mu\partial^\nu \phi \partial_\mu\partial_\nu \phi\,,
\end{align}
or, graphically,
\begin{align}
     &0=
    \begin{gathered}
	\begin{tikzpicture}
	\begin{feynman}[small, baseline=g4]
	\tikzfeynmanset{every vertex={dot,black,minimum size=1mm}}
	\vertex  (g1);
	\vertex [right =0.3cm of g1] (g2) ;
	\tikzfeynmanset{every vertex={dot,black,minimum size=0mm}}
	\vertex [above =2.5mm of g1] (loop1);
	\vertex [below =2.5mm of g1] (loop2);	\tikzfeynmanset{every vertex={dot,black,minimum size=0mm}}
	\vertex [above =0.5cm of g1] ;
	\diagram* {
		(g1) -- [out=45,in=0] (loop1) -- [out=180,in=135] (g1),
		(g1) -- [out=-45,in=0] (loop2) -- [out=-180,in=-135] (g1)
	};
	\end{feynman}
	\end{tikzpicture}
    \end{gathered}
    +
    \begin{gathered}
	\begin{tikzpicture}
	\begin{feynman}[small, baseline=g4]
	\tikzfeynmanset{every vertex={dot,black,minimum size=1mm}}
	\vertex  (g1);
	\vertex [right =0.3cm of g1] (g2) ;
	\tikzfeynmanset{every vertex={dot,black,minimum size=0mm}}
	\vertex [above =2.5mm of g1] (loop1);
	\diagram* {
	    (g1) -- [] (g2),
		(g1) -- [out=45,in=0] (loop1) -- [out=180,in=135] (g1)
	};
	\end{feynman}
	\end{tikzpicture}
    \end{gathered}\\
    &0=
    \begin{gathered}
	\begin{tikzpicture}
	\begin{feynman}[small, baseline=g4]
	\tikzfeynmanset{every vertex={dot,black,minimum size=1mm}}
	\vertex  (g1);
	\vertex [right =0.3cm of g1] (g2) ;
	\tikzfeynmanset{every vertex={dot,black,minimum size=0mm}}
	\vertex [above =2.5mm of g1] (loop1);
	\vertex [above =2.5mm of g2] (loop2);
	\diagram* {
		(g1) -- [out=45,in=0] (loop1) -- [out=180,in=135] (g1),
	    (g2) -- [out=45,in=0] (loop2) -- [out=180,in=135] (g2)
	};
	\end{feynman}
	\end{tikzpicture}
    \end{gathered}
    +
    \begin{gathered}
	\begin{tikzpicture}
	\begin{feynman}[small, baseline=g4]
	\tikzfeynmanset{every vertex={dot,black,minimum size=1mm}}
	\vertex  (g1);
	\vertex [right =0.3cm of g1] (g2) ;
	\tikzfeynmanset{every vertex={dot,black,minimum size=0mm}}
	\vertex [above =2.5mm of g1] (loop1);
	\diagram* {
	    (g1) -- [] (g2),
		(g1) -- [out=45,in=0] (loop1) -- [out=180,in=135] (g1)
	};
	\end{feynman}
	\end{tikzpicture}
    \end{gathered}\\
    &0=
    \begin{gathered}
	\begin{tikzpicture}
	\begin{feynman}[small, baseline=g4]
	\tikzfeynmanset{every vertex={dot,black,minimum size=1mm}}
	\vertex  (g1);
	\vertex [right =0.3cm of g1] (g2) ;
	\tikzfeynmanset{every vertex={dot,black,minimum size=0mm}}
	\vertex [above =2.5mm of g1] (loop1);
	\diagram* {
	    (g1) -- [] (g2),
		(g1) -- [out=45,in=0] (loop1) -- [out=180,in=135] (g1)
	};
	\end{feynman}
	\end{tikzpicture}
    \end{gathered}
    +
\begin{gathered}
	\begin{tikzpicture}
	\begin{feynman}[small, baseline=g4]
	\tikzfeynmanset{every vertex={dot,black,minimum size=1mm}}
	\vertex  (g1);
	\vertex [right =0.3cm of g1] (g2) ;
	\tikzfeynmanset{every vertex={dot,black,minimum size=0mm}}
	\diagram* {
         (g1) -- [out=30,in=150] (g2) -- [out=-150,in=-30] (g1)
	};
	\end{feynman}
	\end{tikzpicture}
\end{gathered}
\end{align}
In this case, these three relations can be used to express operators containing a $\partial^2$ (graphically, a self-loop) in terms of the one without (graphically, no self-loops). At four point, the relation
\begin{align}\label{eq:4pts1edge}
    0=
    \begin{gathered}
	\begin{tikzpicture}
	\begin{feynman}[small, baseline=g4]
	\tikzfeynmanset{every vertex={dot,black,minimum size=1mm}}
	\vertex  (g1);
	\vertex [right =0.3cm of g1] (g2) ;
	\vertex [below =0.3cm of g1] (g3) ;
	\vertex [right =0.3cm of g3] (g4) ;
	\tikzfeynmanset{every vertex={dot,black,minimum size=0mm}}
	\vertex [above =0.3cm of g1] ;	
	\tikzfeynmanset{every vertex={dot,black,minimum size=0mm}}
	\vertex [above =2.5mm of g1] (loop1);
	\diagram* {
		(g1) -- [out=45,in=0] (loop1) -- [out=180,in=135] (g1)
	};
	\end{feynman}
	\end{tikzpicture}
    \end{gathered}
    +3\hspace{2mm}
    \begin{gathered}
	\begin{tikzpicture}
	\begin{feynman}[small, baseline=g4]
	\tikzfeynmanset{every vertex={dot,black,minimum size=1mm}}
	\vertex  (g1);
	\vertex [right =0.3cm of g1] (g2) ;
	\vertex [below =0.3cm of g1] (g3) ;
	\vertex [right =0.3cm of g3] (g4) ;
	\tikzfeynmanset{every vertex={dot,black,minimum size=0mm}}
	\vertex [above =0.3cm of g1] ;
	\tikzfeynmanset{every vertex={dot,black,minimum size=0mm}}
	\vertex [above =2.5mm of g1] (loop1);
	\diagram* {
		(g1) -- (g2) 
	};
	\end{feynman}
	\end{tikzpicture}
    \end{gathered}
\end{align}
which again can be used to remove the self-loop graph operator in favor of the one with no loops. The basis is then reduced to the multigraph basis (without self-loops).

Let us repeat the above exercise in momentum space, using the operator Feynman rules. The six initial operators are (up to a normalization, to avoid numerical factors in the Feynman rules in writing the below)
\begin{align}
    1\,,~~~ \sum_{i<j=1}^4 s_{ij}\,,~~~\sum_{i=1}^4s_{ii}\,,~~~s_{11}^2+s_{22}^2\,,~~~s_{11}s_{22}\,,~~~s_{11}s_{12}+s_{22}s_{12}\,,~~~s_{12}^2\,.
\end{align}
Starting with the statement of momentum conservation for the operators with two $\phi$ fields,
\begin{align}
    p_1^\mu+p_2^\mu=0 \,,
\end{align}
we can take the dot product with both $p_1$ and $p_2$, so as to form the relations
\begin{align}
  X_1=  s_{11}+s_{12}=0\,,~~~X_2=s_{12}+s_{22}=0\,,
\end{align}
which can be used to eliminate $s_{11}$ and $s_{22}$ in the above polynomials, such that the only independent element is $s_{12}^2$. Similarly, the invariants $X_i$, $i=1,\ldots,4$, can be used to eliminate the $s_{ii}$ from the four point operators, such that the only independent element is $\sum_{i<j}s_{ij}$. 

\subsection{Hilbert series and evanescent operators}
\label{sec:hs}
In this subsection, we study Hilbert series for the real and complex scalar EFTs. A Hilbert series is a polynomial that enumerates the independent operators. In \cite{Henning:2017fpj} the formalism was worked out to the level at which we can directly apply the results to obtain Hilbert series for the three objects---$S$-matrix, off-shell correlation functions, and form factors---of Fig.~\ref{fig:qftobjects}. Details of the Hilbert series construction are provided in the appendix \ref{s:hilbert}. 

For the real scalar, the Hilbert series for the $S$-matrix enumerates the physical operator basis of the EFT, providing the independent $S$-matrix contributions at mass dimension $\Delta$ as,
\begin{align}\label{eq:Hilb-real}
  H^{\text{real}}_{S\text{-matrix}}(\Delta)
  =
  &\Delta^6+2\Delta^8+3\Delta^{10}+5\Delta^{12}+9\Delta^{14}+16\Delta^{16}+32\Delta^{18}+65\Delta^{20}+\mathcal{O}(\Delta^{22})
\end{align}
That is, the above Hilbert series tells us there is only one independent operator in the $S$-matrix (or `physical') basis at mass dimension 6, 2 at mass dimension 8, 3 at mass dimension 10, {\it etc.}. To calculate the Hilbert series relevant for off-shell correlation functions we can remove the on-shell condition in the character for the field (see Appendix~\ref{s:hilbert}),
\begin{align}
    H^{\text{real}}_{\text{C.F.}}(\Delta)
    =
    &\ 3\Delta^6+6\Delta^8+12\Delta^{10}+25\Delta^{12}+53\Delta^{14}\nonumber\\
    &\ +120\Delta^{16}+279\Delta^{18}+680\Delta^{20}+\mathcal{O}(\Delta^{22})
    \label{eq:hilbert_real_momonly}
\end{align}
That is, we see there are 3 independent operators in the operator basis for the off-shell correlation functions at mass dimension 6, in agreement with the analysis in the previous section; there are 6 needed at mass dimension 8, {\it etc.}.  Finally, we now look at the Hilbert series relevant for form factors, where we keep the on-shell condition in the character for the field, but remove a factor in the integrand of the matrix integral that accounts for momentum conservation,
\begin{align}
    H^{\text{real}}_{\text{F.F.}}(\Delta)
    =
    &\ 3\Delta^6+6\Delta^8+12\Delta^{10}+25\Delta^{12}+53\Delta^{14}\nonumber\\
    &\ +119\Delta^{16}+275\Delta^{18}+664\Delta^{20}+\mathcal{O}(\Delta^{22})
    \label{eq:hilbert_real_onshellonly}
\end{align}
Comparing eq.~\eqref{eq:hilbert_real_momonly} and eq.~\eqref{eq:hilbert_real_onshellonly}, we see agreement in the number of independent operators up to mass dimension 16, as expected from Eq.~\eqref{eq:multring}. That is, up to mass dimension 16, both Hilbert series are enumerating the number of non-isomorphic graphs that contribute at each mass dimension (note each vertex counts as dimension 1 in $D=4$, being the dimension of $\phi$, and each edge counts as dimension 2).

The same is observed for the complex scalar,
\begin{align}
    H^{\text{complex}}_{S\text{-matrix}}(\Delta)
    &=    
    2\Delta^6+4\Delta^8+8\Delta^{10}+21\Delta^{12}+53\Delta^{14}
    \nonumber\\&\hspace{1.2cm}
    +143\Delta^{16}+415\Delta^{18}+1241\Delta^{20}+\mathcal{O}(\Delta^{22}) \,,\\[2mm]
    H^{\text{complex}}_{\mathrm{C.F.}}(\Delta)
    &=
    5\Delta^6+14\Delta^8+38\Delta^{10}+110\Delta^{12}+325\Delta^{14}
    \nonumber\\&\hspace{1.2cm}
    +1010\Delta^{16}+3266\Delta^{18}+10914\Delta^{20}+\mathcal{O}(\Delta^{22})\,, \\[2mm]
    H^{\text{complex}}_{\mathrm{F.F.}}(\Delta)
    &=
    5\Delta^6+14\Delta^8+38\Delta^{10}+110\Delta^{12}+325\Delta^{14} 
    \nonumber\\&\hspace{1.2cm}
    +1006\Delta^{16}+3236\Delta^{18}+10710\Delta^{20}+\mathcal{O}(\Delta^{22}) \,.
\end{align}
One can further keep information about the number of operators at each mass dimension as broken down by their powers of $\phi$ fields. We include these Hilbert series in appendix \ref{s:hilbert} (for both real and complex scalar theories); again one can see agreement between the off-shell correlation function and form factor bases up to mass dimension 16.

Let us now return to discuss what happens at mass dimension 16. 
This is the point at which we start to see the effect of Gram conditions, which provide relations between the operators in the multigraph basis. The relations stem from the condition that any $(D{+}1)\times (D{+}1)$ minor of the matrix given in eq.~\eqref{eq:gram} vanish. That is, there is a potential non-trivial relation for $N>D$, due to the fact that it is only $D$ vectors that can be independent in $D$ dimensions. However, for the $S$-matrix and off-shell correlation functions, momentum conservation reduces the number of independent vectors by one, and so for these cases the non-trivial Gram relations only appear in the case $N>D+1$. 

We can further pinpoint the mass dimension that the Gram relations becomes important. The determinant of a $(D{+}1)\times (D{+}1)$ submatrix of eq.~\eqref{eq:gram} carries mass dimension $2D+2$ from the kinematics alone. This means that the operators where Gram conditions become important must involve at least $2D+2$ derivatives. We require $N>D$ for the form factor, so minimally $N=D+1$; for the $S$-matrix or off-shell correlation functions, minimally $N=D+2$. Since $\phi$ fields have mass dimension $(D-2)/2$, we expect to first see Gram conditions play a role at mass dimension $2D+2+(D+1)(D-2)/2$ for form factors, and at mass dimension $2D+2+(D+2)(D-2)/2$ for $S$-matrices or off-shell correlation functions---that is, at mass dimension $15$ and 16, respectively, for the case $D=4$. Because of the $\phi \to -\phi$ parity we imposed on the real scalar theory (and charge invariance for the complex scalar), we only consider even mass dimensions, and so the Gram conditions show up for the first time at mass dimension 16.

The above implies that there are operators that exist in high enough spacetime dimensions, but that vanish at some lower value of $D$. This represents a class of evanescent operators, and they have been studied in detail in \cite{Hogervorst:2014rta,Hogervorst:2015akt} (in the basis of operators relevant for form factors). Because the Hilbert series formalism has been worked out in arbitrary $D$, see \cite{Henning:2017fpj}, we can further formulate a systematic way to count the total number of such evanescent operators at any given mass dimension. The technique is to simply take the difference between the number of operators at mass dimension $\Delta$ in a Hilbert series calculated in `high enough' spacetime dimension $D$ and the Hilbert series calculated in $D=4$ (or another choice of low $D$ at which one is interested in counting evanescent operators). From the above, we can further quantify what `high enough' is such that no Gram conditions are present at given mass dimension $\Delta$:
\begin{eqnarray}
  D&>& \frac{1}{2} (-3 + \sqrt{1 + 8 \Delta}) \,, ~~~~~~\text{(F.F.)}\,, \\
  D&>& -2 + 2 \sqrt{1 + \frac{\Delta}{2}} \,,~~~~~~\text{($S$-matrix, C.F.)} \,.
\end{eqnarray}
In future work we plan to return to a detailed study of evanescent operators using the ideas presented in this section, including a study of how they affect calculations in dimensional regularisation at high loop orders.

\subsection{Primary operator construction}
\label{sec:prim}
The previous subsections described the construction of a multigraph basis relevant for off-shell correlation functions in the real and complex scalar theories. In this subsection, we now briefly describe how we construct the conformal primary operators that will be used in our study of the structure of the anomalous dimension matrix in what follows. Note that the number of primary operators is equal to the number of independent physical operators, {\it i.e.}~the number of operators needed to form a basis for the $S$-matrix, as pointed out in~\cite{Henning:2017fpj}.

We find conformal primary operators using the property of annihilation by the special conformal generator, which has the following representation on the polynomial ring we are considering (see {\it e.g.} \cite{Bzowski:2013sza}),
\beq\label{Kcovariant}
K_\mu = \sum_i \left( 2 \frac{\d}{\d p_{i}^\mu} + 2 p_{i}^\nu \frac{\d}{\d p_{i}^\nu} \frac{\d}{\d p_{i}^\mu} - p_{i\mu} \frac{\d}{\d p_{i}^\nu} \frac{\d}{\d p_{i\nu}} \right) \,.
\eeq
If a polynomial in our ring is annihilated by action of $K_\mu$, then the corresponding operator, constructed by the arguments given above, is primary. Practically, we find the primary operators by considering an arbitrary linear combination of the multigraph operators, and solve for the coefficients that ensure annihilation by $K_\mu$. This can be done by evaluating the annihilation condition using independent, random (rationally-valued) sets of momenta.

Let us again consider the example of real scalar multigraphs with four vertices and two edges, \textit{i.e.}~the operators
\vspace{-4mm}
\newcommand{\setbeginlength}[0]{& \hspace{0.7\textwidth} \nonumber \\[-4mm] }
\begin{align}
\setbeginlength
\nonumber
&
%
%
%
\mathcal{O}^{(8)}_{4,1} \hspace{3mm} = \hspace{3mm} \frac{1}{4} \hspace{2mm}
\begin{gathered}
\begin{tikzpicture}
\begin{feynman}[small, baseline=g1]
%
%
\tikzfeynmanset{every vertex={dot,black,minimum size=1mm}}
\vertex  (g1);
\vertex [right =0.3cm of g1] (g2) ;
\vertex [below =0.3cm of g1] (g3) ;
\vertex [right =0.3cm of g3] (g4) ;
\diagram* {
	(g1) -- [out=30,in=150] (g2) -- [out=-150,in=-30] (g1),
};
\end{feynman}
\end{tikzpicture}
\end{gathered}
\hspace{3mm} = \hspace{3mm} \frac{1}{4} \ \phi^2 \ \partial_\mu\partial_\nu\phi \ \partial^\mu\partial^\nu \phi 
=\mathcal{P}_4\left(s_{12}^2\right) \, ,
\\[3mm] 
\nonumber
%
%
&
\mathcal{O}^{(8)}_{4,2} \hspace{3mm} = \hspace{3mm} \frac{1}{2} \hspace{2mm}
\begin{gathered}
\begin{tikzpicture}
\begin{feynman}[small, baseline=g1]
%
%
\tikzfeynmanset{every vertex={dot,black,minimum size=1mm}}
\vertex  (g1);
\vertex [right =0.3cm of g1] (g2) ;
\vertex [below =0.3cm of g1] (g3) ;
\vertex [right =0.3cm of g3] (g4) ;
\diagram* {
	(g1) -- [] (g2),
	(g1) -- [] (g3)
};
\end{feynman}
\end{tikzpicture}
\end{gathered}
\hspace{3mm} = \hspace{3mm} \frac{1}{2} \ \phi \ \partial_\mu\phi \ \partial_\nu \phi \ \partial^\mu \partial^\nu \phi 
= \mathcal{P}_4\left(s_{12}s_{13}\right) \, ,  \\[3mm]
%
%
%
&
\mathcal{O}^{(8)}_{4,3} \hspace{3mm} = \hspace{3mm} \frac{1}{8} \hspace{2mm}
\begin{gathered}
\begin{tikzpicture}
\begin{feynman}[small, baseline=g1]
\tikzfeynmanset{every vertex={dot,black,minimum size=1mm}}
\vertex  (g1);
\vertex [right =0.3cm of g1] (g2) ;
\vertex [below =0.3cm of g1] (g3) ;
\vertex [right =0.3cm of g3] (g4) ;
\diagram* {
	(g1) -- [] (g2),
	(g3) -- [] (g4)
};
\end{feynman}
\end{tikzpicture}
\end{gathered}
\hspace{3mm} = \hspace{3mm} \frac{1}{8} \ \partial_\mu\phi \ \partial^\mu \phi \ \partial_\nu \phi \ \partial^\nu \phi
=\mathcal{P}_4\left(s_{12}s_{34}\right) \, .
\label{eq:IBP8-4} 
\end{align}	
According to the Hilbert series \eqref{realphy} given in appendix \ref{s:hilbert} there should be only one primary (physical) operator with four fields at mass dimension 8. One can easily check that the following combination is annihilated by $K$,
\beq
\mathcal{O}^{(8)c}_{4}=\mathcal{O}^{(8)}_{4,1}-2\mathcal{O}^{(8)}_{4,2}+6\mathcal{O}^{(8)}_{4,3}\,.
\eeq

To give another example, we consider four-point operators of the complex scalar EFT at mass dimension 6,
\begin{align}\label{eq:IBPc6}
\setbeginlength
\nonumber
&
%
%
%
\mathcal{O}^{(6)}_{4,1} \hspace{3mm} = \hspace{3mm} - \hspace{2mm}
\begin{gathered}
\begin{tikzpicture}
\begin{feynman}[small, baseline=g1]
%
%
\tikzfeynmanset{every vertex={dot,black,minimum size=1mm}}
\vertex  (g1);
\tikzfeynmanset{every vertex={empty dot,black,minimum size=1mm}}
\vertex [right =0.3cm of g1] (g2) ;
\tikzfeynmanset{every vertex={dot,black,minimum size=1mm}}
\vertex [below =0.3cm of g1] (g3) ;
\tikzfeynmanset{every vertex={empty dot,black,minimum size=1mm}}
\vertex [right =0.3cm of g3] (g4) ;
\diagram* {
	(g1) -- [] (g2)
};
\end{feynman}
\end{tikzpicture}
\end{gathered}
\hspace{3mm} = \hspace{3mm} - \ \phi \ \phi^{\dagger} \ \partial_\mu\phi \ \partial^\mu\phi^{\dagger} 
= \mathcal{P}_{2\bar{2}}\left(s_{1\bar{1}}\right) \, ,
\\[3mm]
 \nonumber
%
%
&
\mathcal{O}^{(6)}_{4,2} \hspace{3mm} = \hspace{3mm} -\frac{1}{4} \hspace{2mm}
\begin{gathered}
\begin{tikzpicture}
\begin{feynman}[small, baseline=g1]
%
%
\tikzfeynmanset{every vertex={dot,black,minimum size=1mm}}
\vertex  (g1);
\tikzfeynmanset{every vertex={empty dot,black,minimum size=1mm}}
\vertex [right =0.3cm of g1] (g2) ;
\tikzfeynmanset{every vertex={dot,black,minimum size=1mm}}
\vertex [below =0.3cm of g1] (g3) ;
\tikzfeynmanset{every vertex={empty dot,black,minimum size=1mm}}
\vertex [right =0.3cm of g3] (g4) ;
\diagram* {
	(g1) -- [] (g3),
};
\end{feynman}
\end{tikzpicture}
\end{gathered}
\hspace{3mm} = \hspace{3mm} -\frac{1}{4} \ \left(\phi^{\dagger}\right)^2 \ \partial_\mu\phi \ \partial_\mu \phi
=s_{12} \, , \\[3mm]
%
%
%
&
\mathcal{O}^{(6)}_{4,3} \hspace{3mm} = \hspace{3mm} -\frac{1}{4} \hspace{2mm}
\begin{gathered}
\begin{tikzpicture}
\begin{feynman}[small, baseline=g1]
\tikzfeynmanset{every vertex={dot,black,minimum size=1mm}}
\vertex  (g1);
\tikzfeynmanset{every vertex={empty dot,black,minimum size=1mm}}
\vertex [right =0.3cm of g1] (g2) ;
\tikzfeynmanset{every vertex={dot,black,minimum size=1mm}}
\vertex [below =0.3cm of g1] (g3) ;
\tikzfeynmanset{every vertex={empty dot,black,minimum size=1mm}}
\vertex [right =0.3cm of g3] (g4) ;
\diagram* {
	(g2) -- [] (g4)
};
\end{feynman}
\end{tikzpicture}
\end{gathered}
\hspace{3mm} = \hspace{3mm} -\frac{1}{4} \ \phi^2 \ \partial_\mu\phi^{\dagger} \ \partial^\mu \phi^{\dagger}
=  s_{\bar{1}\bar{2}} \, .
\end{align}	
The linear combination of these multigraph operators that is annihilated by $K$ is
\begin{equation}\label{eq:primdim6}
\mathcal{O}^{(6)}_{4,1}-2\left(\mathcal{O}^{(6)}_{4,2}+\mathcal{O}^{(6)}_{4,3}\right)
\end{equation}
and again from the $S$-matrix Hilbert series we know this is the only primary operator.

As a final example, we  generalize the previous conformal primary operator \eqref{eq:primdim6} to arbitrary mass dimension.
For this purpose, consider the  $(n-2)$-point multigraph operators of the complex scalar at mass dimension $n$,
\vspace*{-0.6cm}
\begin{align}\label{eq:IBPcn}
\setbeginlength
\nonumber
&
\mathcal{O}^{(n)}_{n\text{\,-\,}2,1} \hspace{3mm} = \hspace{3mm} - \frac{1}{\left(\frac{n}{2}-2\right)!\left(\frac{n}{2}-2\right)!}\hspace{2mm}
\begin{gathered}
\begin{tikzpicture}
\begin{feynman}[small, baseline=g1]
\tikzfeynmanset{every vertex={dot,black,minimum size=1mm}}
\vertex  (g1);
\tikzfeynmanset{every vertex={dot,black,minimum size=1mm}}
\vertex [right =0.3cm of g1] (g2) ;
\tikzfeynmanset{every vertex={empty dot,black,minimum size=1mm}}
\vertex [below =0.3cm of g1] (g3) ;
\tikzfeynmanset{every vertex={empty dot,black,minimum size=1mm}}
\vertex [right =0.3cm of g3] (g4) ;
\diagram* {
	(g1) -- [] (g3)
};
\end{feynman}
\end{tikzpicture}
\end{gathered}
\cdot\cdot\cdot
\begin{gathered}
\begin{tikzpicture}
\begin{feynman}[small, baseline=g1]
\tikzfeynmanset{every vertex={dot,black,minimum size=1mm}}
\vertex  (g1);
\tikzfeynmanset{every vertex={dot,black,minimum size=1mm}}
\vertex [right =0.3cm of g1] (g2) ;
\tikzfeynmanset{every vertex={empty dot,black,minimum size=1mm}}
\vertex [below =0.3cm of g1] (g3) ;
\tikzfeynmanset{every vertex={empty dot,black,minimum size=1mm}}
\vertex [right =0.3cm of g3] (g4) ;
\end{feynman}
\end{tikzpicture}
\end{gathered}
\hspace{3mm}
= \mathcal{P}_{\frac{n}{2}-1,\overline{\frac{n}{2}-1}}\left(s_{1\bar{1}}\right) =\sum_{i=1}^{\frac{n}{2}-1}\, \sum_{j=\bar{1}}^{\overline{\frac{n}{2}-1}}p_i\cdot p_j\, ,
\\[3mm]
 \nonumber
%
%
&
\mathcal{O}^{(n)}_{n\text{\,-\,}2,2} \hspace{3mm} = \hspace{3mm} -\frac{1}{2\left(\frac{n}{2}-1\right)!\left(\frac{n}{2}-3\right)!} \hspace{2mm}
\begin{gathered}
\begin{tikzpicture}
\begin{feynman}[small, baseline=g1]
%
%
\tikzfeynmanset{every vertex={dot,black,minimum size=1mm}}
\vertex  (g1);
\tikzfeynmanset{every vertex={dot,black,minimum size=1mm}}
\vertex [right =0.3cm of g1] (g2) ;
\tikzfeynmanset{every vertex={empty dot,black,minimum size=1mm}}
\vertex [below =0.3cm of g1] (g3) ;
\tikzfeynmanset{every vertex={empty dot,black,minimum size=1mm}}
\vertex [right =0.3cm of g3] (g4) ;
\diagram* {
	(g1) -- [] (g2),
};
\end{feynman}
\end{tikzpicture}
\end{gathered}
\cdot\cdot\cdot
\begin{gathered}
\begin{tikzpicture}
\begin{feynman}[small, baseline=g1]
%
%
\tikzfeynmanset{every vertex={dot,black,minimum size=1mm}}
\vertex  (g1);
\tikzfeynmanset{every vertex={dot,black,minimum size=1mm}}
\vertex [right =0.3cm of g1] (g2) ;
\tikzfeynmanset{every vertex={empty dot,black,minimum size=1mm}}
\vertex [below =0.3cm of g1] (g3) ;
\tikzfeynmanset{every vertex={empty dot,black,minimum size=1mm}}
\vertex [right =0.3cm of g3] (g4) ;
\end{feynman}
\end{tikzpicture}
\end{gathered}
\hspace{3mm}
= \mathcal{P}_{\frac{n}{2}-1}\left( s_{12}\right)=\sum_{i=1}^{\frac{n}{2}-2}\,\sum_{j>i}^{\frac{n}{2}-1}p_i\cdot p_j \, , \\[3mm]
%
%
%
&
\mathcal{O}^{(n)}_{n\text{\,-\,}2,3} \hspace{3mm} = \hspace{3mm} -\frac{1}{2\left(\frac{n}{2}-1\right)!\left(\frac{n}{2}-3\right)!} \hspace{2mm}
\begin{gathered}
\begin{tikzpicture}
\begin{feynman}[small, baseline=g1]
\tikzfeynmanset{every vertex={dot,black,minimum size=1mm}}
\vertex  (g1);
\tikzfeynmanset{every vertex={dot,black,minimum size=1mm}}
\vertex [right =0.3cm of g1] (g2) ;
\tikzfeynmanset{every vertex={empty dot,black,minimum size=1mm}}
\vertex [below =0.3cm of g1] (g3) ;
\tikzfeynmanset{every vertex={empty dot,black,minimum size=1mm}}
\vertex [right =0.3cm of g3] (g4) ;
\diagram* {
	(g3) -- [] (g4)
};
\end{feynman}
\end{tikzpicture}
\end{gathered}
\cdot\cdot\cdot
\begin{gathered}
\begin{tikzpicture}
\begin{feynman}[small, baseline=g1]
\tikzfeynmanset{every vertex={dot,black,minimum size=1mm}}
\vertex  (g1);
\tikzfeynmanset{every vertex={dot,black,minimum size=1mm}}
\vertex [right =0.3cm of g1] (g2) ;
\tikzfeynmanset{every vertex={empty dot,black,minimum size=1mm}}
\vertex [below =0.3cm of g1] (g3) ;
\tikzfeynmanset{every vertex={empty dot,black,minimum size=1mm}}
\vertex [right =0.3cm of g3] (g4) ;
\end{feynman}
\end{tikzpicture}
\end{gathered}
\hspace{3mm}
= \mathcal{P}_{\overline{\frac{n}{2}-1}}\left(s_{\bar{1}\bar{2}}\right) =\sum_{i=\bar{1}}^{\overline{\frac{n}{2}-2}}\,\sum_{j>i}^{\overline{\frac{n}{2}-1}}p_i\cdot p_j\, .
\end{align}	
As before there is only one primary operator, which we parametrize  as
\beq \label{eq:dim6param}
\mathcal{O}^{(n)c}_{n\text{-}2}= a_1\mathcal{O}^{(n)}_{n\text{-}2,1}+a_2\mathcal{O}^{(n)}_{n\text{-}2,2}+a_3\mathcal{O}^{(n)}_{n\text{-}2,3} \,,
\eeq
and which satisfies
\beq\label{eq:annihilation}
0=K_{\mu}\mathcal{O}^{(n)c}_{n\text{-}2}
=\frac{1}{2}a_1(n-2)
\left(\sum_{i=1}^{\frac{n}{2}-1}p_i+\sum_{i=\bar{1}}^{\overline{\frac{n}{2}-1}}p_i\right)
+\frac{1}{2}a_2(n-4)\left(\sum_{i=1}^{\frac{n}{2}-1}p_i\right)
+\frac{1}{2}a_3(n-4)\left(\sum_{i=\bar{1}}^{\overline{\frac{n}{2}-1}}p_i\right) \,.
\eeq
Since this identity should hold for arbitrary momentum, we require
\beq \label{eq:confnn-2}
a_2=a_3=-\frac{n-2}{n-4}a_1 \,.
\eeq
Therefore, the $(n-2)$-point primary operator at mass dimension $n$ is
\beq
\mathcal{O}^{(n)c}_{n\text{-}2}= \mathcal{O}^{(n)}_{n\text{-}2,1}-\frac{n-2}{n-4}\Big(\mathcal{O}^{(n)}_{n\text{-}2,2}+\mathcal{O}^{(n)}_{n\text{-}2,3}\Big)\,.
\eeq
Appendix \ref{s:zeronn-2} considers the mixing between the general $(n-2)$-point operator \eqref{eq:dim6param} and the $n$-point conformal primary ($\O^{(n)}_{n}=\phi^{n/2}\phi^{\dagger n/2}/[(n/2)!]^2$). It is interesting to note, that the same coefficients \eqref{eq:confnn-2} are uniquely obtained when the mixing between these operators is required to be diagonal. Only in the conformal basis do these operators not mix at arbitrary mass dimension.

We have constructed the primary operators for real scalar theory up to mass dimension 12 and for complex scalar theory up to mass dimension 10, as needed for our results we present in Sec.~\ref{sec:results}; the expressions can be found in appendix \ref{s:IBPbasesConf}.

\section{The $R^*$ method in EFT} \label{sec:rstar}
   
In this section, we review the basics of the $R^*$ method and how it can be applied to calculate the renormalization constants in the MS schemes. More details on the particular R* approach we employ in this work can be found in the following references \cite{Herzog:2017bjx,deVries:2019nsu,Beekveldt:2020kzk}. Concretely we will employ the extension of the R*-approach, which works at the level of uncontracted Feynman rules, and was described to some degree in \cite{deVries:2019nsu}. While we focus mainly on issues related to operator bases, we also provide a brief account of the definition of $R^*$ with some EFT examples.

Let us denote the 1PI Green's functions with $N$ external fields $\phi$ by $\Gamma_N$, \textit{i.e.}
\beq
\Gamma_N(g;p_1,p_2,..,p_N)=
\int \hspace{-1mm} \Big(\,
\prod_{i=1}^N d^{\scaleto{D}{4.5pt}}\!x_i\;e^{ip_i\cdot x_i} \Big)\langle0|\mathrm{T}\{\phi(x_1)...\phi(x_N)\}|0\rangle_{\text{1PI}}
\eeq
To isolate the renormalization constants, in the MS-scheme, we introduce the $\Z$-operation, or UV-counterterm operation, which, in the formalism of Bogoliubov, Parasiuk, Hepp, and Zimmermann (BPHZ), is defined as follows:
\beq
\label{eq:ZRbar}
\Z(\Gamma_N)=-K_\eps\bar R(\Gamma_N)\,.
\eeq
Here $K_\eps$ projects out (simple and higher) pole pieces in the dimensional regulator $\eps$ and the BPHZ $\bar R$-operation subtracts all UV subdivergences. The locality of the UV counterterm ensures that it is a homogeneous polynomial in the external momenta whose degree equals the degree of divergence of $\Gamma_N$, $\omega(\Gamma_N)$. The degree of divergence can be made logarithmic by acting on $\Gamma_n$ with the Taylor expansion operator $\T^{(\omega)}_{\{p_i\}}$, which projects out precisely all terms of order $\omega(\Gamma_N)$,
\beq
\Z(\Gamma_N)=-\T^{(\omega)}_{\{p_i\}} K_\eps\bar R(\Gamma_N)\,.
\eeq
The Taylor expansion operator $\T^{(\omega)}_{\{p_i\}}$ does not naively commute with the $\Z$-operation due to the 
infrared divergences which may be created in the process of differentiation, and setting to zero of external momenta. The additional IR divergences can be subtracted with the $R^*$-operation, or more precisely the $\bar R^*$-operation, which subtracts besides UV-subdivergences also all IR-divergences of the Euclidean (or equivalently off-shell) Green's function. This leads to the following powerful definition of the UV-counterterm in terms of logarithmic scaleless vacuum diagrams: 
\beq
\Z(\Gamma_N)=-K_\eps\bar R^*(\T^{(\omega)}_{\{p_i\}} \Gamma_N)
\eeq
Since scaleless vacuum diagrams vanish in dimensional regularisation, this definition is not particularly useful for calculations. It becomes useful when coupled with an infrared rearrangement. Since the UV counterterm of a log-divergent vacuum is not sensitive to the external momentum configuration or internal masses, arbitrary scales can be introduced by either attaching external momenta to any set of vertices or inserting arbitrary masses into the propagators of the graph. Using a clever infrared rearrangement one can considerably reduce the complexity of the calculation by converting the scaleless vacuum diagram into a simple self energy diagram or a vacuum diagram with a single massive edge.

With the aid of the $\Z$-operation it is straightforward to isolate the renormalization constants of the renormalizable theory as follows:
\beq
\Z(\Gamma_2)=Z_2 \, \O_2^{(4)},\qquad \Z(\Gamma_4)=g Z_g Z_2^2 \, \O_4^{(4)}\,,
\eeq
where the operators have been defined in section \ref{s:dim4background} and appendix \ref{s:IBPbasesConf}.
Going to the effective theory framework brings the complication that the necessary set of operators to renormalize all Green's functions of the theory requires non-physical operators, as discussed in section \ref{section:EFTLandR}. 
To denote 1PI Green's functions with insertions of physical operators we introduce the notation:
\bea
\Gamma_N[\O_i]&&=\Gamma_n(g;p_1,p_2,..,p_N)[\O_i]\nn\\
&&=\int \hspace{-1mm} \Big(\, \prod_{i=1}^N d^{\scaleto{D}{4.5pt}}\!x_i\; 
e^{ip_i\cdot x_i}\Big) 
d^{\scaleto{D}{4.5pt}}\!x\;
\langle0|\mathrm{T}\{\phi(x_1)...\phi(x_N)\O_i(x)\}|0\rangle_{\text{1PI}} \,.
\eea
Since the renormalisation of the Green's functions $\Gamma_n[\O_i]$ requires in general the full off-shell basis (defined in section \ref{sec:background}), the action of the $\Z$-operation leads in general to 
\bea
\label{eq:ZGO}
\Z(\Gamma_N[\O_i])&&=\sum_{j} \tilde Z_{ji} \tilde \O_j^b\Big|_{N} \,,
\eea
where $\tilde \O_j\Big|_{N}$ is an off-shell basis of operators containing $N$-external fields. As discussed in section \ref{sec:EOMandFR}, for single insertions of operators the physical space can be separated from the non-physical by using either EoM-operators or by using explicit field redefinitions. Taking EoM-operators to span the non-physical operators, one can thus write
\bea
\label{eq:ZGOEOM}
\Z(\Gamma_N[\O_i])&&=\sum_{j} Z_{ji} \O_j^b\Big|_{N} + \sum_{j} \widehat Z_{ji} \E_j^b\Big|_{N}\,.
\eea
Note that EoM operators generically contain operator contributions with different numbers of fields. With $\E_j\Big|_{N}$ we thus denote the contribution of all $N$-field operators contributing to $\E_j$. In this R$^*$-based approach the only purpose of the EoM renormalization operators constants is to correctly isolate the physical contributions $Z_{ij}$. This is to be contrasted to a more general off-shell approach where one would try to make the Green's function $\Gamma_N[O_i]$ finite from a renormalized Lagrangian. Then one would also require Green's functions with non-physical operators inserted at loop level in counterterms to correctly eliminate all subdivergences. This is not necessary here, since the $R^*$-operation always \emph{automatically} builds whatever counterterms it requires on a diagram by diagram level. In this approach one thus only has to consider insertions of the physical operators on the left hand side of eqs.\ (\ref{eq:ZGOEOM}) and (\ref{eq:ZGO}).

\subsection{$R^*$: definitions and examples}
In the following we give  some more details on the definition of the $R^*$ and $\bar R^*$ maps. Acting on a graph $\Gamma$, assumed to have non-exceptional off-shell momenta, these maps are defined as follows:
\beq
R^* (\Gamma) = \sum_{\substack{\gamma \subseteq \Gamma,\tilde \gamma \subseteq \Gamma \\\gamma \cap \tilde \gamma = \emptyset}}\;\widetilde \Z(\tilde \gamma)* \Z(\gamma) * \Gamma/\gamma \setminus \tilde \gamma\,,
\eeq
\beq
\bar R^* (\Gamma) = \sum_{\substack{\gamma \subsetneq \Gamma,\tilde \gamma \subseteq \Gamma \\\gamma \cap \tilde \gamma = \emptyset}}\;\widetilde \Z(\tilde \gamma)* \Z(\gamma) * \Gamma/\gamma \setminus \tilde \gamma\,.
\eeq
The different ingredients entering here are (briefly) explained below. The reader should consult \cite{Herzog:2017bjx,Beekveldt:2020kzk} for a more complete presentation of the formalism.
\begin{itemize}
\item $\gamma$ is a set of UV-divergent disjoint 1PI subgraphs.
\item $\tilde \gamma$ is a set of disjoint IR subgraphs, most easily defined via its complement or complementary subgraph. The complementary subgraph has to be mass-momentum spanning, \emph{i.e.} contain all external edges and all massive edges, and motic. The motic condition amounts to it being 1PI if all external edges are joined in a single vertex. Contracting this motic mass-momentum spanning subgraph, $\Gamma_m$, to a point in $\Gamma$ then leads to the IR subgraph $\tilde \gamma=\Gamma/\Gamma_m$; this can be identified as a set of disjoint 1PI vacuum graphs. It is IR divergent as long as $\omega(\tilde \gamma)\leq 0$.
\item The sum goes over all disjoint (no common vertices or edges) IR and UV subgraphs, $\tilde \gamma$ and $\gamma$, of the graph $\Gamma$, leaving out only $\gamma=\Gamma$ in the case of $\bar R^*$.
\item The symbol $/$ means contraction, such that $\Gamma/\gamma$ denotes the graph $\Gamma$ with $\gamma$ contracted to a point.
\item The symbol $\setminus$ means deletion, such that $\Gamma\setminus\tilde\gamma$ denotes the graph $\Gamma$ with $\tilde\gamma$ deleted.
\item The $*$ symbol reduces to the standard multiplication ($\cdot$) for log-divergent graphs. For higher-degree UV-divergences it denotes an insertion. That is it means the insertion of a UV counterterm, which can be thought of as a Feynman rule of a given vertex, into the vertex to which the UV subgraph is contracted. For  higher-degree IR divergences the IR counterterm becomes a Taylor-type differential operator acting on the complementary graph---the $*$ symbol then symbolizes the action of this operator.
\item The IR counterterm is denoted by $\tilde \Z$. We define it via the condition $R^*(\Gamma_0)=0$ with $\Gamma_0$ a log-divergent vacuum graph.
\end{itemize}
\subsubsection*{}
We will now consider some simple examples in the EFT context. Consider the operator $\O_6^{(6)}$ inserted in the following diagram:
$$
\Z\Big(\;
\begin{gathered}
\begin{tikzpicture}
\begin{feynman}[large, baseline=b]

\tikzfeynmanset{every vertex={dot,minimum size=2.5mm}}
\vertex  at (-0.7,0) (eft) ;

\tikzfeynmanset{every vertex={draw,minimum size=0pt, inner sep=0pt}}
\vertex  at (-0.7,0) (l) ;
\vertex at (0.7,0) (r);

\vertex at (-1,0.3) (e4);
\vertex at (-1,0.1) (e5);
\vertex at (-1,-0.1) (e6);
\vertex at (-1,-0.3) (e7);

\vertex at (1,0.3) (e11);
\vertex at (1,-0.3) (e12);

\diagram* {
	(l) -- [half left] (r) -- [half left] (l),
	(l) -- [] (e4),
	(l) -- [] (e5),
	(l) -- [] (e6),
	(l) -- [] (e7),
	(r) -- [] (e11),
	(r) -- [] (e12),
};

\end{feynman}
\end{tikzpicture}
\end{gathered}
\;\Big)
=
\Z\Big(\;
\begin{gathered}
\begin{tikzpicture}
\begin{feynman}[large, baseline=b]

\tikzfeynmanset{every vertex={dot,minimum size=2.5mm}}
\vertex  at (-0.7,0) (eft) ;

\tikzfeynmanset{every vertex={draw,minimum size=0pt, inner sep=0pt}}
\vertex  at (-0.7,0) (l) ;
\vertex at (0.7,0) (r);

\vertex at (-1,0.0) (e1);
\vertex at (1,0.0) (e11);

\diagram* {
	(l) -- [half left] (r) -- [half left] (l),
	(l) -- [] (e1),
	(r) -- [] (e11),
};

\end{feynman}
\end{tikzpicture}
\end{gathered}
\;\Big)
=
-K_\eps \Big    (\;
\begin{gathered}
\begin{tikzpicture}
\begin{feynman}[large, baseline=b]

\tikzfeynmanset{every vertex={dot,minimum size=2.5mm}}
\vertex  at (-0.7,0) (eft) ;

\tikzfeynmanset{every vertex={draw,minimum size=0pt, inner sep=0pt}}
\vertex  at (-0.7,0) (l) ;
\vertex at (0.7,0) (r);

\vertex at (-1,0.0) (e1);

\vertex at (1,0.0) (e11);

\diagram* {
	(l) -- [half left] (r) -- [half left] (l),
	(l) -- [] (e1),
	(r) -- [] (e11),
};

\end{feynman}
\end{tikzpicture}
\end{gathered}
\;\Big)\,.
$$
Since the diagram is log-divergent, the Taylor-expansion reduces to nullifying an arbitrary subset of the external momenta and we can extract the counterterm from a simple one-loop bubble integral. No IR divergences are produced in the nullification here; hence we did not need any extra counterterms. Let us now consider a more complicated example, where we insert the operator $\O_{4,3}^{(8)}$ into the following 2-loop six-point diagram:
$$
\Z\Big(\;
\begin{gathered}
\begin{tikzpicture}
\begin{feynman}[large, baseline=b]

\tikzfeynmanset{every vertex={dot,minimum size=2.5mm}}
\vertex  at (-0.7,0) (eft);

\tikzfeynmanset{every vertex={draw,minimum size=0pt, inner sep=0pt}}
\vertex at (-0.7,0) (l);
\vertex at (0,0.7) (u);
\vertex at (0.7,0) (r);
\vertex at (0,-0.7) (d);

\vertex at (-1,0.1) (e1);
\vertex at (-1,-0.1) (e2);
\vertex at (0,1) (e3);
\vertex at (0,-1) (e4);
\vertex at (1,0.1) (e5);
\vertex at (1,-0.1) (e6);

\diagram* {
	(l) -- [quarter left] (u) -- [quarter left] (r) -- [quarter left] (d) -- [quarter left] (l)  ,
    (u) -- [] (d),
	(l) -- [] (e1),
	(l) -- [] (e2),
	(u) -- [] (e3),
	(d) -- [] (e4),
	(r) -- [] (e5),
	(r) -- [] (e6),
	};

\end{feynman}
\end{tikzpicture}
\end{gathered}
\;\Big)
=
\Z\Big(\;\T^{(2)}_{p_1,..,p_6}
\begin{gathered}
\begin{tikzpicture}
\begin{feynman}[large, baseline=b]

\tikzfeynmanset{every vertex={dot,minimum size=2.5mm}}
\vertex  at (-0.7,0) (eft);

\tikzfeynmanset{every vertex={draw,minimum size=0pt, inner sep=0pt}}
\vertex at (-0.7,0) (l);
\vertex at (0,0.7) (u);
\vertex at (0.7,0) (r);
\vertex at (0,-0.7) (d);

\vertex at (-1,0.1) (e1);
\vertex at (-1,-0.1) (e2);
\vertex at (0,1) (e3);
\vertex at (0,-1) (e4);
\vertex at (1,0.1) (e5);
\vertex at (1,-0.1) (e6);

\diagram* {
	(l) -- [quarter left] (u) -- [quarter left] (r) -- [quarter left] (d) -- [quarter left] (l)  ,
    (u) -- [] (d),
	(l) -- [] (e1),
	(l) -- [] (e2),
	(u) -- [] (e3),
	(d) -- [] (e4),
	(r) -- [] (e5),
	(r) -- [] (e6),
	};

\end{feynman}
\end{tikzpicture}
\end{gathered}
\;\Big)\,.
$$
We thus have to perform a second order Taylor expansion in the external momenta $p_1,...,p_6$, where $p_1$ and $p_2$ flow into the EFT vertex. Usually this would lead to a very large number of terms, since one has to parametrise the momenta by identifying a routing of the external momenta flowing through the diagram, and then differentiate along these paths via the chain/product rule. In this case we are fortunate since the Feynman rule for $\O_{4,3}^{(8)}$ is linear in all four momenta flowing into its vertex. Therefore, after nullification of the external momenta, every term in the Taylor expansion is zero apart from those where both derivatives are acting on the EFT vertex. Thus just one single term will survive,
$$
p_1^\alpha p_2^\beta \;\Z\Big(
\begin{gathered}
\begin{tikzpicture}
\begin{feynman}[large, baseline=b]

\tikzfeynmanset{every vertex={empty dot,minimum size=2.5mm}}
\vertex  at (-0.7,0) (eft);

\tikzfeynmanset{every vertex={dot,minimum size=1.5mm}}
\vertex  at (-0.7,0) (eft);

\tikzfeynmanset{every vertex={draw,minimum size=0pt, inner sep=0pt}}
\vertex at (-0.7,0) (l);
\vertex at (0,0.7) (u);
\vertex at (0.7,0) (r);
\vertex at (0,-0.7) (d);

\vertex at (-1,0.0) (e1);
\vertex at (1,0.0) (e5);

\diagram* {
	(l) -- [quarter left] (u) -- [quarter left] (r) -- [quarter left] (d) -- [quarter left] (l)  ,
    (u) -- [] (d),
	(l) -- [] (e1),
	(r) -- [] (e5),
	};

\end{feynman}
\end{tikzpicture}
\end{gathered}
\;\Big)\,,
$$
where we have reintroduced two arbitrary external momenta after nullifying the initial ones. This has the advantage that the doubly differentiated EFT vertex, depicted as a circled dot, no longer depends on the external momenta of the diagram. Its Feynman rule is given by:
\beq
\begin{gathered}
\begin{tikzpicture}
\begin{feynman}[large, baseline=b]

\tikzfeynmanset{every vertex={empty dot,minimum size=2.5mm}}
\vertex  at (-0.7,0) (eft);

\tikzfeynmanset{every vertex={dot,minimum size=1.5mm}}
\vertex  at (-0.7,0) (eft);

\tikzfeynmanset{every vertex={draw,minimum size=0pt, inner sep=0pt}}
\vertex at (-0.7,0) (l);

\vertex at (-0,0.6) (e1)  {\(q_3\)};
\vertex at (-0,-0.6) (e2) {\(q_4\)};
\vertex at (-1.2,0) (e3);

\diagram* {
	(l) -- [] (e1),
	(l) -- [] (e2),
	(l) -- [] (e3),
	};

\end{feynman}
\end{tikzpicture}
\end{gathered}
= q_3^\alpha q_4^\beta+q_3^\beta q_4^\alpha+\eta^{\alpha\beta}q_3.q_4\,,
\eeq
with $q_3$ and $q_4$ the internal loop momenta to which it is attached. With this IR rearrangement the diagram is free of IR divergences but contains a UV subgraph of logarithmic degree. Unfolding the $\Z$-operation yields:
$$
p_1^\alpha p_2^\beta \;\Z\Big(
\begin{gathered}
\begin{tikzpicture}
\begin{feynman}[large, baseline=b]

\tikzfeynmanset{every vertex={empty dot,minimum size=2.5mm}}
\vertex  at (-0.7,0) (eft);

\tikzfeynmanset{every vertex={dot,minimum size=1.5mm}}
\vertex  at (-0.7,0) (eft);

\tikzfeynmanset{every vertex={draw,minimum size=0pt, inner sep=0pt}}
\vertex at (-0.7,0) (l);
\vertex at (0,0.7) (u);
\vertex at (0.7,0) (r);
\vertex at (0,-0.7) (d);

\vertex at (-1,0.0) (e1);
\vertex at (-1,-0.1) (e2);
\vertex at (0,1) (e3);
\vertex at (0,-1) (e4);
\vertex at (1,0.0) (e5);
\vertex at (1,-0.1) (e6);

\diagram* {
	(l) -- [quarter left] (u) -- [quarter left] (r) -- [quarter left] (d) -- [quarter left] (l)  ,
    (u) -- [] (d),
	(l) -- [] (e1),
	(r) -- [] (e5),
	};

\end{feynman}
\end{tikzpicture}
\end{gathered}
\;\Big)
=
-p_1^\alpha p_2^\beta  \;K_\eps\Big(
\begin{gathered}
\begin{tikzpicture}
\begin{feynman}[large, baseline=b]

\tikzfeynmanset{every vertex={empty dot,minimum size=2.5mm}}
\vertex  at (-0.7,0) (eft);

\tikzfeynmanset{every vertex={dot,minimum size=1.5mm}}
\vertex  at (-0.7,0) (eft);

\tikzfeynmanset{every vertex={draw,minimum size=0pt, inner sep=0pt}}
\vertex at (-0.7,0) (l);
\vertex at (0,0.7) (u);
\vertex at (0.7,0) (r);
\vertex at (0,-0.7) (d);

\vertex at (-1,0.0) (e1);
\vertex at (-1,-0.1) (e2);
\vertex at (0,1) (e3);
\vertex at (0,-1) (e4);
\vertex at (1,0.0) (e5);
\vertex at (1,-0.1) (e6);

\diagram* {
	(l) -- [quarter left] (u) -- [quarter left] (r) -- [quarter left] (d) -- [quarter left] (l)  ,
    (u) -- [] (d),
	(l) -- [] (e1),
	(r) -- [] (e5),
	};

\end{feynman}
\end{tikzpicture}
\end{gathered}
\;
+
\Z\Big(\;
\begin{gathered}
\begin{tikzpicture}
\begin{feynman}[large, baseline=b]

\tikzfeynmanset{every vertex={empty dot,minimum size=2.5mm}}
\vertex  at (-0.5,0) (eft);

\tikzfeynmanset{every vertex={dot,minimum size=1.5mm}}
\vertex  at (-0.5,0) (eft);

\tikzfeynmanset{every vertex={draw,minimum size=0pt, inner sep=0pt}}
\vertex at (-0.5,0) (l);
\vertex at (0,0.5) (u);
\vertex at (0,-0.5) (d);

\vertex at (-0.7,0.0) (e1);
\vertex at (0.2,0.5) (e4);
\vertex at (0.2,-0.5) (e5);

\diagram* {
	(l) -- [] (u) -- [] (d) -- [] (l)  ,
	(l) -- [] (e1),
	(u) -- [] (e4),
	(d) -- [] (e5),
	};

\end{feynman}
\end{tikzpicture}
\end{gathered}
\;\Big)
\cdot
\begin{gathered}
\begin{tikzpicture}
\begin{feynman}[large, baseline=b]


\tikzfeynmanset{every vertex={draw,minimum size=0pt, inner sep=0pt}}
\vertex at (-0.5,0) (l);
\vertex at (0.5,0) (r);

\vertex at (-0.7,0.0) (e1);
\vertex at (0.7,0.0) (e2);

\diagram* {
	(l) -- [half left] (r) -- [half left] (l)  ,
	(l) -- [] (e1),
	(r) -- [] (e2),
	};

\end{feynman}
\end{tikzpicture}
\end{gathered}
\;\Big)\,.
$$
Finally we need to consider the evaluation of the triangle UV-counterterm; since it is of logarithmic degree we can nullify the external momentum flowing into any one vertex to make a self energy. A particularly convenient choice is to nullify the momentum going into the EFT vertex:
\beq
\label{eq:IR1}
\Z\Big(\;
\begin{gathered}
\begin{tikzpicture}
\begin{feynman}[large, baseline=b]

\tikzfeynmanset{every vertex={empty dot,minimum size=2.5mm}}
\vertex  at (-0.5,0) (eft);

\tikzfeynmanset{every vertex={dot,minimum size=1.5mm}}
\vertex  at (-0.5,0) (eft);

\tikzfeynmanset{every vertex={draw,minimum size=0pt, inner sep=0pt}}
\vertex at (-0.5,0) (l);
\vertex at (0,0.5) (u);
\vertex at (0,-0.5) (d);

\vertex at (-0.7,0.0) (e1);
\vertex at (0.2,0.5) (e4);
\vertex at (0.2,-0.5) (e5);

\diagram* {
	(l) -- [] (u) -- [] (d) -- [] (l)  ,
	(l) -- [] (e1),
	(u) -- [] (e4),
	(d) -- [] (e5),
	};

\end{feynman}
\end{tikzpicture}
\end{gathered}
\;\Big)
=
-K_\eps\Big(\;
\begin{gathered}
\begin{tikzpicture}
\begin{feynman}[large, baseline=eft]

\tikzfeynmanset{every vertex={empty dot,minimum size=2.5mm}}
\vertex  at (-0.5,0) (eft);

\tikzfeynmanset{every vertex={dot,minimum size=1.5mm}}
\vertex  at (-0.5,0) (eft);

\tikzfeynmanset{every vertex={draw,minimum size=0pt, inner sep=0pt}}
\vertex at (-0.5,0) (l);
\vertex at (0,0.5) (u);
\vertex at (0,-0.5) (d);

\vertex at (-0.7,0.0) (e1);
\vertex at (0.2,0.5) (e4);
\vertex at (0.2,-0.5) (e5);

\diagram* {
	(l) -- [] (u) -- [] (d) -- [] (l)  ,
	(u) -- [] (e4),
	(d) -- [] (e5),
	};

\end{feynman}
\end{tikzpicture}
\end{gathered}
\;\Big)
\eeq
This avoids an IR divergence since the differentiated EFT vertex is quadratic in the internal momenta.
To give an example of a simple IR counterterm we can alternatively consider the following IR rearrangement:
\beq
\label{eq:IR2}
\Z\Big(\;
\begin{gathered}
\begin{tikzpicture}
\begin{feynman}[large, baseline=b]

\tikzfeynmanset{every vertex={empty dot,minimum size=2.5mm}}
\vertex  at (-0.5,0) (eft);

\tikzfeynmanset{every vertex={dot,minimum size=1.5mm}}
\vertex  at (-0.5,0) (eft);

\tikzfeynmanset{every vertex={dot,minimum size=1mm}}
\vertex  at (0,0.5) (udot);

\tikzfeynmanset{every vertex={draw,minimum size=0pt, inner sep=0pt}}
\vertex at (-0.5,0) (l);
\vertex at (0,0.5) (u);
\vertex at (0,-0.5) (d);

\vertex at (-0.7,0.0) (e1);
\vertex at (0.2,-0.5) (e5);

\diagram* {
	(l) -- [] (u) -- [] (d) -- [] (l)  ,
	(l) -- [] (e1),
	(d) -- [] (e5),
	};

\end{feynman}
\end{tikzpicture}
\end{gathered}
\;\Big)
=
-K_\eps\Big(\;
\begin{gathered}
\begin{tikzpicture}
\begin{feynman}[large, baseline=b]

\tikzfeynmanset{every vertex={empty dot,minimum size=2.5mm}}
\vertex  at (-0.5,0) (eft);

\tikzfeynmanset{every vertex={dot,minimum size=1.5mm}}
\vertex  at (-0.5,0) (eft);

\tikzfeynmanset{every vertex={dot,minimum size=1mm}}
\vertex  at (0,0.5) (udot);

\tikzfeynmanset{every vertex={draw,minimum size=0pt, inner sep=0pt}}
\vertex at (-0.5,0) (l);
\vertex at (0,0.5) (u);
\vertex at (0,-0.5) (d);

\vertex at (-0.7,0.0) (e1);
\vertex at (0.2,-0.5) (e5);

\diagram* {
	(l) -- [] (u) -- [] (d) -- [] (l)  ,
	(l) -- [] (e1),
	(d) -- [] (e5),
	};

\end{feynman}
\end{tikzpicture}
\end{gathered}
+
\tilde \Z\Big(
\begin{gathered}
\begin{tikzpicture}
\begin{feynman}[large, baseline=b]

\tikzfeynmanset{every vertex={dot,minimum size=1mm}}
\vertex  at (0,0.5) (udot);
\vertex  at (0,-0.5) (ddot);

\tikzfeynmanset{every vertex={draw,minimum size=0pt, inner sep=0pt}}
\vertex at (0,0.5) (u);
\vertex at (0,-0.5) (d);

%

\diagram* {
	(u) -- [half left] (d) -- [half left] (u)  ,
	};

\end{feynman}
\end{tikzpicture}
\end{gathered}
\Big)
\cdot
\begin{gathered}
\begin{tikzpicture}
\begin{feynman}[large, baseline=b]

\tikzfeynmanset{every vertex={empty dot,minimum size=2.5mm}}
\vertex  at (-0.5,0) (eft);

\tikzfeynmanset{every vertex={dot,minimum size=1.5mm}}
\vertex  at (-0.5,0) (eft);

\tikzfeynmanset{every vertex={dot,minimum size=1mm}}

\tikzfeynmanset{every vertex={draw,minimum size=0pt, inner sep=0pt}}
\vertex at (-0.5,0) (l);
\vertex at (0,-0.5) (d);

\vertex at (-0.7,0.0) (e1);
\vertex at (0.2,-0.5) (e5);

\diagram* {
	(l) -- [] (d) ,
	(l) -- [] (e1),
	(d) -- [] (e5),
	};

\end{feynman}
\end{tikzpicture}
\end{gathered}
\;\Big)\,.
\eeq
The IR counterterm can be evaluated using
$$
R^*\Big(\;
\begin{gathered}
\begin{tikzpicture}
\begin{feynman}[large, baseline=b]

\tikzfeynmanset{every vertex={dot,minimum size=1mm}}
\vertex  at (0,0.5) (udot);
\vertex  at (0,-0.5) (ddot);

\tikzfeynmanset{every vertex={draw,minimum size=0pt, inner sep=0pt}}
\vertex at (0,0.5) (u);
\vertex at (0,-0.5) (d);

%

\diagram* {
	(u) -- [half left] (d) -- [half left] (u)  ,
	};

\end{feynman}
\end{tikzpicture}
\end{gathered}
\Big)=0\quad \Rightarrow \quad
\tilde \Z\Big(\;
\begin{gathered}
\begin{tikzpicture}
\begin{feynman}[large, baseline=b]

\tikzfeynmanset{every vertex={dot,minimum size=1mm}}
\vertex  at (0,0.5) (udot);
\vertex  at (0,-0.5) (ddot);

\tikzfeynmanset{every vertex={draw,minimum size=0pt, inner sep=0pt}}
\vertex at (0,0.5) (u);
\vertex at (0,-0.5) (d);

%

\diagram* {
	(u) -- [half left] (d) -- [half left] (u)  ,
	};

\end{feynman}
\end{tikzpicture}
\end{gathered}
\;\Big)=-\Z\Big(\;
\begin{gathered}
\begin{tikzpicture}
\begin{feynman}[large, baseline=b]


\tikzfeynmanset{every vertex={draw,minimum size=0pt, inner sep=0pt}}
\vertex at (-0.5,0) (l);
\vertex at (0.5,0) (r);

\vertex at (-0.7,0.0) (e1);
\vertex at (0.7,0.0) (e2);

\diagram* {
	(l) -- [half left] (r) -- [half left] (l)  ,
	(l) -- [] (e1),
	(r) -- [] (e2),
	};

\end{feynman}
\end{tikzpicture}
\end{gathered}
\;\Big)=K_\eps\Big( \begin{gathered}
\begin{tikzpicture}
\begin{feynman}[large, baseline=b]


\tikzfeynmanset{every vertex={draw,minimum size=0pt, inner sep=0pt}}
\vertex at (-0.5,0) (l);
\vertex at (0.5,0) (r);

\vertex at (-0.7,0.0) (e1);
\vertex at (0.7,0.0) (e2);

\diagram* {
	(l) -- [half left] (r) -- [half left] (l)  ,
	(l) -- [] (e1),
	(r) -- [] (e2),
	};

\end{feynman}
\end{tikzpicture}
\end{gathered}  \Big)\,.
$$
Of course both eqs. (\ref{eq:IR1}) and (\ref{eq:IR2}) yield the same result after integration. We remark that even though it is formally required, the IR-counterterm gives a vanishing contribution since its complementary graph vanishes (due to the circled vertex being linear in the IR momentum at leading order). In a convenient normalisation one then gets
\beq
\Z\Big(
\begin{gathered}
\begin{tikzpicture}
\begin{feynman}[large, baseline=b]

\tikzfeynmanset{every vertex={empty dot,minimum size=2.5mm}}
\vertex  at (-0.7,0) (eft);

\tikzfeynmanset{every vertex={dot,minimum size=1.5mm}}
\vertex  at (-0.7,0) (eft);

\tikzfeynmanset{every vertex={draw,minimum size=0pt, inner sep=0pt}}
\vertex at (-0.7,0) (l);
\vertex at (0,0.7) (u);
\vertex at (0.7,0) (r);
\vertex at (0,-0.7) (d);

\vertex at (-1,0.0) (e1);
\vertex at (-1,-0.1) (e2);
\vertex at (0,1) (e3);
\vertex at (0,-1) (e4);
\vertex at (1,0.0) (e5);
\vertex at (1,-0.1) (e6);

\diagram* {
	(l) -- [quarter left] (u) -- [quarter left] (r) -- [quarter left] (d) -- [quarter left] (l)  ,
    (u) -- [] (d),
	(l) -- [] (e1),
	(r) -- [] (e5),
	};

\end{feynman}
\end{tikzpicture}
\end{gathered}
\;\Big) = -c_{4,3}^{(8)}g^3  \eta^{\alpha\beta} \frac{3}{2\eps^2}\,,
\eeq
leading to the final result:
\beq
\Z\Big(\;
\begin{gathered}
\begin{tikzpicture}
\begin{feynman}[large, baseline=b]

\tikzfeynmanset{every vertex={dot,minimum size=2.5mm}}
\vertex  at (-0.7,0) (eft);

\tikzfeynmanset{every vertex={draw,minimum size=0pt, inner sep=0pt}}
\vertex at (-0.7,0) (l);
\vertex at (0,0.7) (u);
\vertex at (0.7,0) (r);
\vertex at (0,-0.7) (d);

\vertex at (-1,0.1) (e1);
\vertex at (-1,-0.1) (e2);
\vertex at (0,1) (e3);
\vertex at (0,-1) (e4);
\vertex at (1,0.1) (e5);
\vertex at (1,-0.1) (e6);

\diagram* {
	(l) -- [quarter left] (u) -- [quarter left] (r) -- [quarter left] (d) -- [quarter left] (l)  ,
    (u) -- [] (d),
	(l) -- [] (e1),
	(l) -- [] (e2),
	(u) -- [] (e3),
	(d) -- [] (e4),
	(r) -- [] (e5),
	(r) -- [] (e6),
	};

\end{feynman}
\end{tikzpicture}
\end{gathered}
\;\Big)
= -c_{4,3}^{(8)}g^3  p_1.p_2 \frac{3}{2\eps^2}\,.
\eeq
We note that this counterterm is not yet expressible in the multigraph operator basis; only after summing over all permuations of the external momenta will the operator $\O_6^{(8)}$ emerge. We have thus demonstrated that with the aid of IR rearrangement and the $R^*$ operation one can conveniently extract UV counterterms of multi-scale multi-loop Feynman diagrams from self-energy type diagrams.

\subsection{Example calculation with the real scalar}
\label{s:excalc}
\subsubsection*{Dimension 6}

To renormalize the Green's functions one has to introduce a basis of 3 operators,%
\footnote{In this section we often suppress the superscripts ($n$) on the operators and their couplings, as we restrict to $n=6$ everywhere}%
\beq
\label{eq:L6OSbasis}
\Lambda^2 \, \L^{(6)}=\tilde c_6^{b} \, \tilde \O_6^{b}+\tilde c_4^{b} \, \tilde \O_4^{b}
+\tilde c_2^{b} \, \tilde \O_2^{b} \,,
\eeq   
where one choice of basis---the multigraph basis---would be 
\beq
\tilde \O_6^{b}=\frac{(g^b)^2 Z_2^3}{6!}\p^6,\qquad
\tilde \O_4^{b}=-\frac{g^b Z_2^2}{4}(\d^\mu\phi)(\d_\mu\phi)\phi^2,\qquad
\tilde \O_2^{b}=\frac{Z_2}{2}(\d_\mu\d_\nu \p)^2,
\eeq
and in general the bare couplings are 
$\tilde c^b_i=\tilde Z_{ij}\tilde c_j=(\delta_{ij}+\delta \hspace{-0.5mm} \tilde Z_{ij})\tilde c_j$.
However, from the Hilbert series \eqref{eq:Hilb-real}, we know that the physical Lagrangian is spanned by the single operator $ \O_6^{}=\tilde \O_6$. This also follows from the field redefinition 
  $$\phi \stackrel{\textsc{fr}}{\longrightarrow} \phi + \frac{\tilde c^{b}_2}{2\Lambda^2} \, \partial^2 \phi 
  + \frac{\tilde c^{b}_4
  -\tilde c^{b}_2}{12\Lambda^2} \, g^b \, Z_2 \, \phi^3 $$
which sets
\beq \label{eq:FRdim6x}
\L^{(6)} \stackrel{\textsc{fr}}{\longrightarrow} \l( \tilde c^{b}_6 
  - 10 \tilde c^{b}_4 
  + 10 \tilde c^{b}_2 \r) \tilde \O_6^{b} 
  \equiv c^{b}_6 \, \O_6^{b} \, ,
\eeq
where we see explicitly that there is only one physically independent parameter, which we called $c^{b}_6$ (without tilde). At tree level (\textit{i.e.} not for the counterterms), we are therefore free to neglect contributions from $\tilde c^{}_4$ and  
$\tilde c^{}_2$ and work with the Lagrangian
$$
\L^{(6)}= c_6^{} \ \tilde Z^{}_{6,6} \, , \tilde \O_6^{b}
+ c_6^{}\,\tilde Z^{}_{4,6} \, \tilde \O_4^{b} 
+ c_6^{}\,\tilde Z^{}_{2,6} \,  \tilde \O_2^{b} \,,
$$
with 
$$\tilde Z^{}_{6,6}=1+\delta \hspace{-0.5mm} \tilde Z_{6,6}(g) \,, \quad 
\tilde Z^{}_{4,6}=\delta \hspace{-0.5mm}\tilde Z_{4,6}(g) \,, \quad 
\tilde Z^{}_{2,6}= \delta \hspace{-0.5mm} \tilde Z_{2,6}(g) \,.
$$
Here we have made explicit that the non-physical operators are only required as counterterms. As stated before, these do not explicitly have to be included in the R$^*$ approach, so that one can restrict to Green's functions with insertions of only physical operators.
From \eqref{eq:FRdim6x}, we also learn that the renormalization constant of the physical coupling in the basis is given by
\beq \label{eq:dim6renrel}
Z^{}_{6,6} = \tilde Z^{}_{6,6} - 10 \tilde Z^{}_{4,6} + 10 \tilde Z^{}_{2,6} \, .
\eeq
The renormalization constants $\tilde Z_i$ are conveniently extracted using the $\Z$-operation,
\begin{align*}
\Z(\Gamma_2[c_6^{}\O_6^{}])&
\stackrel{\textsc{ibp}}{=}
\tilde Z_{2,6}^{} \, c_6 \, \tilde \O_2^{b}\,,\\
\Z(\Gamma_4[c_6^{}\O_6^{}])&
\stackrel{\textsc{ibp}}{=}
\tilde Z_{4,6}^{} \, c_6 \, \tilde \O_4^{b}\,,
\\
\Z(\Gamma_6[c_6^{}\O_6^{}])&
\stackrel{\textsc{ibp}}{=}
\tilde Z_{6,6}^{} \, c_6 \, \tilde \O_6^{b}
\,.
\end{align*}

\vspace{3mm}
A common alternative approach to extract the physical renormalization constants uses the EoM operators to span the unphysical space. By construction, this has the advantage that the field redefinition to move to the physical basis is trivial. 
At dimension six the two EoM operators can be chosen as:
\begin{align}
\label{eq:EoMdim6}
\E_1^{b(6)}&=\frac{1}{2}E_4[\p](\partial^2\phi)
\stackrel{\textsc{ibp}}{=}
\tilde \O_2^{b}+ \tilde \O_4^{b},\qquad\quad\\
\E_2^{b(6)}&=\frac{1}{12}E_4[\p]g^b Z_2\p^3 
\stackrel{\textsc{ibp}}{=}
\tilde \O_4^{b}+10\tilde \O_6^{b}\,.\nn
\end{align}
with the EoM, obtained from the bare Lagrangian, given by:%
\footnote{Since the lowest term in the EoM operator is proportional to $\partial^2\phi$, it seems convenient to use another choice of off-shell basis,
\beq
-\frac{g^b Z_2^2}{3!}\phi^3(\d^2\phi),\qquad \text{and} \qquad 
\frac{g_b \, Z_2}{2}(\d^2 \p)^2 \,.
\eeq
However, implementing momentum conservation is most conveniently done 
by projecting onto the multigraph basis, which is accomplished by 
$$ s_{ii} \stackrel{\textsc{ibp}}{=} -\sum_{j\neq i} s_{ij} \,. $$
}
\beq
E_4[\p] \coloneqq Z_2\d^2\p +\frac{g}{3!}Z_g Z_2^2\p^3\,.
\eeq
The dimension six Lagrangian in this basis is then given by:
\beq
\Lambda^2\, \L^{(6)}=c_6^{b} \O_6^{b}+\hat c_1^{b}\E_1^{b}+\hat c_2^{b}\E_2^{b} \,,
\eeq
 with the operator $\O_6^{}$ spanning the physical Lagrangian. We have indicated with the hatted notation that the couplings of the EoM operators are unphysical.

The renormalization constants $\tilde Z_i$ can be extracted as before,
\begin{align}
\Z(\Gamma_2[c_6^{}\O_6^{}])&
\stackrel{\textsc{ibp}}{=}
\tilde Z_{2,6}^{} \, c_6 \, \tilde \O_2^{b}
=
\hat Z_{1,6}^{} \, c_6 \, \E_1^{b}\Big|_2\,,\nn\\
 \label{eq:rendim6}
\Z(\Gamma_4[c_6^{}\O_6^{}])&
\stackrel{\textsc{ibp}}{=}
\tilde Z_{4,6}^{} \, c_6 \, \tilde \O_4^{b}
=
\hat Z_{1,6}^{} \, c_6 \, \E_1^{}\Big|_4+ \hat Z_{2,6}^{} \, c_6 \, \E_2^{b}\Big|_4\,,\\
\Z(\Gamma_6[c_6^{}\O_6^{}])&
\stackrel{\textsc{ibp}}{=}
\tilde Z_{6,6}^{} \, c_6 \, \tilde \O_6^{b}
=
Z_{6,6}^{} \, c_6 \, \O_6^{b}+ \hat Z_{2,6}^{} \, c_6 \, \E_2^{b}\Big|_6\,,\nn
\end{align}
where by $\E_i^{}\Big|_n$ we denote the contribution to the $n$-point vertex of the operator $\E_i$. 
From these equations and eq.\ (\ref{eq:EoMdim6}) we can establish the relations:
\beq
\tilde Z_{2,6}^{} =\hat Z_{1,6}^{}\,,\qquad 
\tilde Z_{4,6}^{} = \hat Z_{1,6}^{}+ \hat Z_{2,6}^{} \,,\qquad
\tilde Z_{6,6}^{} =Z_{6,6}^{}+10 \hat Z_{2,6}^{}\,.
\eeq
Inverting this system one finds the desired relation for $Z_{6,6}^{}$ as a function of the $\tilde Z_{ij}$s:
\beq \label{RelationZs}
Z_{6,6}^{}=\tilde Z_{6,6}^{}-10\tilde Z_{6,4} +10\tilde Z_{6,2}\,,
\eeq
which agrees with \eqref{eq:dim6renrel}.

\subsubsection*{Dimension 8}
A physical basis at mass dimension 8 can be constructed by choosing two EoM-independent operators from the off-shell basis. A convenient choice is given by $\O^{(8)}_8$ and $\O^{(8)}_{4,3}$. The following four EoM operators span the non-physical space, 
$$E_4[\phi] \cdot
\left\{
\frac{(g^b)^2Z_2^2}{240} \, \p^5, \ 
\frac{g_bZ_2}{4}\, \phi^2\d^2\p, \ 
\minus \frac{g_bZ_2}{4} \, \p(\d_\mu\p)(\d^\mu\p), \
\frac{1}{2} \, \d^4\phi 
\right\}.$$ 
Expressing the EoM operators in terms of the off-shell basis, we obtain:%
\footnote{In this section we again suppress most superscripts ($n$) since we restrict to $n=8$ everywhere}
\begin{align}
\E^{b}_1&=\tilde \O_6^{b}+28\tilde \O_8^{b}\,,\nn\\
\E^{b}_2&=\tilde \O_{4,1}^{b}+3\tilde \O_{4,2}^{b}+4\tilde \O_{4,3}^{b}+10\O_6^{b}\,,\nn\\
\E^{b}_3&=\tilde \O_{4,2}^{b}+2\tilde \O_{4,3}^{b}+2\tilde \O_{6}^{b}\,,\\
\E^{b}_4&=\tilde \O_2^{b}+\tilde \O_{4,1}^{b}+\tilde \O_{4,2}^{b}
\,,\nn
\end{align}
where the operators are defined in appendix \ref{s:IBPbasesConf}. 
Given these definitions we obtain the relations corresponding to eqs.\ (\ref{eq:rendim6}) at dimension 8:%
\footnote{ To avoid cumbersome notation, we suppress the dependence on the couplings, \textit{i.e.}~$Z_2 = Z_{2,(4,3)}\, c_{4,3} + Z_{2,8} \, c_8$, and similarly for the other renormalization constants.
} 
\begin{align}
\Z(\Gamma_2[c_8^{}\O_8^{}+c_{4,3}^{}\O_{4,3}^{}])&
\stackrel{\textsc{ibp}}{=}
\tilde Z^{}_2\tilde \O_{2}^{b(8)}=\hat Z^{b(8)}_4\E^{b(8)}_{4}\Big|_2, \nn\\
\Z(\Gamma_4[c_8^{}\O_8^{}+c_{4,3}^{}\O_{4,3}^{}])&
\stackrel{\textsc{ibp}}{=}
\tilde Z^{}_{4,1}\tilde \O_{4,1}^{b(8)}+\tilde Z^{}_{4,2}\tilde \O_{4,2}^{b(8)}
+\tilde Z^{}_{4,3}\tilde \O_{4,3}^{b(8)}\nn\\
\label{eq:rendim8}
&
=
\hat Z^{}_2 \E^{b}_{2}\Big|_4+\hat Z^{}_4 \E^{b}_{3}\Big|_4+\hat Z^{}_4 \E^{b}_{4}\Big|_4+ Z^{}_{4,3} \O_{4,3}^{b}\\
\Z(\Gamma_6[c_8^{}\O_8^{}+c_{4,3}^{}\O_{4,3}^{}])&
\stackrel{\textsc{ibp}}{=}
\tilde Z^{}_6 \tilde \O_{6}^{b} = 
\hat Z^{}_1 \E^{b}_{1}\Big|_6 +\hat Z^{}_2 \E^{b}_{2}\Big|_6 +\hat Z^{}_3 \E^{b}_{3}\Big|_6, \nn\\
\Z(\Gamma_8[c_8^{}\O_8^{}+c_{4,3}^{}\O_{4,3}^{}])&
\stackrel{\textsc{ibp}}{=}
\tilde Z^{}_8\tilde \O_{8}^{b}= Z^{}_8\O_{8}^{b}+\hat Z^{}_1 \E^{b}_{1}\Big|_8, \nn
\end{align}
This leads to the following system: 
\begin{align}
\tilde Z^{}_2-\hat Z^{}_4&=0, \nn\\
 \tilde Z^{}_{4,1}-\hat Z^{}_2-\hat Z^{}_4&=0, \nn\\
\label{eq:Zreldim8}
\tilde Z^{}_{4,2}-3\hat Z^{}_2-\hat Z^{}_3-\hat Z^{}_4&=0 \\
\tilde Z^{}_{4,3}-Z^{}_{4,3}-4\hat Z^{}_2-2\hat Z^{}_3&=0, \qquad\phantom{.} \nn\\
 \tilde Z^{}_6-\hat Z^{}_1-10\hat Z^{}_2-2\hat Z^{}_3&=0    , \nn\\
\tilde Z^{}_8-Z^{}_8-28\hat Z^{}_1&=0. \nn
\end{align}
This system can be solved easily to obtain the solution for the physical renormalization constants:  
\begin{align}
Z^{}_8 &= \tilde Z^{}_8-168\tilde Z^{}_{2}+112 
\tilde Z^{}_{4,1}+56\tilde Z^{}_{4,2}
-28 \tilde Z^{}_6 \,, \\
Z^{}_{4,3} &= 2 \tilde Z^{}_{4,1}-2\tilde Z^{}_{4,2}
+\tilde Z^{}_{4,3} \,,
\end{align}
together with similar relations that express the non-physical (hatted) renormalization constants in terms of the renormalization constants for Green's functions (tildes), which we are not interested in.

\section{Anomalous dimensions: results and zeros}\label{sec:results}

After discussing the technical details of our calculation, 
we will present the obtained anomalous dimensions of the massless complex scalar EFT in the minimal subtraction scheme. The anomalous dimensions will be provided and analyzed using a physical basis of conformal primary operators, for which we have collected our conventions in appendix \ref{s:IBPbasesConf}. 
The results for the massless $Z_2$-symmetric real scalar EFT exhibit a similar structure, and the results are provided in appendix \ref{s:resreal}. 
We summarize and comment on the general structures in the anomalous dimensions in section \ref{s:structure}.

\subsection{Details of the calculation}

The calculation of counterterms in this work is split into three parts: diagram generation, graph manipulations of the $R^*$-operation, and the final evaluation of the expressions after substitution of Feynman rules. 
First, all relevant 1PI loop diagrams are generated using the graph generator by Kaneko \cite{Kaneko:1994fd} via the new interface of \textsc{Form} \cite{ Vermaseren:2000nd,Ruijl:2017dtg,Ueda:2020wqk}. Only vertices corresponding to operators in a physical operators basis need to be included, as discussed in section \ref{sec:rstar}. Importantly, the diagram generator is equipped with a `Symmetrize' option. When employed, the program outputs only one representative diagram for all diagrams that are related by permutations of the external lines. The final result can thus be obtained by summing over all permutations of each representative. 
This realizes a large speed-up for the calculation of  higher-multiplicity correlators. For example, the real-scalar 12-point correlator with a single insertion of dimension 12 operators at one loop involves 4,407,546 distinct 1PI graphs of only 4 different representatives:
$$
%
\begin{gathered}
\begin{tikzpicture}
\begin{feynman}[large, baseline=b]

\tikzfeynmanset{every vertex={dot,minimum size=2.5mm}}
\vertex  at (-0.7,0) (eft) ;

\tikzfeynmanset{every vertex={draw,minimum size=0pt, inner sep=0pt}}
\vertex  at (-0.7,0) (l) ;
\vertex at (0.7,0) (r);

\vertex at (-2,0.9) (e1);
\vertex at (-2,0.7) (e2);
\vertex at (-2,0.5) (e3);
\vertex at (-2,0.3) (e4);
\vertex at (-2,0.1) (e5);
\vertex at (-2,-0.1) (e6);
\vertex at (-2,-0.3) (e7);
\vertex at (-2,-0.5) (e8);
\vertex at (-2,-0.7) (e9);
\vertex at (-2,-0.9) (e10);

\vertex at (1.2,0.3) (e11);
\vertex at (1.2,-0.3) (e12);

\diagram* {
	(l) -- [half left] (r) -- [half left] (l),
	(l) -- [] (e1),
	(l) -- [] (e2),
	(l) -- [] (e3),
	(l) -- [] (e4),
	(l) -- [] (e5),
	(l) -- [] (e6),
	(l) -- [] (e7),
	(l) -- [] (e8),
	(l) -- [] (e9),
	(l) -- [] (e10),
	(r) -- [] (e11),
	(r) -- [] (e12),
};

\end{feynman}
\end{tikzpicture}
\end{gathered}
\hspace{3mm} ,\hspace{5mm}
\begin{gathered}
\begin{tikzpicture}
\begin{feynman}[large, baseline=b]

\tikzfeynmanset{every vertex={dot,minimum size=2.5mm}}
\vertex  at (-0.7,0) (eft) ;

\tikzfeynmanset{every vertex={draw,minimum size=0mm, inner sep=0pt}}
\vertex  at (-0.7,0) (l) ;
\vertex at (0.7,0) (r);
\vertex at (0,0.6) (r1);
\vertex at (0.7,0) (r2);
\vertex at (0,-0.6) (r3);

\vertex at (-1.8,0.75) (e3);
\vertex at (-1.8,0.45) (e4);
\vertex at (-1.8,0.15) (e5);
\vertex at (-1.8,-0.15) (e6);
\vertex at (-1.8,-0.45) (e7);
\vertex at (-1.8,-0.75) (e8);
%
\vertex at (-0.16,0.97) (r1e1);
\vertex at (0.16,0.97) (r1e2);
\vertex at (1.07,0.16) (r2e1);
\vertex at (1.07,-0.16) (r2e2);
\vertex at (-0.16,-0.97) (r3e1);
\vertex at (0.16,-0.97) (r3e2);

\diagram* {
	(l) -- [half left] (r) -- [half left] (l),
	(l) -- [] (e3),
	(l) -- [] (e4),
	(l) -- [] (e5),
	(l) -- [] (e6),
	(l) -- [] (e7),
	(l) -- [] (e8),
	(r1) -- [] (r1e1),
	(r1) -- [] (r1e2),
	(r2) -- [] (r2e1),
	(r2) -- [] (r2e2),
		(r3) -- [] (r3e1),
		(r3) -- [] (r3e2),
};

\end{feynman}
\end{tikzpicture}
\end{gathered}
\hspace{3mm} ,\hspace{5mm}
%
\begin{gathered}
\begin{tikzpicture}
\begin{feynman}[large, baseline=b]

\tikzfeynmanset{every vertex={dot,minimum size=2.5mm}}
\vertex  at (-0.7,0) (eft) ;

\tikzfeynmanset{every vertex={draw,minimum size=0mm, inner sep=0pt}}
\vertex  at (-0.7,0) (l) ;
\vertex at (0.7,0) (r);
\vertex at (-0.18,0.6) (r1);
\vertex at (0.55,0.38) (r2);
\vertex at (0.55,-0.38) (r3);
\vertex at (-0.18,-0.6) (r4);

\vertex at (-1.5,0.45) (e4);
\vertex at (-1.5,0.15) (e5);
\vertex at (-1.5,-0.15) (e6);
\vertex at (-1.5,-0.45) (e7);
%
\vertex at (-0.43,0.88) (r1e1);
\vertex at (-0.15,0.96) (r1e2);
\vertex at (0.75,0.7) (r2e1);
\vertex at (0.93,0.5) (r2e2);
\vertex at (-0.43,-0.88) (r4e1);
\vertex at (-0.15,-0.96) (r4e2);
\vertex at (0.75,-0.7) (r3e1);
\vertex at (0.89,-0.46) (r3e2);

\diagram* {
	(l) -- [half left] (r) -- [half left] (l),
	(l) -- [] (e4),
	(l) -- [] (e5),
	(l) -- [] (e6),
	(l) -- [] (e7),
%
	(r1) -- [] (r1e1),
	(r1) -- [] (r1e2),
	(r2) -- [] (r2e1),
	(r2) -- [] (r2e2),
	(r3) -- [] (r3e1),
	(r3) -- [] (r3e2),
	(r4) -- [] (r4e1),
	(r4) -- [] (r4e2),
};

\end{feynman}
\end{tikzpicture}
\end{gathered}
\hspace{3mm} ,\hspace{5mm}
%
\begin{gathered}
\begin{tikzpicture}
\begin{feynman}[scale=0.9, baseline=b]

\tikzfeynmanset{every vertex={dot,minimum size=2.5mm}}
\vertex  at (-0.8,0) (l) ;

\tikzfeynmanset{every vertex={draw,minimum size=0mm, inner sep=0pt}}
\vertex at (0.8,0) (r);
\vertex at (-0.4,0.69) (r1);
\vertex at (0.4,0.69) (r2);
\vertex at (0.8,0) (r3);
\vertex at (0.4,-0.69) (r4);
\vertex at (-0.4,-0.69) (r5);

\vertex at (-1.27,0.16) (e5);
\vertex at (-1.27,-0.16) (e6);
%
\vertex at (-0.77,1.02) (r1e1);
\vertex at (-0.50,1.18) (r1e2);
\vertex at (0.50,1.18) (r2e1);
\vertex at (0.77,1.02) (r2e2);
\vertex at (1.27,0.16) (r3e1);
\vertex at (1.27,-0.16) (r3e2);
\vertex at (0.50,-1.18) (r4e1);
\vertex at (0.77,-1.02) (r4e2);
\vertex at (-0.77,-1.02) (r5e1);
\vertex at (-0.50,-1.18) (r5e2);

\diagram* {
	(l) -- [half left, line width=0.25mm] (r) -- [half left, line width=0.25mm] (l),
	(l) -- [line width=0.25mm] (e5),
	(l) -- [line width=0.25mm] (e6),
%
	(r1) -- [line width=0.25mm] (r1e1),
	(r1) -- [line width=0.25mm] (r1e2),
	(r2) -- [line width=0.25mm] (r2e1),
	(r2) -- [line width=0.25mm] (r2e2),
	(r3) -- [line width=0.25mm] (r3e1),
	(r3) -- [line width=0.25mm] (r3e2),
	(r4) -- [line width=0.25mm] (r4e1),
	(r4) -- [line width=0.25mm] (r4e2),
		(r5) -- [line width=0.25mm] (r5e1),
		(r5) -- [line width=0.25mm] (r5e2),
};
\end{feynman}
\end{tikzpicture}
\end{gathered}
\hspace{3mm}.
$$
The number of calculated diagrams for each correlator are listed in Table \ref{t:numberreal} for the real scalar and in Table \ref{t:numbercomplex} for the complex scalar.\footnote{As a technical trick to avoid diagrams with multiple insertions of EFT vertices, we label the operator by a `fake' additional external particle. This has the (unwanted) consequence that some scaleless diagrams are not recognized as such. For example, 
$$
%
\begin{gathered}
\begin{tikzpicture}
\begin{feynman}[large, baseline=b]

\tikzfeynmanset{every vertex={dot,minimum size=2.5mm}}
\vertex  at (0,0.52) (eft) ;

\tikzfeynmanset{every vertex={draw,minimum size=0pt, inner sep=0pt}}
\vertex  at (-0.6,0) (l) ;
\vertex at (0.6,0) (r);
\vertex  at (0,-0.52) (b) ;

\vertex at (-1.2,0.2) (e1);
\vertex at (-1.2,-0.2) (e2);
\vertex at (1,0.75) (ext);

\diagram* {
	(l) -- [half left] (r) -- [half left] (l),
	(eft)    -- [out=-50,in=50] (b) -- [out=130,in=-130] (eft),
	(l) -- [] (e1),
	(l) -- [] (e2),
	(eft) --[scalar] (ext)
};

\end{feynman}
\end{tikzpicture}
\end{gathered}
\hspace{3mm} ,
$$
where the fake particle is shown with a dashed line, and which integrates to zero, is included in the calculation (and in the counting in Tables \ref{t:numberreal} and \ref{t:numbercomplex}) of $\Gamma_2^{(8)}$.}  

Example diagrams for $\Gamma_8^{(8)}$ at three loops with insertions of the 8-point operator include
{
$$
\begin{gathered}
\begin{tikzpicture}
\begin{feynman}[large, baseline=b]

\tikzfeynmanset{every vertex={dot,minimum size=2.5mm}}
\vertex  at (-0.7,0) (l) ;

\tikzfeynmanset{every vertex={draw,minimum size=0mm, inner sep=0pt}}
\vertex at (0.7,0) (r);
\vertex at (0,0.67) (r1);
\vertex at (0.7,0) (r2);
\vertex at (0,-0.67) (r3);

\vertex at (-1.8,0.75) (e3);
\vertex at (-1.8,0.45) (e4);
\vertex at (-1.8,0.15) (e5);
\vertex at (-1.8,-0.15) (e6);
\vertex at (-1.8,-0.45) (e7);
\vertex at (-1.8,-0.75) (e8);
%
\vertex at (-0.16,1.27) (r1e1);
\vertex at (0.16,1.27) (r1e2);
\vertex at (1.07,0.16) (r2e1);
\vertex at (1.07,-0.16) (r2e2);
\vertex at (-0.16,-1.27) (r3e1);
\vertex at (0.16,-1.27) (r3e2);

\diagram* {
	(l) -- [half left] (r) -- [half left] (l),
	(l) -- [] (e3),
	(l) -- [] (e4),
	(l) -- [] (e5),
	(l) -- [] (e6),
	(l) -- [] (e7),
	(l) -- [] (e8),
	(r1) -- [out=-55,in=55] (r3) -- [out=125,in=-125] (r1),
	(r2) -- [] (r2e1),
	(r2) -- [] (r2e2),
};

\end{feynman}
\end{tikzpicture}
\end{gathered}
\hspace{3mm} ,\hspace{5mm}
%
%
\begin{gathered}
\begin{tikzpicture}
\begin{feynman}[large, baseline=b]

\tikzfeynmanset{every vertex={dot,minimum size=2.5mm}}
\vertex  at (-0.8,0) (l) ;

\tikzfeynmanset{every vertex={draw,minimum size=0mm, inner sep=0pt}}
\vertex at (0.8,0) (r);
\vertex at (0,0.375) (r2);
\vertex at (0,0.8) (r3);

\vertex at (-1.8,0.6) (e4);
\vertex at (-1.8,0.3) (e5);
\vertex at (-1.8,0) (e6);
\vertex at (-1.8,-0.3) (e7);
\vertex at (-1.8,-0.6) (e8);
%
%
\vertex at (1.4,0) (re1);
\vertex at (-0.16,1.27) (r2e1);
\vertex at (0.16,1.27) (r2e2);
\vertex at (-0.16,-1.27) (r3e1);
\vertex at (0.16,-1.27) (r3e2);

\diagram* {
	(l) -- [] (re1),
	(l) -- [out=45,in=135] (r) -- [out=-135,in=-45] (l),
	(l) -- [] (e4),
	(l) -- [] (e5),
	(l) -- [] (e6),
	(l) -- [] (e7),
	(l) -- [] (e8),
%
	(r2) -- [half left] (r3) -- [half left] (r2),
		(r3) -- [] (r2e1),
		(r3) -- [] (r2e2),
};

\end{feynman}
\end{tikzpicture}
\end{gathered}
\hspace{3mm} ,\hspace{5mm}
%
%
\begin{gathered}
\begin{tikzpicture}
\begin{feynman}[large, baseline=b]

\tikzfeynmanset{every vertex={dot,minimum size=2.5mm}}
\vertex  at (-0.6,0) (l) ;

\tikzfeynmanset{every vertex={draw,minimum size=0mm, inner sep=0pt}}
\vertex at (0.6,0) (r);
\vertex at (0,0.6) (r2);
\vertex at (0,-0.6) (r3);

\vertex at (-1.5,0.45) (e4);
\vertex at (-1.5,0.15) (e5);
\vertex at (-1.5,-0.15) (e6);
\vertex at (-1.5,-0.45) (e7);
%
%
\vertex at (-0.16,1.07) (r2e1);
\vertex at (0.16,1.07) (r2e2);
\vertex at (-0.16,-1.07) (r3e1);
\vertex at (0.16,-1.07) (r3e2);

\diagram* {
	(l) -- [half left] (r) -- [half left] (l),
	(l) -- [out=35,in=145] (r) -- [out=-145,in=-35] (l),
	(l) -- [] (e4),
	(l) -- [] (e5),
	(l) -- [] (e6),
	(l) -- [] (e7),
%
	(r2) -- [] (r2e1),
	(r2) -- [] (r2e2),
	(r3) -- [] (r3e1),
	(r3) -- [] (r3e2),
};

\end{feynman}
\end{tikzpicture}
\end{gathered}
\hspace{3mm} ,
$$\vspace{-4mm}$$
%
\begin{gathered}
\begin{tikzpicture}
\begin{feynman}[large, baseline=b]

\tikzfeynmanset{every vertex={dot,minimum size=2.5mm}}
\vertex  at (-0.8,0) (l) ;

\tikzfeynmanset{every vertex={draw,minimum size=0mm, inner sep=0pt}}
\vertex at (0.6,0.4) (rt);
\vertex at (0.6,-0.4) (rb);
\vertex at (0,0.42) (r3);

\vertex at (-1.8,0.45) (e4);
\vertex at (-1.8,0.15) (e5);
\vertex at (-1.8,-0.15) (e6);
\vertex at (-1.8,-0.45) (e7);
%
%
\vertex at (1,0.65) (rte1);
\vertex at (1,-0.65) (rbe1);
\vertex at (-0.16,0.87) (r3e1);
\vertex at (0.16,0.9) (r3e2);

\diagram* {
	(l) -- [out=35,in=165] (rt) -- [out=-150,in=0] (l),
	(l) -- [out=-35,in=-165] (rb) -- [out=150,in=0] (l),
	(rt) -- [] (rb),
		(rt) -- [] (rte1),
			(rbe1) -- [] (rb),
	(l) -- [] (e4),
	(l) -- [] (e5),
	(l) -- [] (e6),
	(l) -- [] (e7),
		(r3) -- [] (r3e1),
			(r3) -- [] (r3e2),
};

\end{feynman}
\end{tikzpicture}
\end{gathered}
\hspace{3mm} ,\hspace{5mm}
%
%
%
\begin{gathered}
\begin{tikzpicture}
\begin{feynman}[large, baseline=b]

\tikzfeynmanset{every vertex={dot,minimum size=2.5mm}}
\vertex  at (0,0) (c) ;

\tikzfeynmanset{every vertex={draw,minimum size=0mm, inner sep=0pt}}
\vertex at (0,0.6) (t);
\vertex at (0.6,0) (r);
\vertex at (-0.6,0) (l);

\vertex at (-0.15,-0.5) (e1);
\vertex at (0.15,-0.5) (e2);
\vertex at (0.15,0.9) (te1);
\vertex at (-0.15,0.9) (te2);
\vertex at (0.9,0.15) (re1);
\vertex at (0.9,-0.15) (re2);
\vertex at (-0.9,0.15) (le1);
\vertex at (-0.9,-0.15) (le2);

\diagram* {
	(c) -- [out=135,in=45] (l) -- [out=-45,in=-135] (c),
	(c) -- [out=45,in=135] (r) -- [out=-135,in=-45] (c),
	(c) -- [out=45,in=-45] (t) -- [out=-135,in=135] (c),
	(c) -- [] (e1),
	(c) -- [] (e2),
	(t) -- [] (te1),
	(t) -- [] (te2),
	(r) -- [] (re1),
	(r) -- [] (re2),
	(l) -- [] (le1),
	(l) -- [] (le2),
};

\end{feynman}
\end{tikzpicture}
\end{gathered}
\hspace{3mm}.
$$
} 

As a final example, diagrams for $\Gamma_4^{(6)}$ at five loops include,
{
$$
%
\begin{gathered}
\begin{tikzpicture}
\begin{feynman}[large, baseline=b]

\tikzfeynmanset{every vertex={dot,minimum size=2mm}}
\vertex  at (0,0) (l1) ;

\tikzfeynmanset{every vertex={draw,minimum size=0mm, inner sep=0pt}}
\vertex  at (0,0) (l) ;
\vertex at (0.8,0) (r1);
\vertex at (1.6,0) (r2);
\vertex at (2.4,0) (r3);
\vertex at (0.4,0.35) (v);

\vertex at (-0.4,0.2) (e1);
\vertex at (-0.4,-0.2) (e2);
\vertex at (2.7,0.2) (re1);
\vertex at (2.7,-0.2) (re2);

\diagram* {
	(l) -- [half left] (r1) -- [half left] (r2)
		-- [half left] (r3) -- [half left] (r2)
		-- [half left] (r1) -- [half left] (l),
	(l) -- [out=-10,in=-60] (v) -- [] (l),
	(e1) -- [] (l) --[](e2),
	(re1) -- [] (r3) --[](re2),
};

\end{feynman}
\end{tikzpicture}
\end{gathered}
\hspace{3mm} ,\hspace{5mm}
%
%
\begin{gathered}
\begin{tikzpicture}
\begin{feynman}[large, baseline=b]

\tikzfeynmanset{every vertex={dot,minimum size=2mm}}
\vertex  at (0,0) (l1) ;

\tikzfeynmanset{every vertex={draw,minimum size=0mm, inner sep=0pt}}
\vertex  at (0,0) (l) ;
\vertex at (0.8,-0.5) (r1);
\vertex at (1.6,0) (r2);
\vertex at (0,0.6) (r3);
\vertex at (1.6,0.6) (r4);
\vertex at (0.8,1.1) (r5);

\vertex at (-0.4,0.2) (e1);
\vertex at (-0.4,-0.2) (e2);
\vertex at (0.65,-0.8) (re1);
\vertex at (0.95,-0.8) (re2);

\diagram* {
	(l) -- [out=-90,in=180] (r1) --[ out=0,in=-90] (r2), 
	(l) -- [out=-30,in=-150] (r2) --[out=150,in=30] (l),
	(l) -- (r3),
	(r2) -- (r4),
	(r3) -- [out=-30,in=-150] (r4) --[out=150,in=30] (r3),
	(r3) -- [out=90,in=180] (r5) --[ out=0,in=90] (r4), 
	(e1) --[](l)--[](e2),
	(re1) --[](r1)--[](re2),
};

\end{feynman}
\end{tikzpicture}
\end{gathered}
\hspace{3mm} ,\hspace{5mm}
%
\begin{gathered}
\begin{tikzpicture}
\begin{feynman}[large, baseline=b]

\tikzfeynmanset{every vertex={dot,minimum size=1.7mm}}
\vertex  at (0,-0.52) (eft) ;

\tikzfeynmanset{every vertex={draw,minimum size=0mm, inner sep=0pt}}
\vertex  at (-0.6,0) (l) ;
\vertex at (0.6,0) (r) ;
\vertex at (0.56,-0.2) (rb);
\vertex at (-0.56,-0.2) (lb);
\vertex at (0.56,0.2) (rt);
\vertex at (-0.56,0.2) (lt);

\vertex at (0.8,0.4) (e1);
\vertex at (0.85,0.1) (e2);
\vertex at (-0.8,0.4) (le1);
\vertex at (-0.85,0.1) (le2);

\diagram* {
	(l) --[half left] (r) --[half left] (l),
	(eft) --[out=80,in=160] (rb) --[out=-155,in=35] (eft),
	(eft) --[out=100,in=20] (lb) --[out=-25,in=145] (eft),
	(e1) --[] (rt) --[](e2),
	(le1) --[] (lt) --[](le2)
};

\end{feynman}
\end{tikzpicture}
\end{gathered}
\hspace{3mm} ,\hspace{5mm}
%
\begin{gathered}
\begin{tikzpicture}
\begin{feynman}[large, baseline=b]

\tikzfeynmanset{every vertex={dot,minimum size=1.7mm}}
\vertex  at (0,0) (eft) ;

\tikzfeynmanset{every vertex={draw,minimum size=0mm, inner sep=0pt}}
\vertex  at (0,0) (c) ;
\vertex at (0,0.5) (t);
\vertex at (0,-0.5) (b);
\vertex at (-0.5,-0.5) (bl);
\vertex at (0.5,-0.5) (br);

\vertex at (0.2,0.8) (e1);
\vertex at (-0.2,0.8) (e2);
\vertex at (-0.8,-0.5) (e3);
\vertex at (0.8,-0.5) (e4);

\diagram* {
	(c) --[half left] (t) --[half left] (c),
	(c) --[half left] (b) --[half left] (c),
	(c) --[out=180,in=75] (bl) --[](br),
	(c) --[out=0,in=105] (br) -- [out=-135,in=-45](bl),
	(e1) --[] (t) --[] (e2),
	(e3) --[] (e4),
};

\end{feynman}
\end{tikzpicture}
\end{gathered}
\hspace{3mm}.
$$
} The flow of charge in the complex scalar theory may generate multiple inequivalent representatives from these examples.

\begin{table}[t]
\begin{center}
\begin{tabular}{||c |c |c||}
\hline
\textbf{Correlator} & \textbf{Diagrams} & \textbf{Representatives} \\\hline\hline&&\\[-4mm]
$\Gamma_{2}^{(4)}$	&$0+1+1+4+11$		&$0+1+1+4+11$ 
\\\hline&&\\[-4mm]
$\Gamma_{4}^{(4)}$	&$3+9+40+204+1165$	& $1+2+8+26+124$ 
\\\hline&&\\[-4mm]
$\Gamma_{2}^{(6)}$	&$0+0+0+3+15$		&$0+0+0+3+9$ 
\\\hline&&\\[-4mm]
$\Gamma_{4}^{(6)}$	&$0+4+28+184+1307$		&$0+1+4+21+130$ 
\\\hline&&\\[-4mm]
$\Gamma_{6}^{(6)}$	&$15+135+1355+13680+136532$		&$1+4+21+132+942$ 
\\\hline&&\\[-4mm]
$\Gamma_{2}^{(8)}$	&$0+2+4$		&$0+1+3$ 
\\\hline&&\\[-4mm]
$\Gamma_{4}^{(8)}$	&$6+27+157$		&$1+4+18$ 
\\\hline&&\\[-4mm]
$\Gamma_{6}^{(8)}$	&$45+726+8031$		&$1+8+60$ 
\\\hline&&\\[-4mm]
$\Gamma_{8}^{(8)}$	&$1288+35609+617925$		&$2+18+166$ 
\\\hline&&\\[-4mm]
$\Gamma_{2,4,6,8,10}^{(10)}$	&$0,\ 6,\ 60,\ 1470,\ 66195$		&$0,\ 1,\ 2,\ 2,\ 3$ \\\hline&&\\[-4mm]
$\Gamma_{2,4,6,8,10,12}^{(12)}$	
			&$0,\ 6,\ 60,\ 1498,\ \ 66780,\ \ 4407546$		
			&$0,\ 1,\ 2,\ 3,\ 3,\ 4$ \\\hline
\end{tabular}
\vspace{-3mm}
\end{center}
\caption{The number of calculated 1PI diagrams for the real scalar in a natural basis,
where the `$+$' explicitly show the contributions of each considered loop order. 
$\Gamma_{N}^{(n)}$ refers to the correlator with $N$ external legs at mass dimension $n$.
When there exist multiple independent operators with the same field content, these are accounted for by a single vertex.
\label{t:numberreal}  
}
\end{table}
\begin{table}[h]
\begin{center}
\begin{tabular}{||c |c |c||}
\hline
\textbf{Correlator} & \textbf{Diagrams} & \textbf{Representatives} \\\hline\hline&&\\[-4mm]
$\Gamma_{2}^{(4)}$	& $0+1+2+10+54	$	&$0+ 1 +2+10+54$
\\\hline&&\\[-4mm]
$\Gamma_{4}^{(4)}$	& $3+13+79+564+4604	$	&$2+ 6 +30+183+1390$
\\\hline&&\\[-4mm]
$\Gamma_{2}^{(6)}$	& $0+2+8+52$	& $0+2+8+52$
\\\hline&&\\[-4mm]
$\Gamma_{4}^{(6)}$	& $6+43+357+3237$	& $3+17+120+1007$
\\\hline&&\\[-4mm]
$\Gamma_{6}^{(6)}$	& $78+1332+18874+256452	$&$ 7+78+915+11161$
\\\hline&&\\[-4mm]
$\Gamma_{2}^{(8)}$	&$ 0+2+8$	& $0+2+8$
\\\hline&&\\[-4mm]
$\Gamma_{4}^{(8)}$	& $3+43+357$	& $6+17+120 $
\\\hline&&\\[-4mm]
$\Gamma_{6}^{(8)}$	& $78+1338+19000$	& $7+80+933$
\\\hline&&\\[-4mm]
$\Gamma_{8}^{(8)}$	&$1780+60840+1373395$	& $17+329+5721$
\\\hline&&\\[-4mm]
$\Gamma_{2}^{(10)} $	& $0+2  $	& $0+2  $
\\\hline&&\\[-4mm]
$\Gamma_{4}^{(10)}$	& $ 3+43 $	& $ 6+17 $
\\\hline&&\\[-4mm]
$\Gamma_{6}^{(10)}$	&	$78+1338 $  & $7+80 $
\\\hline&&\\[-4mm]
$\Gamma_{8}^{(10)}$	&	$1780+60848$  & $17+331$
\\\hline&&\\[-4mm]
$\Gamma_{10}^{(10)}$	&$65954+3860505$	& $33+1135$
\\\hline
\end{tabular}
\vspace{-3mm}
\end{center}
\caption{The same as Table \ref{t:numberreal} for the complex scalar. \label{t:numbercomplex}}
\end{table}

Subsequently, the generated diagrams are manipulated through an implementation of the $R^*$-operation in Maple \cite{Maple} and \textsc{Form}, as explained in some more detail in section \ref{sec:rstar}. The result of this procedure is that all diagrams are rewritten in terms of massless self-energy diagrams, which can be evaluated as products of at most $L-1$ loop massless propagator integrals in \textsc{Form} via the \textsc{Forcer} program \cite{Ruijl:2017cxj}.

Evaluation times were of the order of a few days for the more computationally demanding calculations. With refinements of the computer code, it would  be possible to extend the results to higher mass dimension and loop number ($\leq5$) with the current setup. However, we did observe a fairly fast growth of the generated expression sizes due mostly to the high-order derivatives which have to be applied to render the degree of divergence to be logarithmic at higher mass dimension; this may be seen as the main bottleneck of the method. The problem clearly becomes increasingly worse with higher loops as well, since to apply the derivatives the product rule is used along momentum flows through the diagram. Hence the more edges in the graph the more terms are generated by each derivative.
Without further optimizations, pushing to higher loops and mass-dimension would  require considerably longer computing times, potentially of the order of weeks or months.

\subsubsection*{Checks on the calculation}

Our results have been checked in various ways. Where available, the anomalous dimensions have been confirmed against existing literature and general results on the renormalization structure to be discussed below.
In addition, our set-up allows for self-consistency checks. Firstly, all calculated renormalization constants are independent of logarithms $\log(\mu^2)$, which indicates that subdivergences have been subtracted correctly. Furthermore, at the level of counterterms, all $1/\eps$ poles (up to $1/\eps^L$ for $L$ loops) were retained, which conspire in a non-trivial way to generate finite anomalous dimensions. This constitutes an important check at higher loops. In essence, this checks relations between the higher poles and lower poles at the lower loop level. Finally, at one loop level and beyond, consistent results were obtained by calculations with different choices of operator basis. This involves an independent calculation with different Feynman rules and potentially even different diagrams.

\subsection{Results}
\subsubsection*{Mass dimension 4}

For the massless complex scalar, defined in \eqref{Lagr4},
the 5-loop $\beta$-function is determined to be 
\begin{align*}
\beta(g(\mu)) &= -2\epsilon g
+5g^2-15g^3
+\Big(48\zeta_3+ \frac{617}{8}\Big)g^4-\Big(514\zeta_3 -120\zeta_4 +760\zeta_5 +\frac{7975}{16}\Big)g^5
\nonumber\\
&\hspace{1.2cm} +\Big(363\zeta_3^2+\frac{84353\zeta_3 }{16}-\frac{13005\zeta_4 }{8}+10355\zeta_5-\frac{7125\zeta_6 }{2} +13230\zeta_7 +\frac{931401}{256}\Big)g^6 \, ,
\end{align*}
and the anomalous dimension of the field is
	\begin{align*}	
		\gamma_\phi & = (\textcolor{black}{0g}) 
			-\frac{1}{4}g^2 
			+ \frac{5}{16}g^3 
			- \frac{165}{64}g^4 
			+ \left(- \frac{37}{16}\zeta_3 + 6\zeta_4 + \frac{5169}{256}\right) g^5 \, .
\end{align*}
These results agree with \cite{Kleinert:1991rg}; the results for a complex scalar are equivalent to those in the O(2) model.

\subsubsection*{Mass dimension 6}

In the basis of conformal primary operators at mass dimension 6, 
the anomalous dimension matrix up to 4-loops is
\begin{adjustwidth}{-0.7cm}{0cm}
\begin{equation} \label{eq:AD6}
%
\gamma_\text{\textsc{c}}^{(6)} = 
\begin{pNiceMatrix}[first-row]
\begin{matrix}\textcolor{gray}{g^2\,\O^{(6)c}_6}\\[3mm]\end{matrix}
&\begin{matrix}\textcolor{gray}{g\,\O^{(6)c}_4}\\[3mm]\end{matrix}\\
\begin{matrix}
	14g-\frac{297g^2}{2}+\Big(816\zeta_3 +\frac{14981}{8}\Big)g^3\\-\Big(\frac{32087\zeta_3 }{2}-2892\zeta_4 +23320\zeta_5 +\frac{888983}{32}\Big)g^4
\end{matrix} \ \ \ \ 
& \begin{matrix}
	\textcolor{black}{0 g}+\frac{45g^2}{2}
	-\Big(216\zeta_3 +\frac{6153}{16}\Big)g^3\\
	+\Big(783\zeta_3 +8100\zeta_5 +\frac{74079}{16}\Big)g^4  
\end{matrix} 
	\\[1cm]
%
%
\textcolor{black}{0 g}\textcolor{black}{+0 g^2}\textcolor{black}{+0 g^3}
-\frac{5g^4}{6}
& \begin{matrix}
-g+\frac{13g^2}{2}-\Big(36\zeta_3 +\frac{383}{12}\Big)g^3\\+\Big(\frac{769\zeta_3 }{2}-123\zeta_4 +560\zeta_5 +\frac{7893}{32}\Big)g^4
\end{matrix}
%
\end{pNiceMatrix}
\,,
\end{equation}
\end{adjustwidth}
\vspace{2mm}
which multiplies the couplings of the operators 
$$
\left\{ \O^{(6)c}_6 \ , \ \O^{(6)c}_4   \right\},
$$
defined in appendix \ref{s:IBPbasesConf}. These operators appear above the columns in the matrix for the convenience of reading. Here and below, the anomalous dimensions encode the scale dependence of a vector of couplings in the same order. 
Recall that the bare operators (and their associated couplings) involve a scaling by factors of $g$, as in \eqref{bareoperators}. This aligns the power in $g$ of each term in the anomalous dimensions with its loop order.

\subsubsection*{Mass dimension 8}

At mass dimension 8, we work in the conformal primary basis. We parameterize the general four-point conformal operator as a function of two parameters $x$ and $y$; a basis is given by setting $(x,y)=(1,0)$ and $(0,1)$. The anomalous dimension matrix up to 3 loops is

{
\begin{adjustwidth}{-1.7cm}{0cm}
\begin{align*}
\gamma_\text{\textsc{c}}^{(8)} = 
\setlength\arraycolsep{1mm}
\begin{pNiceMatrix}[first-row]
\begin{matrix}\textcolor{gray}{g^3\,\O^{(8)c}_8}\\[3mm]\end{matrix}
&\begin{matrix}\textcolor{gray}{g^2\,\O^{(8)c}_6}\\[3mm]\end{matrix}
&\begin{matrix}\textcolor{gray}{g\,\O^{(8)c}_4\scriptstyle{(1,0)}}\\[3mm]\end{matrix}
&\begin{matrix}\textcolor{gray}{\ g \, \O^{(8)c}_4\scriptstyle{(0,1)} }\\[3mm]\end{matrix}\\
%
%
\matrixx{29g-409g^2\\+\Big(2352\zeta_3 +\frac{57765}{8}\Big)g^3}
	&\matrixx{\textcolor{black}{0 g}
		+240g^2\\
		-\Big(2304\zeta_3 +5882\Big)g^3}
	&\matrixx{\frac{54216g}{5}-\frac{958314g^2}{5}\\+\Big(\frac{4083264\zeta_3}{5}+\frac{55731313}{15}\Big)g^3}
	&\matrixx{-\frac{8856g}{5}+\frac{159894g^2}{5}\\-\Big(\frac{713664\zeta_3 }{5}+\frac{9281093}{15}\Big)g^3}
	\\[7mm] 
%
%
\textcolor{black}{0g}\textcolor{black}{+0 g^2} \textcolor{black}{ + 0g^3}
	&\matrixx{4g-\frac{122g^2}{3}\\+\Big(216\zeta_3 +\frac{4559}{12}\Big)g^3}
	&\matrixx{\frac{679g}{5}
	-\frac{14209g^2}{12}\\
	+\Big(\frac{29216\zeta_3}{5}+\frac{3248605}{324}\Big)g^3}
	&\matrixx{-\frac{164g}{5}
		+\frac{9233g^2}{36}\\
		-\Big(\frac{6056\zeta_3 }{5}+\frac{2740291}{1296}\Big)g^3}
	\\[7mm]
%
%
\textcolor{black}{0g+0g^2+0g^3}
	&\textcolor{black}{0g}\textcolor{black}{+0 g^2}
		-\frac{5g^3}{108}
	&\matrixx{\frac{11g}{3}-\frac{29g^2}{180}\\[1mm]-\Big(\frac{32\zeta_3}{3}+\frac{97091}{9720}\Big)g^3}
	&\matrixx{-\frac{4g}{3}+\frac{1057g^2}{540}\\[1mm]-\Big(\frac{16\zeta_3}{3}+\frac{10868}{1215}\Big)g^3}
	\\[7mm]	
%
%
\textcolor{black}{0g+0g^2+0g^3}
	&\textcolor{black}{0g} \textcolor{black}{+0 g^2}
		-\frac{115g^3}{324}
	&\matrixx{\frac{46g}{3}-\frac{14557g^2}{540}\\+\Big(96\zeta_3 +\frac{198001}{2430}\Big)g^3}
	&\matrixx{-\frac{19g}{3}+\frac{2809g^2}{180}\\-\Big(\frac{176\zeta_3}{3}+\frac{718739}{9720}\Big)g^3}
			\end{pNiceMatrix} ,
		\end{align*}
\end{adjustwidth} 
}

\noindent
which multiplies the couplings of 
$$
\left\{
\O^{(8)c}_8 \ , \ 
\O^{(8)c}_6 \ , \ 
\O^{(8)c}_4{\scriptstyle{(1,0)}} \ , \ 
\O^{(8)c}_4{\scriptstyle{(0,1)}} \right\} \,.
$$
It is important to note that the conformal primary basis is not unique, since other choices for $\O^{(8)c}_{4}\scriptstyle{(x,y)}$ can be made. 
Any other set of primary operators consists of a linear combination of the presently chosen operators. Therefore the anomalous dimension for any other choice of conformal primary basis can be obtained by a simple similarity transform, without the need to invoke field redefinitions or integration by parts (see section \ref{s:alpha}).
The operators in a different basis are constructed by
\begin{equation}
\mathcal{O}^{(8)c}_{4}{\scriptstyle{(a,b)}} = a \, \mathcal{O}^{(8)c}_{4}{\scriptstyle{}(1,0)}+ b \, \mathcal{O}^{(8)c}_{4}{\scriptstyle{}(0,1)}\,,
\label{eq:choice8a}
\end{equation}
\begin{equation}
\mathcal{O}^{(8)c}_{4}{\scriptstyle{(a',b')}} = a' \, \mathcal{O}^{(8)c}_{4}{\scriptstyle{}(1,0)}+ b' \, \mathcal{O}^{(8)c}_{4}{\scriptstyle{}(0,1)}\,,
\label{eq:choice8b}
\end{equation}
which implies the relations on their associated couplings,
\begin{equation}
c^{(8)c}_{4}{\scriptstyle{(1,0)}} = a \, 
c^{(8)c}_{4}{\scriptstyle{(a,b)}}+ a' \, c^{(8)c}_{4}{\scriptstyle{(a',b')}} \,,
\end{equation}
\begin{equation}
c^{(8)c}_{4}{\scriptstyle{(0,1)}} = b \, 
c^{(8)c}_{4}{\scriptstyle{(a,b)}}+ b' \, c^{(8)c}_{4}{\scriptstyle{(a',b')}} \,.
\end{equation}
We thus find that the anomalous dimension matrix for the basis $\{\mathcal{O}^{(8)c}_{8}, \mathcal{O}^{(8)c}_{6}, \mathcal{O}^{(8)c}_{4}{\scriptstyle{(a,b)}}, \mathcal{O}^{(8)c}_{4}{\scriptstyle{(a',b')}} \}$ is given by 
$B^{-1} \, \gamma_\text{\textsc{c}}^{8} \, B$, with
\begin{equation}
B_{4\times 4}=\begin{pmatrix}
\hspace{1.5mm}\begin{matrix}\mathbbm{1}\end{matrix} 
& \hspace{2.5mm}\begin{matrix}{0}\end{matrix}\\[1mm]
\hspace{1.5mm}\begin{matrix}{0}\end{matrix} \ 
& \hspace{2.5mm} \begin{matrix} a & a' \\ b & b' \end{matrix} 
\end{pmatrix} \, ,
\label{eq:choiceBmat}
\end{equation}
where $B$ needs to be invertible, i.e.\ $\det B \neq 0$. A particular choice of $B$ that symmeterizes the ADM is presented in appendix~\ref{s:ortho}.

{

\begin{table}[H]

\subsection*{Mass dimension 10}
At mass dimension 10 in the conformal primary basis, the anomalous dimension matrix up to two loops is

\begin{adjustwidth}{-1cm}{0cm}
\hskip-0.9cm 
\setlength\arraycolsep{4mm}
\begin{align*}
&
\hspace{1cm}
  \begin{NiceMatrix}
  {\begin{matrix}\textcolor{gray}{\hspace{-2mm} g^4\,\O^{(10)c}_{10}}\\[2mm]\end{matrix}}
  &\hspace{5mm}\begin{matrix}\textcolor{gray}{g^3\,\O^{(10)c}_8}\\[2mm]\end{matrix}
  &\hspace{4mm}\begin{matrix}\textcolor{gray}{g^2\,\O^{(10)c}_6\scriptstyle{(1,0,0,0)}}\\[2mm]\end{matrix}
  &\hspace{2mm}{\begin{matrix}\textcolor{gray}{g^2\,\O^{(10)c}_6\scriptstyle{(0,1,0,0)}}\\[2mm]\end{matrix}}\\
  \end{NiceMatrix}
  \\[1.5mm]
 \setlength\arraycolsep{3mm}
\gamma^{(10)}_\text{\textsc{c}} = 
&\left(
\begin{NiceMatrix}[]
%
%
 50g-\frac{1725g^2}{2}
&             \textcolor{black}{0g} +1125g^2
 &     \frac{412500g}{7}-\frac{61613665g^2}{42}
 &      \frac{3300g}{7}-\frac{19655g^2}{6}
&\\[2mm]
\textcolor{black}{0g}\textcolor{black}{+0g^2}
&           15g-\frac{385g^2}{2}
 &      \frac{1474g}{7}-\frac{83689g^2}{45}
 &     \frac{234g}{7}-\frac{386467g^2}{315}
  &\\[2mm]
\textcolor{black}{0g+0g^2}
&               \textcolor{black}{0g}\textcolor{black}{+0g^2}
 &      \frac{35g}{6}-\frac{678389g^2}{15120}
 &      -\frac{g}{6}+\frac{24379g^2}{15120}
  & \\[2mm]
      \textcolor{black}{0g+0g^2}
&          \textcolor{black}{0g}\textcolor{black}{+0g^2}
&        \frac{4g}{3}-\frac{331g^2}{630}
 &       -\frac{2g}{3}-\frac{5401g^2}{945}
 & (...) \\[2mm]
      \textcolor{black}{0g+0g^2}
&       \textcolor{black}{0g}\textcolor{black}{+0g^2}
 &      \frac{15g}{2}-\frac{247109g^2}{5040}
 &      -\frac{g}{2}-\frac{4127g^2}{1680}
  & \\[2mm]
                0
&               0
 &               0
 &              0
  & \\[2mm]
         \textcolor{black}{0g+0g^2}
&       \textcolor{black}{0g+0g^2}
 &            \textcolor{black}{0g}\textcolor{black}{+0g^2}
 &       \textcolor{black}{0g}\textcolor{black}{+0g^2}
 & \\[2mm]  
%
        \textcolor{black}{0g+0g^2}
&       \textcolor{black}{0g+0g^2}
 &            \textcolor{black}{0g}\textcolor{black}{+0g^2}
 &       \textcolor{black}{0g}\textcolor{black}{+0g^2}
  & \\[2mm]
\end{NiceMatrix}
\right.\\[7mm]
%
%
%
%
&
\hspace{21mm}
\begin{NiceMatrix}
{\begin{matrix}\textcolor{gray}{\hspace{-2mm}g^2\,\O^{(10)c}_6\scriptstyle{(0,0,1,0)}}\\[2mm]\end{matrix}}
&\hspace{3mm}\begin{matrix}\textcolor{gray}{g^2\,\O^{(10)c}_6\scriptstyle{(0,0,0,i)}}\\[2mm]\end{matrix}
&\hspace{0mm}\begin{matrix}\textcolor{gray}{g\,\O^{(10)c}_4\scriptstyle{(1,0)}}\\[2mm]\end{matrix}
&\hspace{10mm}{\begin{matrix}\textcolor{gray}{g\,\O^{(10)c}_4\scriptstyle{(0,1)}}\\[2mm]\end{matrix}}
\end{NiceMatrix}
\\[2mm]
& 
\left. 
\begin{NiceMatrix}
%
  &    \frac{141900g}{7}-\frac{6734025g^2}{14}
  &             0
   &   \frac{500460g}{7}-\frac{31376515g^2}{21}
   &     64740g-\frac{16819700g^2}{21}\\[2mm]
 &       -\frac{326g}{7}+\frac{4577g^2}{7}
  &             0
   &   -\frac{43732g}{35}+\frac{5863532g^2}{315}
   &   \frac{14272g}{5}-\frac{15354406g^2}{315}\\[2mm]
 &              \frac{7g}{6}-\frac{28265g^2}{3024}
  &             0
   &   -\frac{1006g}{105}-\frac{1249571g^2}{15120}
   &   \frac{967g}{30}-\frac{1903001g^2}{3780}\\[2mm]
 (...)&         
         2g-\frac{14965g^2}{756}
  &              0
  &   \frac{4159g}{105}-\frac{3146033g^2}{7560}
   &    -\frac{914g}{15}+\frac{103037g^2}{315}\\[2mm]
 &         
      \frac{7g}{2}-\frac{31333g^2}{1008}
  &             0
   &   \frac{1769g}{35}-\frac{243011g^2}{336}
   &   -\frac{573g}{10}-\frac{37049g^2}{1260}\\[2mm]
 &         
               0
  &      \frac{10g}{3}-\frac{770g^2}{27}
   &             0
   &            0\\[2mm]
  &         
     \textcolor{black}{0g}\textcolor{black}{+0g^2}
  &    0
   &        -2g+\frac{223g^2}{24}
   &        -g+\frac{7g^2}{24}\\[2mm] 
   &         
     \textcolor{black}{0g}\textcolor{black}{+0g^2}
  &    0
   &         -g+\frac{7g^2}{12}
   &       -3g+\frac{251g^2}{24}\\[2mm]
\end{NiceMatrix}
\right)
%
\end{align*}

\end{adjustwidth}

\end{table}

}

\noindent
Here, we note that the Wilson coefficient that carries a factor of $i$, which is necessary to give a Hermitian term in the Lagrangian, does not mix into any of the real couplings at all loop orders. 
The operators with imaginary Wilson coefficients are C-odd, while operators with real couplings are C-even, which prevents their mixing.

\subsection{Structure of the ADM}\label{s:structure}

\begin{figure}
  \centering
  \includegraphics[clip, trim=0cm 1.6cm 7.5cm 4.2cm, width=11cm]{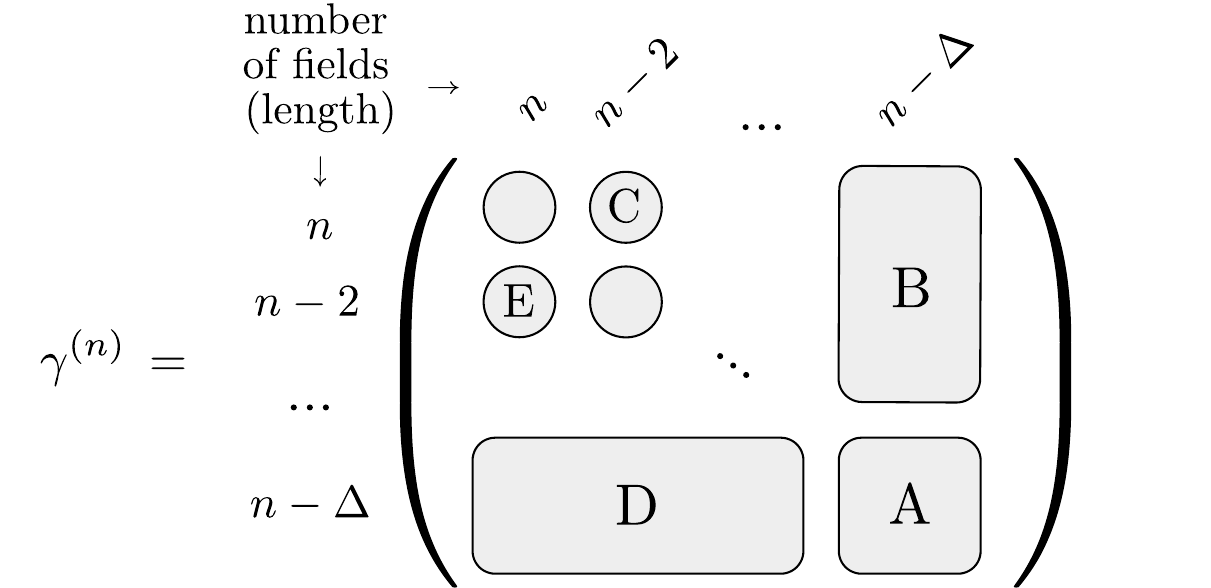}
  \caption{Schematic representation of the structure in the anomalous dimension matrix of the complex scalar at mass dimension $n$ in a natural basis. 
  A: sub-matrices on the diagonal encode intra-length mixing. They depend only on the conformal primary part of operators and become symmetric at one-loop for orthonormal operators. 
  B: Sub-matrices on the upper triangle encode inter-length mixing of a smaller operator into a larger operator -- generally non-zero in a basis of primaries.
  C: Entry corresponding to the mixing of $\O_{n\text{-}2}^{(n)}$ into $\O_{n}^{(n)}$ -- one-loop zero only if $\O_{n\text{-}2}^{(n)}$ is conformal primary.
  D: Sub-matrices on the lower triangle encode inter-length mixing of a larger operator into a smaller operator -- zeros up to $L=\Delta$, explained by \cite{Bern2020}.
  E: Entry corresponding to the mixing of $\O_{n}^{(n)}$ into $\O_{n\text{-}2}^{(n)}$, three-loop zero for any $\O_{n\text{-}2}^{(n)}$. This result is stronger than the minimal non-renormalization result predicted by \cite{Bern2020}.
  \label{fig:adm_pic}}
\end{figure}

In commenting on the structure of the ADM, we distinguish the cases of \emph{inter-length} and $\text{\emph{intra-length}}$ mixing, in which the considered operators have different or the same length (number of fields), respectively. This allows us to split the ADMs into sub-matrices depending on the type of mixing that is considered. On the diagonal, the various blocks encode the intra-length mixing, while the off-diagonal blocks describe the mixing of operators with different numbers of fields, 
as shown in figure \ref{fig:adm_pic}. In our conventions, blocks in the upper triangle describe the mixing of a smaller operator (\textit{i.e.}~with a smaller number of fields) into a larger operator and vice versa for the lower triangle.

We will also make use of the terminology `natural' when describing operators or bases  to mean an operator of field content that matches the leading on-shell amplitude to which it contributes (as a contact term). Equivalently, a natural operator does not vanish when all its legs are taken on-shell. A natural basis is a basis containing only operators with the least number of derivatives possible.

\subsubsection*{The general structure at one loop}

We make the following points on the general structure of the ADM, see Fig.~\ref{fig:adm_pic},
\begin{itemize}
\item \textbf{At one-loop, the inter-length blocks on the lower triangle are zero.} This is simply because the diagrams that potentially contribute are scaleless, if they exist. 

\item \textbf{Inter-length mixing on the upper triangle is generically non-zero for a basis of conformal primaries}. 
More specifically, these entries are dependent on both primary components (that is annihilated by $K$) and non-primary components of operators in the basis. If a pure primary basis is chosen, then the blocks generically contain non-zero entries. 
That is, to zero these blocks one has to move away from a basis of conformal primaries. Note that this is a full one-loop statement; at leading order in the coupling, these upper triangle blocks are zero simply because no diagrams exist.
However, in our results, we do observe that at mass dimension $n$ the mixing of the $n-2$ point operator into the $n$ point operator is zero \emph{only} if the $n-2$ point operator is conformal primary (entry {\bf C} in Fig.~\ref{fig:adm_pic}). 
In appendix \ref{s:zeronn-2}, we explicitly prove that this particular entry is zero in the basis of conformal primaries for any mass dimension. We observe that in the off-shell method, contributions to this entry from $\Gamma_{n-2}[\O^{(n)c}_{n-2}]$ (after field redefinitions) and $\Gamma_{n}[\O^{(n)c}_{n-2}]$ cancel, but an underlying explanation has not been found. Also in an on-shell method, the calculation would involve cancellations between contributions from different diagrams

\item {\bf Intra-length mixing is controlled only by the conformal primary part of the operator basis.} This was argued in Section~\ref{s:alpha}.
That is, non-primary components of operators in a chosen basis do not affect the intra-length mixing at one loop. Symmeterizing this block is possible by choosing an orthonormal basis for the conformal components, a fact proven in \cite{Craigie1985,Rychkov:2015naa}; we provide a proof using on-shell methods in appendix \ref{s:ortho}. One particular choice of orthogonal operators then always exists to diagonalize any block of intra-length mixing. 
\end{itemize}

\subsubsection*{The structure at two loops and beyond}

At two loops and beyond, the ADM becomes scheme dependent, so a discussion of its structure must assume some particular scheme, which we have chosen to be minimal subtraction (MS).
Within the MS scheme, 
the zeros in the ADM have the following structure
\begin{itemize}
  \item \textbf{Inter-length mixing of an $l(\O_\ell)$-point operator into an $l(\O_s)$-point operator is zero up to at least $L=l(\O_\ell)-l(\O_s)$ loops}. This is the theorem of \cite{Bern2020}. 
\end{itemize}
Here $l(\O)$ refers to the length of $\O$, its number of fields, and we take $l(\O_\ell)>l(\O_s)$.  
We note that this theorem minimally explains all observed zeros we find with exception of one three-loop zero.
That is, we observe most entries in the ADM are in fact non-zero at $L=l(\O_\ell)-l(\O_s)+1$. However, at three loops for the 6-point into the 4-point at mass dimension 6 and the 8-point into the 6-point at mass dimension 8, we do observe a zero. Furthermore, these zeros cannot be explained by the more general rule of~\cite{Bern2020}, because the diagrams that contribute are not scaleless bubbles. The zeros occur irrespective of the choice of ($n$-2)-point operator in a natural basis. The entry is labelled {\bf E}  in Fig.~\ref{fig:adm_pic}.

To get insight in this zero entry, let us consider the diagrams contributing to the correlator $\Gamma_4[\O^{(6)}_6]$ at mass dimension 6 and at three loops. The correlator $\Gamma_2[\O^{(6)}_6]$ is zero at three loops.
We distinguish two types of diagrams. Firstly, diagrams of the form 
$$ \begin{gathered}
\begin{tikzpicture}
\begin{feynman}[large, baseline=b]
\tikzfeynmanset{every vertex={dot,minimum size=2mm}}
\vertex at (0,0) (EFT);
\tikzfeynmanset{every vertex={draw,minimum size=0pt, inner sep=0pt}}
%
\vertex at (-1,0) (left);
\vertex at (1,0) (right);
\vertex at (1.7,0) (p) {$p_i$};
\vertex at (0,-0.6) (bottom);
\vertex at (-1.4, 0.2) (lt);
\vertex at (-1.4, -0.2) (lb);

\diagram* {
(left) --[out = 60, in = 135] (EFT) --[out= -135, in=-60] (left),
(EFT) --[] (right),
(EFT) --[out=50, in=120] (right) --[out=-120, in=-50] (EFT),
(right) --[] (p),
(EFT) --[] (bottom),
(lt) --[] (left) --[] (lb)
};
\end{feynman}
\end{tikzpicture}
\end{gathered} $$
are proportional to $p_i^2$ for any choice of charge flow. That is because the integral factorizes on the EFT vertex, and $p_i$ is the only momentum scale flowing through the diagram 
$$\gamma \ = \ 
\begin{gathered}
\begin{tikzpicture}
\begin{feynman}[large, baseline=b]
\tikzfeynmanset{every vertex={dot,minimum size=2mm}}
\vertex at (0,0) (EFT);
\tikzfeynmanset{every vertex={draw,minimum size=0pt, inner sep=0pt}}
%
\vertex at (1,0) (right);
\vertex at (1.7,0) (p) {$p_i$};
\vertex at (-0.4, 0.2) (lt);
\vertex at (-0.4, -0.2) (lb);
\vertex at (-0.4,-0.07) {.};
\vertex at (-0.4,0.07) {.};
\vertex at (0,0.5);

\diagram* {
(EFT) --[] (right),
(EFT) --[out=50, in=120] (right) --[out=-120, in=-50] (EFT),
(right) --[] (p),
(lt) --[] (EFT) --[] (lb)
};
\end{feynman}
\end{tikzpicture}
\end{gathered} \, ,$$ 
which has superficial degree of divergence ${\omega(\gamma)=2}$. When using field redefinitions to go from the operator basis for off-shell correlation functions to a physical basis, a 4-point counterterm proportional to ${\sum_i p_i^2}$ is absorbed by the coefficient of a 6-point operator. 
Therefore, this type of diagram actually contributes to the mixing of the 6-point operator into itself instead of to the mixing of the 6-point operator into a 4-point operator.
In contrast, there are also diagrams which are individually non-zero and not proportional to any $p_i^2$. These have the form
\begin{equation*}\label{eq:diags3loop}
\begin{gathered}
\begin{tikzpicture}
\begin{feynman}[large, baseline=b]
\tikzfeynmanset{every vertex={dot,minimum size=2mm}}
\vertex at (0,0) (EFT);
\tikzfeynmanset{every vertex={draw,minimum size=0.5mm, inner sep=0pt}}
\vertex at (0,1) (top);
\tikzfeynmanset{every vertex={draw,minimum size=0pt, inner sep=0pt}}
\vertex at (0.28,0.5) (right);
\vertex at (0.5,0.65) (r1);
\vertex at (0.5, 0.35) (r2);
\vertex at (-0.2,-0.3) (d1);
\vertex at (0.2,-0.3) (d2);
\diagram* {
(EFT) -- [out=145, in = -145] (top)
-- [out = -107, in=107] (EFT)
-- [out = 73, in = -73] (top)
-- [out = -35, in=90] (right) 
-- [out = -90, in=45] (EFT),
(r1) -- [] (right) --[] (r2),
(d1) -- [] (EFT) --[] (d2)
};
\end{feynman}
\end{tikzpicture}
\end{gathered} \ \, ,
\hspace{0.7cm}
\begin{gathered}
\begin{tikzpicture}
\begin{feynman}[large, baseline=b]
\tikzfeynmanset{every vertex={dot,minimum size=2mm}}
\vertex at (0,0) (EFT);
\tikzfeynmanset{every vertex={draw,minimum size=0pt, inner sep=0pt}}
\vertex at (0.4,0.5) (right);
\vertex at (-0.4,0.5) (left);
\vertex at (-0.6,0.7) (lt);
\vertex at (0.6,0.7) (rt);
\vertex at (-0.2,-0.3) (d1);
\vertex at (0.2,-0.3) (d2);
\diagram* {
(EFT) --[out=15, in=-70] (right) --[out=180, in=80] (EFT),
(EFT) --[out=165, in=-110] (left) --[out=0, in=100] (EFT),
(left) --[out= 35, in=145] (right),
(left) --[] (lt),
(right) --[] (rt),
(d1) -- [] (EFT) --[] (d2)
};
\end{feynman}
\end{tikzpicture}
\end{gathered}\ .
\end{equation*}
We have observed that the necessary counterterm to the divergence of these diagrams is proportional to ${\sum_i p_i^2}$ only after summing the permutations and choices of charge flow of both graphs.

It follows that the same entry will be zero at every mass dimension $n$, as the contributing graphs will always have the same form, simply having more external legs  directly attached to the EFT vertex, { which does not depend on momenta}. 
We further note that the 3-loop entry of the 6-point into the 4-point at mass dimension 8 is found nonzero, while the same diagrams contribute. Therefore, something special seems to occur at ${n \leftrightarrow n-2}$, where the operators are relatively simple. It would be interesting to test the 5-loop mixing of the 8-point operator into a 4-point operator at mass dimension 8.

\section{Discussion and Conclusions}\label{sec:conclusions}

In this paper, we investigated and performed the renormalization of massless scalar effective field theories at higher orders in both the number of loops and mass dimension, working at linear order in the EFT operators. For this purpose, we established a powerful method to compute the ADM, based on the R* renormaliztion method. To facilitate this, we detailed the use of Hilbert series, commutative algebra, and graphical methods, as well as conformal representation theory in the construction of relevant off-shell and on-shell operator bases. We obtained results ranging from five-loop order at mass dimension 6 in the real scalar field case, and the full one-loop corrections up to mass dimension 12; Fig.~\ref{fig1} summarises the full set of results. 

We further explored the ADM structure using conformal primary bases, detailing the one-loop breakdown of the known structure at leading order in the perturbative coupling \cite{Craigie1985}.
We identified two types of unexpected zeros in the ADM: a zero that occurs in the mixing of $n$ into $n-2$ point operators in the complex scalar EFT at 3 loops (at mass dimension $n$), which is at higher loop than what could be expected by the minimal application of the non-renormalization theorem/rule of~\cite{Bern2020}; and, a zero that occurs for one-loop mixing of $n-2$ point into $n$ point operators only for the choice of conformal primaries. It would be interesting to shed further light on these zeros, beyond what we discussed in the present paper.

While the R* method, in its automated form \cite{Herzog:2017bjx}, and further developed in \cite{deVries:2019nsu}, could be readily used to compute the ADMs of correlation functions with higher dimensional operators, a complication enters due to the mixing with unphysical operators. Since the local R* operation automatically builds the required off-shell counterterms, we could avoid explicitly computing the full ADM in the off-shell basis. We thus established a path where only physical operator insertions are required. Only the interpretation of the final results for the local UV counterterms required the knowledge of the full off-shell basis. To project these results onto a physical basis we deployed both field redefinitions and, alternatively, equation-of-motion operators. The method based on field redefinitions, while being more involved, has the advantage of being more general: it also works for multiple insertions of operators. We fully automated general field redefinitions, demonstrating that this approach can be used efficiently even at high mass dimensions. 

We have checked our results in several ways. While there are several consistency checks build into the R* framework, such as the locality (absence of kinematic logarithms in the result), we were also able to find agreement with the literature. For instance, we reproduced the known block-diagonal structure of the ADM at leading order in the perturbative coupling \cite{Craigie1985}, a structure dictated by conformal primary operators. At higher loop order, we confirm the appearance of zeros in line with the theorem of \cite{Bern2020}.  

In our analysis of operator bases, we studied the structure of polynomial rings in kinematic variables for three different objects of interest in QFT---$S$-matrices, off-shell correlation functions, and form factors. We established a ring isomorphism between the latter two
which is valid in dimensional regularisation and holds in four dimensional spacetime at low enough mass dimension (below mass dimension 16). We identified a non-redundant basis for both off-shell correlation functions and form factors to be the set of operators in correspondence with non-isomorphic multigraphs (two-colored for the complex scalar case). Furthermore, we provided a systematic way of enumerating evanescent operators that appear at mass dimension 16 and above due to finite rank conditions in four dimensional spacetime. Further study of the effect of these operators in higher-loop calculations within dimensional regularisation is left for future work.

While we focused on scalar theories here, the methods (R*, Hilbert series, polynomial ring) are generalizable to theories with particles of higher spin, that transform under internal symmetries {\it etc.} Indeed, the R* method has, for instance, already been widely applied to gauge theories. Also Hilbert series technology can encompass spin~\cite{Henning:2015alf,Henning:2017fpj,Henning:2019enq,Henning:2019mcv}, non-linearly realized internal symmetries~\cite{Henning:2017fpj,Graf:2020yxt}, gravity~\cite{Ruhdorfer:2019qmk} and non-relativistic EFTs~\cite{Kobach:2017xkw,Kobach:2018pie}, and the systematic construction of operators that involve particles of higher spin via the study of polynomial rings and conformal representation theory has recently been studied in detail~\cite{Henning:2019enq,Henning:2019mcv}, and by related methods~\cite{Christensen:2018zcq,Shadmi:2018xan,Ma:2019gtx,Durieux:2019eor,Aoude:2019tzn,Christensen:2019mch,Durieux:2019siw,Durieux:2020gip,Li:2020gnx,Jiang:2021tqo,Dong:2021yak}. We expect that such on-shell techniques, and their development to off-shell objects of interest that was presented here, could also be of use in studying full loop amplitudes in EFT, {\it i.e.} beyond the renormalization of the theory. 
The above statements apply to non-evanescent operators; it would be of further interest to explore whether Hilbert series technology can be applied (or developed) to systematically identify and enumerate other classes of evanescent operators that appear in theories where particles have spin, for instance those associated with four-fermion interactions. The $R^*$ operation should be applicable in situations with evanescent operators, but subtleties may arise with $\gamma_5$ schemes, operator basis choices etc., which would have to be carefully addressed depending on the context.
Remaining within the scope of scalar field theories, the extension to calculate beyond linear order in EFT operators, and to study $O(n)$ symmetric theories at higher loop order and higher mass dimension, can be achieved with minimal modification to our methods.

\section*{Acknowledgements}

We  thank J. Vermaseren for providing us with a private version of \textsc{Form} which includes an interface to the graph generator by T. Kaneko. We thank J. Gracey for comments and suggestions on the manuscript. T.M. thanks Brian Henning for valuable conversations regarding the impact of orthonormality and conformal operator bases on the ADM structure, and Xiaochuan Lu for valuable discussions about evanescent operators. W.C. thanks Zhiyuan Ding and Chenghan Zha for instructive discussions about the ring isomorphism.
W.C. is supported by  the Global Science Graduate Course (GSGC) program of the University of Tokyo, the World Premier International Research Center Initiative (WPI) and acknowledges support from JSPS KAKENHI grant number JP19H05810. F. H. is supported by the NWO Vidi grant 680-47-551 and the UKRI FLF Mr/S03479x/1. 
T.M. is supported by the World Premier International Research Center Initiative (WPI) MEXT, Japan, and by JSPS KAKENHI grants JP18K13533, JP19H05810, JP20H01896, and JP20H00153. J.R.N.\ is supported by the Deutsche Forschungsgemeinschaft (DFG, German Research Foundation) - Projektnummer 417533893/GRK2575 “Rethinking Quantum Field Theory”.
J.R.N.\ thanks Kavli IPMU for hospitality in the course of the work on this paper.
 F.H. and J.R.N. would like to thank the Nikhef theory group, where much of this work was carried out, for hospitality.

\appendix
\section{Review of the Hilbert series method}\label{s:hilbert}
 
In this appendix, we review the Hilbert series technique developed in Ref.\ \cite{Henning:2017fpj} by appealing to the representation theory of the conformal group.
In summary, the space of local operators modulo EoM can be written in terms of a set of primary operators and a tower of descendants generated by total derivatives acting on the primaries.
IBP relations can further be used to remove total derivatives, leaving only the primary operators as a physical basis for the $S$-matrix.
Using these insights one can build a Hilbert series as a partition function that enumerates the operator basis of the $S$-matrix. 
For off-shell correlation functions or form factors, it is possible to relax EoM or IBP respectively, based on results of Ref.\ \cite{Henning:2017fpj}. Furthermore, imposing (spacetime) parity conditions and working in arbitrary spacetime dimensions provide additional handles that are useful for counting independent operators.
This method has been applied {\it e.g.} to count the number of operators in the Standard Model Effective Field Theory up to high mass dimension \cite{Henning:2015alf}.

Following \cite{Henning:2015alf}, let us list
the formulas that are relevant for a scalar theory.
The main structure of the Hilbert series is given by
\beq \label{eq:hilbertmain}
H(t,\{\phi_a\})\ =\ \int d\mu(x,y)\frac{1}{P(t,x)}Z(t,\{\phi_a\},x,y)\,.
\eeq
We will first briefly explain the components of this expression in order, after which we discuss how to further impose spacetime parity. 
\begin{itemize}
	\item The Haar integral
	\begin{equation}
	\int d\mu(x,y)\ =\ \int d\mu_{\text{spacetime}}(x)\int d\mu_{\text{internal}}(y)
	\end{equation}
	selects the singlet of the spacetime group and the internal group, 
where $x$ and $y$ are (sets of) variables for the character under the spacetime and internal symmetries, respectively. 
	When parity is not imposed, the Haar measure of four dimensional spacetime ($SO(4)$) symmetry is
	\begin{align}
  	\int d\mu_{\text{spacetime}}(x)\ &=\ \int d\mu_{SO(4)}(x)\nonumber\\
  	&\hspace{-15mm}=\ \frac{1}{4}\oint_{|x_1|=1}\frac{dx_1}{2\pi ix_1}\oint_{|x_2|=1}\frac{dx_2}{2\pi ix_2}\left(1-			x_1x_2\right)\left(1-\frac{1}{x_1x_2}\right)\left(1-\frac{x_2}{x_1}\right)\left(1-\frac{x_1}{x_2}\right)\,.
	\end{align}
	The Haar measure of a $U(1)$ internal symmetry group is
	\beq
	\int d\mu_{U(1)}\ =\ \oint_{|y|=1}\frac{dy}{2\pi iy} \,,
	\eeq
	which is relevant for the complex scalar theory.
	\item The factor $1/P(t,x)$ imposes translation invariance (namely IBP redundancies, or momenta sum to zero). For an $SO(4)$ spacetime symmetry group, 
	\beq
	P(t,x)\ 
	=\ \frac{1}{(1-tx_1)(1-t/x_1)(1-tx_2)(1-t/x_2)}\,.
	\eeq
	The variable $t$ is the so-called ``spurion'' which keeps track of the number of partial derivatives $(\partial)$ in operators.
	When calculating the Hilbert series for form factors, IBP should not be imposed, and the $1/P$ factor should not appear in the integrand of eq.~\eqref{eq:hilbertmain}.
\item $Z(t,\{\phi_a\},x,y)$ is the partition function---called the plethystic exponential---for characters $\chi_a$ of what in Ref.\ \cite{Henning:2017fpj} were termed single particle modules (a field and a tower of traceless derivatives acting on it),
	\beq
	Z(t,\{\phi_a\},x,y)\ =\ \exp\left(\sum_a\sum_{n=1}^{\infty}\frac{1}{n}\phi_a^n\ \chi_a(t^n,x^n,y^n)\right)\,.
	\eeq
	Similarly to $t$, $\{\phi_a\}$ are the spurions 
	for the operators $\{\phi_a\}$.
	The character $\chi_a$ is given by
	\beq
	\chi_a(t,x,y)\ =\ \chi_{\mathrm{scalar}}(t,x) \, 
	\chi_a^{\mathrm{internal}}(y) \,,
	\eeq
	where $\chi_{\mathrm{scalar}}$ is the character of the conformal group in four dimensions for a representation of the scalar field,
	\beq\label{scalar}
	\chi_{\text{scalar}}(t,x)
	=(1-t^2) P(t,x)\,.
	\eeq
	Here, the factor $(1-t^2)$ accounts for the on-shell, or EoM, condition. When calculating the Hilbert series for off-shell correlation functions, one should not include this factor in the scalar character.

	\item The $\mathbb{Z}_2$ symmetry of the real scalar theory does not affect the character of the internal symmetry group 
	($\chi_a^{\text{internal, real}}=1$), but can be imposed by removing all terms that are odd in the spurion $\phi$ from the Hilbert series,
	\beq
	H^{\mathbb{Z}_2,\ \text{real}}(\phi,t)\ =\ \frac{1}{2}\left(H^{\text{real}}(\phi,t)+H^{\text{real}}(-\phi,t)		\right)\,.
	\eeq
	We will omit the $\mathbb{Z}_2$ superscript for simplicity below.
	For the complex scalar theory, the character for a $U(1)$ internal symmetry is
	\beq
	\chi_a^{\mathrm{internal}}\ =\ \chi^{U(1)}_{Q}\ =\ y^Q
	\eeq
	where the $U(1)$ charge $Q=1$ for $\phi$ and $Q=-1$ for $\phi^{\dagger}$.
\end{itemize}

\vspace{5mm}

\noindent
Imposing spacetime parity via the Hilbert series has been derived in Appendix C of \cite{Henning:2017fpj} (see also the application of charge conjugation symmetry to the QCD chiral Lagrangian in \cite{Graf:2020yxt} that contains many parallels). We will summarize the results here.
The Lorentz symmetry $SO(4)$ can be promoted to $O(4)$ by parity P: ${O(4)=SO(4)\rtimes\{1,-1\}=\{O_+(4),O_-(4)\}}$. Then the $P$-even Hilbert series is given by an average over the two disconnected branches:
\begin{align}\label{eq:p-even}
  H^{\text{P-even}}\ 
  =\ \int d\mu_{\text{internal}}(y)\frac{1}{2}&\left(\int d\mu_{O_+(4)}(x)\frac{1}{P_+(t,x)}Z^{\text{P}^+}(t,\{\phi_a\},x,y)+\right.\nonumber\\
  &\left. \int d\mu_{O_-(4)}(\tilde{x})\frac{1}{P_-(t,\tilde{x})}Z^{\text{P}^-}(t,\{\phi_a\},\tilde{x},y)\right)
\end{align}
where $\tilde{x}=(x_1)$ has one less parameter than $x=(x_1,x_2)$ and we will also use $\bar{x}=(x_1,x_2=1)$. The components of \eqref{eq:p-even} are:
\begin{align}
d\mu_{O_+(4)}(x)\ &=\ d\mu_{SO(4)}(x)\\
d\mu_{O_-(4)}(\tilde{x})\ &=\ d\mu_{Sp(2)}(\tilde{x})=\frac{1}{2}\oint_{|x_1|=1}\frac{dx_1}{2 \pi ix_1}\left(1-x_1^2\right)\left(1-\frac{1}{x_1^2}\right)\\
P_+(t,x)\ &=\ \frac{1}{(1-tx_1)(1-t/x_1)(1-tx_2)(1-t/x_2)}\\
P_-(t,\tilde{x})\ &=\ \frac{1}{(1-tx_1)(1-t/x_1)(1-t^2)}
\end{align}
\begin{align}
  Z^{\text{P}^{\pm}}(t,\{\phi_a\},x,y)=\ &\exp\left[\sum_a\sum_{k=0}^{\infty}\frac{\phi_a^{2k+1}}{2k+1}\chi^{\text{P}^{\pm}}_{a,\ \text{odd-power}}\left(t^{2k+1},x^{2k+1},y^{2k+1}\right)\right.\nonumber\\
  \ &\left. +\sum_a\sum_{k=1}^{\infty}\frac{\phi_a^{2k}}{2k}\chi^{\text{P}^{\pm}}_{a,\ \text{even-power}}\left(t^{2k},x^{2k},y^{2k}\right)\right]
\end{align}
with, for a scalar field,
\beq
	\chi^{\text{P}^{+}}_{a,\ \text{odd-power}}(t,x,y)\ =\ \chi^{\text{P}^{+}}_{a,\ \text{even-power}}(t,x,y)
	\ =\ (1-t^2)\ P_+(t,x)\ \chi_a^{\text{internal}}(y)
\eeq
\begin{align}
  &\chi_{a,\ \text{odd-power}}^{\text{P}^{-}}\left(t,\tilde{x},y\right)\ =\ (1-t^2)\ P_-(t,\tilde{x})\ \chi_a^{\text{internal}}(y)\\
  &\chi_{a,\ \text{even-power}}^{\text{P}^{-}}\left(t,\bar{x},y\right)\ =\ (1-t^2)\ P_+(t,x_1,1)\ \chi_a^{\text{internal}}(y)\,.
\end{align}
Eq.~\eqref{eq:p-even} is the Hilbert series for the $S$-matrix of a parity-even theory. Similarly to the situation without parity, for form factors, the $1/P_+$ and $1/P_-$ should be removed; for off-shell correlation functions, all $(1-t^2)$ factors in the characters should be removed. 

Let us now compute the Hilbert series in various situations. 
To enumerate operators at different orders in the EFT expansion separately, we rescale the spurions according to their mass dimension
$
\left(t \to \Delta\, t, \ \phi \to \Delta\, \phi,\ \phi^{\dagger} \to \Delta\,\phi^{\dagger}\right)
$
and expand the Hilbert series in $\Delta$.
For the real scalar, the Hilbert series for the $S$-matrix is
\begin{align}\label{realphy}
  H^{\text{real}}_{S\text{-matrix}}(\Delta,t,\phi)
  =
  &\ \Delta^6\, \phi^6+\Delta^8\left(\phi^4t^4+\phi^8\right)+\Delta^{10}\left(\phi^4t^6+\phi^6t^4+\phi^{10}\right)\nonumber\\
  +&\ \Delta^{12}\left(\phi^4t^8+2\phi^6t^6+\phi^8t^4+\phi^{12}\right)\nonumber\\
  +&\ \Delta^{14}\left(\phi^4t^{10}+4\phi^6t^8+2\phi^8t^6+\phi^{10}t^4+\phi^{14}\right)\nonumber\\
  +&\ \Delta^{16}\left(2\phi^4t^{12}+5\phi^6t^{10}+5\phi^8t^8+2\phi^{10}t^6+\phi^{12}t^4+\phi^{16}\right)\nonumber\\
  +&\ \Delta^{18}\left(\phi^4t^{14}+13\phi^6t^{12}+9\phi^8t^{10}+5\phi^{10}t^8+2\phi^{12}t^6+\phi^{14}t^4+\phi^{18} \right)\nonumber\\
  +&\ \Delta^{20}\left(2\phi^4t^{16}+18\phi^6t^{14}+26\phi^8t^{12}+10\phi^{10}t^{10}+5\phi^{12}t^8+2\phi^{14}t^6+\phi^{16}t^4+\phi^{20} \right)\nonumber\\
  +&\ \mathcal{O}(\Delta^{22})
\end{align}
That is, the Hilbert series tells us there is only one independent operator with six $\phi$ fields in the physical basis at mass dimension 6, one with four derivatives and four $\phi$ fields and one with eight $\phi$ fields at mass dimension 8, \textit{etc.} Setting the spurions to 1 leads to the counting in Section \ref{sec:operatorbasis}. For complex scalar, the Hilbert series for the $S$-matrix is
{\allowdisplaybreaks
\begin{align}
  H^{\text{complex}}_{S\text{-matrix}}(\Delta,t,\phi,\phi^{\dagger})
  =
  &\ \Delta^6\left(\phi^2\phi^{\dagger\, 2}t^2+\phi^3\phi^{\dagger\, 3}\right)+\Delta^8 \left(2\phi^2\phi^{\dagger\, 2}t^4+\phi^3\phi^{\dagger\, 3}t^2+\phi^4\phi^{\dagger\, 4}\right)\nonumber\\
  +&\ \Delta^{10}\left(2\phi^2\phi^{\dagger\, 2}t^6+4\phi^3\phi^{\dagger\, 3}t^4+\phi^4\phi^{\dagger\, 4}t^2+\phi^5\phi^{\dagger\, 5}\right)\nonumber\\
  +&\ \Delta^{12}\left(3\phi^2\phi^{\dagger\, 2}t^8+10\phi^3\phi^{\dagger\, 3}t^6+6\phi^4\phi^{\dagger\, 4}t^4+\phi^5\phi^{\dagger\, 5}t^2+\phi^6\phi^{\dagger\, 6}\right)\nonumber\\
  +&\ \Delta^{14}\left(3\phi^2\phi^{\dagger\, 2}t^{10}+24\phi^3\phi^{\dagger\, 3}t^8+18\phi^4\phi^{\dagger\, 4}t^6+6\phi^5\phi^{\dagger\, 5}t^4+\phi^6\phi^{\dagger\, 6}t^2+\phi^7\phi^{\dagger\, 7}\right)\nonumber\\
  +&\ \Delta^{16}\left(4\phi^2\phi^{\dagger\, 2}t^{12}+50\phi^3\phi^{\dagger\, 3}t^{10}+61\phi^4\phi^{\dagger\, 4}t^8+20\phi^5\phi^{\dagger\, 5}t^6+6\phi^6\phi^{\dagger\, 6}t^4\right.\nonumber\\
  &\ \left.+\phi^7\phi^{\dagger\, 7}t^2+\phi^8\phi^{\dagger\, 8}\right)\nonumber\\
  +&\ \Delta^{18}\left(4\phi^2\phi^{\dagger\, 2}t^{14}+133\phi^3\phi^{\dagger\, 3}t^{12}+187\phi^4\phi^{\dagger\, 4}t^{10}+81\phi^5\phi^{\dagger\, 5}t^8+22\phi^6\phi^{\dagger\, 6}t^6\right.\nonumber\\
  &\ \left.+6\phi^7\phi^{\dagger\, 7}t^4+\phi^8\phi^{\dagger\, 8}t^2+\phi^9\phi^{\dagger\, 9}\right)\nonumber\\
  +&\ \Delta^{20}\left(5\phi^2\phi^{\dagger\, 2}t^{16}+215\phi^3\phi^{\dagger\, 3}t^{14}+604\phi^4\phi^{\dagger\, 4}t^{12}+296\phi^5\phi^{\dagger\, 5}t^{10}\right. \nonumber\\
  &\left.+91\phi^6\phi^{\dagger\, 6}t^8+22\phi^7\phi^{\dagger\, 7}t^6+6\phi^8\phi^{\dagger\, 8}t^4+\phi^{9}\phi^{\dagger\, 9}+\phi^{10}\phi^{\dagger\, 10}\right)\nonumber\\
  +&\ \mathcal{O}(\Delta^{22})
\end{align}
}

\noindent
The Hilbert series of the off-shell correlation functions and form factors for the real scalar share a common part until mass dimension 16 (as discussed in Sec.~\ref{sec:operatorbasis}, Appendix \ref{s:isoapp}), and they are given by
\begin{align}
  H_{\text{F.F.}}^{\text{real}}(\Delta,t,\phi)=
  &\ \Delta^6\left(\phi^2t^4+\phi^4t^2+ \phi ^6\right)+\Delta^8\left(\phi^2t^6+3\phi^4t^4+\phi^6t^2+\phi^8\right)\nonumber\\
  &+\ \Delta^{10}\left(\phi^2t^8+6\phi^4t^6+3\phi^6t^4+\phi^8t^2+\phi^{10}\right)\nonumber\\
  &+\ \Delta^{12}\left(\phi^{2}t^{10}+11\phi^4t^8+8\phi^6t^6+3\phi^8t^4+\phi^{10}t^2+\phi^{12}\right)\nonumber\\
  &+\ \Delta^{14}\left(\phi^{2}t^{12}+18\phi^4t^{10}+21\phi^6t^8+8\phi^8t^6+3\phi^{10}t^4+\phi^{12}t^2+\phi^{14}\right)\nonumber\\
  &+\ \Delta^{16}\left( \phi^{2}t^{14}+32\phi^4t^{12}+50\phi^6t^{10}+23\phi^8t^8+8\phi^{10}t^6+3\phi^{12}t^4+\phi^{14}t^2+\phi^{16}\right)\nonumber\\
  &+\ \Delta^{18}\left(\phi ^2t^{16}+ 48 \phi ^4t^{14} +128 \phi ^6t^{12}+ 62\phi ^8 t^{10} +23\phi ^{10} t^8 +8\phi ^{12} t^6 +3 \phi ^{14}t^4 \right.\nonumber\\
  &\left.\hspace{3mm}+\phi ^{16}t^2 +\phi ^{18}\right)\nonumber\\
  &+\ \Delta^{20}\phi ^2\left(t^{18} +75\phi ^4 t^{16} +299 \phi ^6t^{14} +189\phi ^8 t^{12} +64 \phi ^{10}t^{10} +23 \phi ^{12}t^8 +8\phi ^{14} t^6 \right.\nonumber\\
  &\left.\hspace{3mm}+3 \phi ^{16}t^4 +\phi ^{18}t^2 +\phi ^{20} \right)\nonumber\\
  &+\ \mathcal{O}(\Delta^{22})\\
  H_{\text{C.F.}}^{\text{real}}(\Delta,t,\phi)
  =
  &\ H_{\text{F.F.}}^{\text{real}}(\Delta,t,\phi)+\Delta^{16}(\phi^6t^{10})+\Delta^{18}\left( 3\phi^6t^{12}+\phi^8t^{10}\right)\nonumber\\
  &+\Delta^{20}\left( 11\phi^6t^{14}+4\phi^8t^{12}+\phi^{10}t^{10}\right)
  + \mathcal{O}(\Delta^{22})
\end{align}
The Hilbert series of the off-shell correlation functions and form factors for the real scalar also share a common part until mass dimension 16, and they are given by
{\allowdisplaybreaks
\begin{align}
  H_{\text{F.F.}}^{\text{complex}}(\Delta,t,\phi,\phi^{\dagger})=
  &\ \Delta^6\left(\phi\phi^{\dagger}t^4+3\phi^2\phi^{\dagger\, 2}t^2+\phi^3\phi^{\dagger\, 3}\right)\nonumber\\
  +&\ \Delta^8 \left(\phi\phi^{\dagger}t^6+9\phi^2\phi^{\dagger\, 2}t^4+3\phi^3\phi^{\dagger\, 3}t^2+\phi^4\phi^{\dagger\, 4}\right)\nonumber\\
  +&\ \Delta^{10}\left(\phi\phi^{\dagger}t^8+20\phi^2\phi^{\dagger\, 2}t^6+13\phi^3\phi^{\dagger\, 3}t^4+3\phi^4\phi^{\dagger\, 4}t^2+\phi^5\phi^{\dagger\, 5}\right)\nonumber\\
  +&\ \Delta^{12}\left(\phi\phi^{\dagger}t^{10}+42\phi^2\phi^{\dagger\, 2}t^8+48\phi^3\phi^{\dagger\, 3}t^6+15\phi^4\phi^{\dagger\, 4}t^4+3\phi^5\phi^{\dagger\, 5}t^2+\phi^6\phi^{\dagger\, 6}\right)\nonumber\\
  +&\ \Delta^{14}\left(\phi\phi^{\dagger}t^{12}+78\phi^2\phi^{\dagger\, 2}t^{10}+163\phi^3\phi^{\dagger\, 3}t^8+64\phi^4\phi^{\dagger\, 4}t^6+15\phi^5\phi^{\dagger\, 5}t^4\right. \nonumber\\
  &\left.+3\phi^6\phi^{\dagger\, 6}t^2+\phi^7\phi^{\dagger\, 7}\right)\nonumber\\
  +&\ \Delta^{16}\left(\phi\phi^{\dagger}t^{14}+138\phi^2\phi^{\dagger\, 2}t^{12}+506\phi^3\phi^{\dagger\, 3}t^{10}+274\phi^4\phi^{\dagger\, 4}t^8+68\phi^5\phi^{\dagger\, 5}t^6\right. \nonumber\\
  &\left.+15\phi^6\phi^{\dagger\, 6}t^4+3\phi^7\phi^{\dagger\, 7}t^2+\phi^8\phi^{\dagger\, 8}\right)\nonumber\\
  +&\ \Delta^{18}\left(\phi\phi^{\dagger}t^{16}+228\phi^2\phi^{\dagger\, 2}t^{14}+1487\phi^3\phi^{\dagger\, 3}t^{12}+1109\phi^4\phi^{\dagger\, 4}t^{10}+322\phi^5\phi^{\dagger\, 5}t^8\right. \nonumber\\
  &\left.+70\phi^6\phi^{\dagger\, 6}t^6+15\phi^7\phi^{\dagger\, 7}t^4+3\phi^8\phi^{\dagger\, 8}t^2+\phi^{9}\phi^{\dagger\, 9}\right)\nonumber\\
  +&\ \Delta^{20}\left(\phi\phi^{\dagger}t^{18}+363\phi^2\phi^{\dagger\, 2}t^{16}+4028\phi^3\phi^{\dagger\, 3}t^{14}+4411\phi^4\phi^{\dagger\, 4}t^{12}+1478\phi^5\phi^{\dagger\, 5}t^{10}\right. \nonumber\\
  &\left.+340\phi^6\phi^{\dagger\, 6}t^8+70\phi^7\phi^{\dagger\, 7}t^6+15\phi^8\phi^{\dagger\, 8}t^4+3\phi^{9}\phi^{\dagger\, 9}+\phi^{10}\phi^{\dagger\, 10}\right)\nonumber\\
   +&\ \mathcal{O}(\Delta^{22})
\end{align}
}
\begin{align}
  H_{\text{C.F.}}^{\text{complex}}(\Delta,t,\phi,\phi^{\dagger})
  =
  &\ H_{\text{F.F.}}^{\text{complex}}(\Delta,t,\phi,\phi^{\dagger})+\Delta^{16}\left(4\phi^3\phi^{\dagger\, 3}t^{10}\right)+\Delta^{18}\left( 20\phi^3\phi^{\dagger\, 3}t^{12}+10\phi^4\phi^{\dagger\, 4}t^{10}\right)\nonumber\\
  &+\Delta^{20}\left(100\phi^3\phi^{\dagger\, 3}t^{14}+90\phi^4\phi^{\dagger\, 4}t^{12}+14\phi^5\phi^{\dagger\, 5}t^{10} \right)
  + \mathcal{O}(\Delta^{22})
\end{align}

 \section{A ring isomorphism}\label{s:isoapp}
  
Here we prove the ring isomorphism defined in eq.~\eqref{eq:multring}. Suppose we have a ring ${R=\mathbb{C}[\{s_{ii},s_{ij}\}]}$ ($i,j=1,\dots, N$, $s_{ij}=s_{ji}$) equipped with an action $S_N$ and its quotient rings 
\[R_1=\mathbb{C}[\{s_{ii},s_{ij}\}]/I_1,\quad R_2=\mathbb{C}[\{s_{ii},s_{ij}\}]/I_2\]
where
\[ I_1=\langle\{s_{ii}\}_{i=1,\dots, N}\rangle,\quad I_2=\langle\{X_{i}\}_{i=1,\dots, N}\rangle,\qquad X_i=\sum_{j=1}^N s_{ij}. 
\]
We see that both $I_1$ and $I_2$ are invariant under the $S_N$-action. Let us further define the two quotient maps $q_1:R\rightarrow R_1$ and $q_2:R\rightarrow R_2$.

We now define a ring homomorphism $f^0:R\to R$ by 
\begin{equation}
        f^0(s_{ij})=
        \begin{cases}
            s_{ij} &\text{if} \quad i\neq j\, \\
            X_i & \text{if} \quad i=j\,.
        \end{cases}
\end{equation}
It has an inverse $h$ given by 
\begin{equation}
        h(s_{ij})=
        \begin{cases}
            s_{ij} &\text{if} \quad i\neq j\, \\
            s_{ii}-\sum_{j \neq i}s_{ij} & \text{if} \quad i=j\,.
        \end{cases}
\end{equation}
Therefore, $f^0$ is an automorphism of the ring $R$. Since $f^0$ is surjective and the kernel of the composition $R\xrightarrow{f^0}R\xrightarrow{q_2} R_2$ equals $I_1$, by the First Isomorphism Theorem $f^0$ induces an isomorphism $f:R_1\xrightarrow{\sim} R_2$, satisfying $f\circ q_1= q_2\circ f^0$.

Since $f^0$ is $S_N$-equivariant, so is $f$. Therefore, $f$ induces an isomorphism $f': R_1^{S_N}\xrightarrow{\sim}R_2^{S_N}$. The same argument holds for complex scalar theory where the permutation group is $S_{N/2}\times S_{N/2}$.

This equivalence breaks down when we meet Gram conditions. For example, in $d=4$, $N=5$, the Gram determinant for the form factor, as a constraint in the ideal $I_1$, doesn't vanish,\\
\[ \begin{vmatrix} 
  0  & s_{12} & s_{13} &s_{14} & s_{15}\\
  s_{12} & 0 & s_{23} &s_{24} & s_{25}\\
  s_{13} & s_{23} & 0 &s_{34} & s_{35}\\
  s_{14} & s_{24} & s_{34} &0 & s_{45}\\
  s_{15} & s_{25} & s_{35} &s_{45} & 0
  \end{vmatrix}
  \neq 0\,,
\]\\
while the Gram determinant for an off-shell correlation function, as a constraint in the ideal $I_2$, vanishes
\[ \begin{vmatrix} 
  -(\sum_{i\neq 1} s_{1i}) & s_{12} & s_{13} &s_{14} & s_{15}\\
  s_{12} & -(\sum_{i\neq 2} s_{2i}) & s_{23} &s_{24} & s_{25}\\
  s_{13} & s_{23} & -(\sum_{i\neq 3} s_{3i}) &s_{34} & s_{35}\\
  s_{14} & s_{24} & s_{34} &-(\sum_{i\neq 4} s_{4i}) & s_{45}\\
  s_{15} & s_{25} & s_{35} &s_{45} & -(\sum_{i\neq 5} s_{5i})
  \end{vmatrix}
  =0 \,.
\]
\\

\newpage
\section{Multigraph bases and primary operators} \label{s:IBPbasesConf}

{\allowdisplaybreaks
\newcommand{\setbeginlengthTwo}[0]{ & \hspace{0.9\textwidth} \\[-4mm] }

In this appendix, we state our conventions for the multigraph bases and the conformal primary operators, as introduced in section \ref{sec:operatorbasis}. Except for $\mathcal{O}^{\scaleto{(4)}{7pt}}_2$, the prefactor of each operator is chosen to give rise to a Feynman rule with coefficient $+1$. 
Whenever more than one primary operator exists at a particular order, we label different choices by the parameters $x,\,y,...$. These generate independent operators up to an overall multiplicative constant that has not been taken out. 

\begin{multicols}{2}

\subsection*{Real scalar multigraphs}
\noindent
\begin{align*}
\setbeginlengthTwo
	\mathcal{O}^{\scaleto{(4)}{7pt}}_4 \hspace{2mm} &= \hspace{2mm} \frac{1}{4!} \hspace{2mm} 
			\begin{gathered}

			\end{gathered}
\end{align*}

\vspace{-0.0cm}

\newpage
\subsection*{Complex scalar multigraphs}
\noindent

\begin{align*}
\setbeginlengthTwo
	\mathcal{O}^{\scaleto{(4)}{7pt}}_4 \hspace{2mm} &= \hspace{2mm} \frac{1}{4} \hspace{2mm} 
			\begin{gathered}
			%
			\end{gathered}
\end{align*}
}

\end{multicols}

\noindent\rule{\textwidth}{0.4pt}

\vspace{1cm}

\subsection*{Real scalar primaries}

\vspace{-9mm}

{\begin{adjustwidth}{-0.52cm}{1cm}
\allowdisplaybreaks
\begin{align*}
\setbeginlengthTwo
	\mathcal{O}^{\scaleto{(6)}{7pt}c}_{\scaleto{6}{5pt}} &= \mathcal{O}^{\scaleto{(6)}{7pt}}_{\scaleto{6}{5pt}}\\[3mm]
	\mathcal{O}^{\scaleto{(8)}{7pt}c}_8 &= \mathcal{O}^{\scaleto{(8)}{7pt}}_8\\
	\mathcal{O}^{\scaleto{(8)}{7pt}c}_4 &= 	\mathcal{O}^{\scaleto{(8)}{7pt}}_{\scaleto{4,1}{6pt}} -2 \mathcal{O}^{\scaleto{(8)}{7pt}}_{\scaleto{4,2}{6pt}} + 6 \mathcal{O}^{\scaleto{(8)}{7pt}}_{\scaleto{4,3}{6pt}}				\\[4mm]
	\mathcal{O}^{\scaleto{(10)}{7pt}c}_{\scaleto{10}{5pt}} &= \mathcal{O}^{\scaleto{(10)}{7pt}}_{\scaleto{10}{5pt}}\\
	\mathcal{O}^{\scaleto{(10)}{7pt}c}_{\scaleto{6}{5pt}} &=	\mathcal{O}^{\scaleto{(10)}{7pt}}_{\scaleto{6,1}{6pt}}-\mathcal{O}^{\scaleto{(10)}{7pt}}_{\scaleto{6,2}{6pt}}+\mathcal{O}^{\scaleto{(10)}{7pt}}_{\scaleto{6,3}{6pt}}	\\
	\mathcal{O}^{\scaleto{(10)}{7pt}c}_4 &=  2\mathcal{O}^{\scaleto{(10)}{7pt}}_{\scaleto{4,1}{6pt}} 
							-9 \mathcal{O}^{\scaleto{(10)}{7pt}}_{\scaleto{4,2}{6pt}}
							+72\mathcal{O}^{\scaleto{(10)}{7pt}}_{\scaleto{4,3}{6pt}}
							+72\mathcal{O}^{\scaleto{(10)}{7pt}}_{\scaleto{4,4}{6pt}}
							+126\mathcal{O}^{\scaleto{(10)}{7pt}}_{\scaleto{4,5}{6pt}}
							-108\mathcal{O}^{\scaleto{(10)}{7pt}}_{\scaleto{4,6}{6pt}}		\\[4mm]
	\mathcal{O}^{\scaleto{(12)}{7pt}c}_{\scaleto{12}{5pt}} &= \mathcal{O}^{\scaleto{(12)}{7pt}}_{\scaleto{12}{5pt}}\\
	\mathcal{O}^{\scaleto{(12)}{7pt}c}_8 &=	15\mathcal{O}^{\scaleto{(12)}{7pt}}_{\scaleto{8,1}{6pt}}
							-10\mathcal{O}^{\scaleto{(12)}{7pt}}_{\scaleto{8,2}{6pt}}
							+6\mathcal{O}^{\scaleto{(12)}{7pt}}_{\scaleto{8,3}{6pt}}	\\[1mm]
	\mathcal{O}^{\scaleto{(12)}{7pt}c}_{\scaleto{6}{5pt}}(x,y) &=	
							4x\mathcal{O}^{\scaleto{(12)}{7pt}}_{\scaleto{6,1}{6pt}}
							-9x\mathcal{O}^{\scaleto{(12)}{7pt}}_{\scaleto{6,2}{6pt}}
							+y\mathcal{O}^{\scaleto{(12)}{7pt}}_{\scaleto{6,3}{6pt}}
							+(48x-2y)\mathcal{O}^{\scaleto{(12)}{7pt}}_{\scaleto{6,4}{6pt}} \\ &\hspace{4mm}
							+(3y-54x)\mathcal{O}^{\scaleto{(12)}{7pt}}_{\scaleto{6,5}{6pt}}
							+(2y-84x)\mathcal{O}^{\scaleto{(12)}{7pt}}_{\scaleto{6,6}{6pt}}
							+24x\mathcal{O}^{\scaleto{(12)}{7pt}}_{\scaleto{6,7}{6pt}}
							+(504x-12y)\mathcal{O}^{\scaleto{(12)}{7pt}}_{\scaleto{6,8}{6pt}}	\\[2mm]
	\mathcal{O}^{\scaleto{(12)}{7pt}c}_4 &= 	\mathcal{O}^{\scaleto{(12)}{7pt}}_{\scaleto{4,1}{6pt}}
							-8\mathcal{O}^{\scaleto{(12)}{7pt}}_{\scaleto{4,2}{6pt}}
							+64\mathcal{O}^{\scaleto{(12)}{7pt}}_{\scaleto{4,3}{6pt}}
							+18\mathcal{O}^{\scaleto{(12)}{7pt}}_{\scaleto{4,4}{6pt}}
							+72\mathcal{O}^{\scaleto{(12)}{7pt}}_{\scaleto{4,6}{6pt}}\\ &\hspace{4mm}
							-144\mathcal{O}^{\scaleto{(12)}{7pt}}_{\scaleto{4,8}{6pt}}
							+234\mathcal{O}^{\scaleto{(12)}{7pt}}_{\scaleto{4,9}{6pt}}
							+216\mathcal{O}^{\scaleto{(12)}{7pt}}_{\scaleto{4,10}{6pt}}
							+72\mathcal{O}^{\scaleto{(12)}{7pt}}_{\scaleto{4,11}{6pt}}
\end{align*}
\end{adjustwidth}}

\vspace{-2mm}
\subsection*{Complex scalar primaries}

\vspace{-9mm}

{\begin{adjustwidth}{-1.3cm}{1cm}
\allowdisplaybreaks
\begin{align*}
\setbeginlengthTwo
	\mathcal{O}^{\scaleto{(6)}{7pt}c}_{\scaleto{6}{5pt}} &= 
	\mathcal{O}^{\scaleto{(6)}{7pt}}_{\scaleto{6}{5pt}}\\
	\mathcal{O}^{\scaleto{(6)}{7pt}c}_4 &= \minus\frac{1}{2}\mathcal{O}^{\scaleto{(6)}{7pt}}_{\scaleto{4,1}{6pt}} + \left( \mathcal{O}^{\scaleto{(6)}{7pt}}_{\scaleto{4,2}{6pt}}+\mathcal{O}^{\scaleto{(6)}{7pt}}_{\scaleto{4,3}{6pt}}\right) \\[4mm]
	\mathcal{O}^{\scaleto{(8)}{7pt}c}_8 &= \mathcal{O}^{\scaleto{(8)}{7pt}}_8\\	
	\mathcal{O}^{\scaleto{(8)}{7pt}c}_{\scaleto{6}{5pt}} &= 	\minus\frac{2}{3}\mathcal{O}^{\scaleto{(8)}{7pt}}_{\scaleto{6,1}{6pt}} + \left( 
							\mathcal{O}^{\scaleto{(8)}{7pt}}_{\scaleto{6,2}{6pt}}+\mathcal{O}^{\scaleto{(8)}{7pt}}_{\scaleto{6,3}{6pt}}\right)				\\
\mathcal{O}^{\scaleto{(8)}{7pt}c}_{4}(x,y) &= 	
								(5x-y)\mathcal{O}^{\scaleto{(8)}{7pt}}_{\scaleto{4,1}{6pt}}	
								+2x\mathcal{O}^{\scaleto{(8)}{7pt}}_{\scaleto{4,2}{6pt}}		
								+2x\mathcal{O}^{\scaleto{(8)}{7pt}}_{\scaleto{4,3}{6pt}}		
								-4x\mathcal{O}^{\scaleto{(8)}{7pt}}_{\scaleto{4,4}{6pt}}	
								-4x\mathcal{O}^{\scaleto{(8)}{7pt}}_{\scaleto{4,5}{6pt}} \\		&\hspace{4mm}	
								+(4y-16x)\mathcal{O}^{\scaleto{(8)}{7pt}}_{\scaleto{4,6}{6pt}}	
								+(4y-16x)\mathcal{O}^{\scaleto{(8)}{7pt}}_{\scaleto{4,7}{6pt}}
								+4y\mathcal{O}^{\scaleto{(8)}{7pt}}_{\scaleto{4,8}{6pt}}	
								+(36x-8y)\mathcal{O}^{\scaleto{(8)}{7pt}}_{\scaleto{4,9}{6pt}}		\\[4mm]
	\mathcal{O}^{\scaleto{(10)}{7pt}c}_{\scaleto{10}{5pt}} &= \mathcal{O}^{\scaleto{(10)}{7pt}}_{\scaleto{10}{5pt}}\\
	\mathcal{O}^{\scaleto{(10)}{7pt}c}_8 &=	\minus\frac{3}{4}\mathcal{O}^{\scaleto{(10)}{7pt}}_{\scaleto{8,1}{6pt}} + \left( 							
								\mathcal{O}^{\scaleto{(10)}{7pt}}_{\scaleto{8,2}{6pt}}+\mathcal{O}^{\scaleto{(10)}{7pt}}_{\scaleto{8,3}{6pt}}\right)	\\[1mm]
	\mathcal{O}^{\scaleto{(10)}{7pt}c}_{\scaleto{6}{5pt}}(x,y,z,w) &= 	
								4x\mathcal{O}^{\scaleto{(10)}{7pt}}_{\scaleto{6,1}{6pt}}
								+(9x+y-3z)\mathcal{O}^{\scaleto{(10)}{7pt}}_{\scaleto{6,2}{6pt}}
								+4y\mathcal{O}^{\scaleto{(10)}{7pt}}_{\scaleto{6,3}{6pt}}
								+(2z+3w)\mathcal{O}^{\scaleto{(10)}{7pt}}_{\scaleto{6,4}{6pt}} 
								+(2z-3w)\mathcal{O}^{\scaleto{(10)}{7pt}}_{\scaleto{6,5}{6pt}}\\		&\hspace{4mm}
								-(x+y+z+2w)\mathcal{O}^{\scaleto{(10)}{7pt}}_{\scaleto{6,6}{6pt}}
								-(x+y+z-2w)\mathcal{O}^{\scaleto{(10)}{7pt}}_{\scaleto{6,7}{6pt}}
								+(5z-3x-3y+6w)\mathcal{O}^{\scaleto{(10)}{7pt}}_{\scaleto{6,8}{6pt}} \\		&\hspace{4mm}
								+(5z-3x-3y-6w)\mathcal{O}^{\scaleto{(10)}{7pt}}_{\scaleto{6,9}{6pt}}
								+(z-7x+y-2w)\mathcal{O}^{\scaleto{(10)}{7pt}}_{\scaleto{6,10}{6pt}}
								+(z-7x+y+2w)\mathcal{O}^{\scaleto{(10)}{7pt}}_{\scaleto{6,11}{6pt}}\\		&\hspace{4mm}
								+(3x+3y-5z-6w)\mathcal{O}^{\scaleto{(10)}{7pt}}_{\scaleto{6,12}{6pt}} 
								+(3x+3y-5z+6w)\mathcal{O}^{\scaleto{(10)}{7pt}}_{\scaleto{6,13}{6pt}} 	\\[2mm]
	\mathcal{O}^{\scaleto{(10)}{7pt}c}_{4}(x,y) &= 	
							x	\mathcal{O}^{\scaleto{(10)}{7pt}}_{\scaleto{4,1}{6pt}}
							+(51x+24y)	\mathcal{O}^{\scaleto{(10)}{7pt}}_{\scaleto{4,2}{6pt}}
							-(18x+72y)	\mathcal{O}^{\scaleto{(10)}{7pt}}_{\scaleto{4,3}{6pt}}
							-(42x+24y)	\mathcal{O}^{\scaleto{(10)}{7pt}}_{\scaleto{4,4}{6pt}}	\\		&\hspace{4mm}
							+(-6x+3y)	\left( \mathcal{O}^{\scaleto{(10)}{7pt}}_{\scaleto{4,5}{6pt}} + 
								\mathcal{O}^{\scaleto{(10)}{7pt}}_{\scaleto{4,6}{6pt}} \right)	
								+2y\left( \mathcal{O}^{\scaleto{(10)}{7pt}}_{\scaleto{4,7}{6pt}}+
									\mathcal{O}^{\scaleto{(10)}{7pt}}_{\scaleto{4,8}{6pt}} \right)
							-(3x+3y) \left(	\mathcal{O}^{\scaleto{(10)}{7pt}}_{\scaleto{4,9}{6pt}} + 
								\mathcal{O}^{\scaleto{(10)}{7pt}}_{\scaleto{4,10}{6pt}} \right)\\		&\hspace{4mm}
							-9y	\left( \mathcal{O}^{\scaleto{(10)}{7pt}}_{\scaleto{4,11}{6pt}} + 
								\mathcal{O}^{\scaleto{(10)}{7pt}}_{\scaleto{4,12}{6pt}} \right)		
							+(24x+78y)\left( 	\mathcal{O}^{\scaleto{(10)}{7pt}}_{\scaleto{4,13}{6pt}}+
								\mathcal{O}^{\scaleto{(10)}{7pt}}_{\scaleto{4,14}{6pt}} \right)
							-(48x+12y)	\left( \mathcal{O}^{\scaleto{(10)}{7pt}}_{\scaleto{4,15}{6pt}}+
								\mathcal{O}^{\scaleto{(10)}{7pt}}_{\scaleto{4,16}{6pt}} \right)\\		&\hspace{4mm}
							+(24x+24y) \left(	\mathcal{O}^{\scaleto{(10)}{7pt}}_{\scaleto{4,17}{6pt}} + 
								\mathcal{O}^{\scaleto{(10)}{7pt}}_{\scaleto{4,18}{6pt}} \right)
							+(24x+24y) \left(	\mathcal{O}^{\scaleto{(10)}{7pt}}_{\scaleto{4,19}{6pt}} + 
								\mathcal{O}^{\scaleto{(10)}{7pt}}_{\scaleto{4,20}{6pt}}\right) 
\end{align*}
\end{adjustwidth}
}

}

\newpage
\section{Results for real scalar EFT}\label{s:resreal}

Here we present the anomalous dimension matrices in the conformal basis for the real scalar field. 
The dimension 4 Lagrangian has been defined in \eqref{Lagr4}, and the conformal bases are defined in appendix \ref{s:IBPbasesConf}.

\noindent
The five-loop beta function is
\begin{align*}
\beta(g,\epsilon) &= -2\epsilon g + 3{g}^{2}
-{\frac{17}{3}}{g}^{3} 
+ \left( 12\,{ \zeta_3}+{\frac{145}{8}}\right) {g}^{4}
+ \left( -78\,{\zeta_3}+18\,{ \zeta_4}-120\,{ \zeta_5}-{ \frac{3499}{48}} \right) {g}^{5}\\
& \hspace{1.5cm} + \left(45\zeta_3^2+\frac{7965}{16}\zeta_3-\frac{1189}{8}\zeta_4+987\zeta_5-\frac{675}{2}\zeta_6+1323\zeta_7+\frac{764621}{2304}\right) g^6 \, ,
\end{align*}
and the anomalous dimension of the field is 
	\begin{align*}
		\gamma_\phi &= \textcolor{black}{0g} 
			- \frac{1}{12} g^2 
			+ \frac{1}{16}g^3 
			- \frac{65}{192}g^4 
			+ \left(-\frac{3}{16}\zeta_3 + \frac{1}{2}\zeta_4 + \frac{3709}{2304}\right)g^5 ,
\end{align*}
as previously found in {\it e.g.}~\cite{Chetyrkin:1981jq,Kleinert:1991rg,Gorishnii:1983gp}.
At mass dimension 6, the anomalous dimension of the coupling associated to $\O_6^{(6)c}$ is up to five loops
\begin{align*}
	\gamma^{(6)}_\textsc{c} = 						
		9g-&\frac{359g^2}{6}+\Big(216\zeta_3+\frac{5773}{12}\Big)g^3
			-\Big(\frac{5283\zeta_3}{2}-459\zeta_4+3960\zeta_5+\frac{1312907}{288}\Big)g^4\\
		&+\left( 1755 \zeta_3^2 + \frac{510193}{16}\zeta_3 -\frac{25483}{4}\zeta_4 		+54621\zeta_5 -\frac{29025}{2}\zeta_6 +73584\zeta_7 +\frac{333811367}{6912}	\right)g^5 \, .
\end{align*}
At mass dimension 8 in the conformal basis,
$$
\left\{ \O^{(8)c}_8 \ , \ \O^{(8)c}_4 \right\},
$$
the AD matrix up to three loops is
$$
\gamma_\text{\textsc{c}}^{(8)} = 
\begin{pNiceMatrix}[first-row]
\begin{matrix}\textcolor{gray}{g^3\,\O^{(8)c}_8}\\[3mm]\end{matrix}
&\begin{matrix}\textcolor{gray}{g\,\O^{(8)c}_4}\\[3mm]\end{matrix}\\[2mm]
19g-169g^2+\Big(636\zeta_3 +\frac{46255}{24}\Big)g^3 \ \ \ \ 
& 2352g-\frac{76720g^2}{3}+\Big(65856\zeta_3 +\frac{8676248}{27}\Big)g^3\\[2mm]
\textcolor{black}{0g+0g^2+0g^3}
& \frac{g}{3}+\frac{49g^2}{27}-\Big(\frac{12017}{1944}+\frac{20\zeta_3 }{3}\Big)g^3
\end{pNiceMatrix} .
$$
At mass dimension 10 in the conformal basis, 
$$
\left\{
\O^{(10)c}_{10} \ , \
\O^{(10)c}_{6} \ , \ 
\O^{(10)c}_{4}
\right\},
$$
we find at one-loop
\begin{align*}
\gamma_\text{\textsc{c}}^{(10)} = 
\begin{pNiceMatrix}[first-row]
\textcolor{gray}{\hspace{2mm} g^4\,\O^{(10)c}_{10}}
&\textcolor{gray}{g^2\,\O^{(10)c}_{6}}
&\textcolor{gray}{g \O^{(10)c}_{4}}
\\[3mm]
	 33g
	&23100g
	&233520g
	\\[2mm]
		\textcolor{black}{0g}
	&5g
	&\frac{112g}{5}
	\\[2mm]
		\textcolor{black}{0g}
	&	\textcolor{black}{0g}
	&-2g
\end{pNiceMatrix} \, .
\end{align*}
Finally, at mass dimension 12 in the conformal basis,
$$
\left\{
\O^{(12)c}_{12} \ , \ 
\O^{(12)c}_{8} \ , \ 
{\O^{(12)c}_{6}\scriptstyle{(1,0)} }\ , \ 
{\O^{(12)c}_{6}\scriptstyle{(0,1)}} \ , \ 
\O^{(12)c}_{4}
\right\},
$$
we find at one-loop
\begin{align*}
\gamma_\text{\textsc{c}}^{(12)} = 
\begin{pNiceMatrix}[first-row]
\textcolor{gray}{g^5\,\O^{(12)c}_{12}}  
& \quad \textcolor{gray}{g^3\,\O^{(12)c}_{8}}  
& \quad \textcolor{gray}{g^2\,\O^{(12)c}_{6}\scriptstyle{(1,0)}} 
& \quad \textcolor{gray}{g^2\,\O^{(12)c}_{6}\scriptstyle{(0,1)}} 
& \quad \textcolor{gray}{g\,\O^{(12)c}_{4}}\\[3mm]
                             51g
&                      \frac{1121120g}{9}
 &                     \frac{477400g}{3}
  &                    \frac{284900g}{3}
   &                \frac{7037295100g}{147}\\[1mm]
                              \textcolor{black}{0g}
&                        \frac{41g}{3}
 &                      -\frac{4675g}{2}
  &                      \frac{395g}{4}
   &                 \frac{45981307g}{1372}\\[1mm]
                              \textcolor{black}{0g}
&                              \textcolor{black}{0g}
 &                             2g
  &                            \textcolor{black}{0g}
   &                  -\frac{26779g}{490}\\[1mm]
                              \textcolor{black}{0g}
&                              \textcolor{black}{0g}
 &                             \textcolor{black}{0g}
  &                            2g
   &                  -\frac{445574g}{245}\\[1mm]
                              \textcolor{black}{0g}
&                             \textcolor{black}{0g} 
 &                             \textcolor{black}{0g}
  &                            \textcolor{black}{0g}
   &                      -\frac{g}{5}
\end{pNiceMatrix}\ .
\end{align*}
where one should note that the intra-$N$ mixing between the two 6-point operators is diagonal in any natural basis (\textit{i.e.}~a basis with the least number of derivatives possible), since the two entries are exactly (by chance?) the same.

\section{One-loop mixing between $\mathcal{O}_n^{(n)}$ and $\mathcal{O}_{n-2}^{(n)}$  for the complex scalar}\label{s:zeronn-2}

In the following we sketch the computation of the one-loop mixing of the length-$(n-2)$ operator given in \eqref{eq:dim6param},
 $$\O^{(n)}_{n\text{-}2}(a) \equiv a\, \O^{(n)}_{n\text{-}2,1} + \O^{(n)}_{n\text{-}2,2} + \O^{(n)}_{n\text{-}2,3}\,,$$ 
 into the length-$n$ operator $\O^{(n)}_{n}$ for the complex scalar at arbitrary mass dimension $n$. 
Here we have conveniently set $a_2=a_3=1$, which amounts to taking only the C-even (real) part
 and fixing an overall prefactor in \eqref{eq:dim6param}.
Let us consider the $m$-point (off-shell) correlator with an insertion of this operator:
$$
\Gamma_{m}[\O^{(n)}_{n\text{-}2}(a)]
 = 
a\, \Gamma_{m}[\O^{(n)}_{n\text{-}2,1}]
+
\Gamma_{m}[\O^{(n)}_{n\text{-}2,2}]
+
\Gamma_{m}[\O^{(n)}_{n\text{-}2,3}]\,.
$$
The contributing one-loop diagrams are bubbles, of tensor rank at most one, for the case of $\Gamma_{n\text{-}2}[\O^{(n)}_{n\text{-}2}(a)]$,
$$\begin{gathered}
\begin{tikzpicture}
\begin{feynman}[large, baseline=b]

\tikzfeynmanset{every vertex={dot,minimum size=2mm}}
\vertex  at (-0.5,0) (eft) ;

\tikzfeynmanset{every vertex={draw,minimum size=0pt, inner sep=0pt}}
\vertex  at (-0.5,0) (l) ;
\vertex at (0.5,0) (r);

\vertex at (-1,0.2) (e1);
\vertex at (-1,-0.2) (e2);
\vertex at (-0.97,0.06) {.};
\vertex at (-0.97,-0.06) {.};

\vertex at (0.9,0.17) (e11);
\vertex at (0.9,-0.17) (e12);
\vertex at (-0.5,-0.5) {\tiny $\O^{(n)}_{n-2}(a)$};

\diagram* {
	(l) -- [out=65,in=115] (r) -- [out=-115,in=-65] (l),
	(l) -- [] (e1),
	(l) -- [] (e2),
	(e11) --[] (r) --[] (e12)
 };

\end{feynman}
\end{tikzpicture}
\end{gathered}\ ,
$$
and scalar triangles for $\Gamma_n[\O^{(n)}_{n\text{-}2}(a)]$,
$$
\begin{gathered}
\begin{tikzpicture}
\begin{feynman}[large, baseline=b]

\tikzfeynmanset{every vertex={dot,minimum size=2mm}}
\vertex  at (-0.5,0) (eft) ;

\tikzfeynmanset{every vertex={draw,minimum size=0pt, inner sep=0pt}}
\vertex  at (-0.5,0) (l) ;
\vertex at (0.5,0) (r);
\vertex at (0.3,0.25) (rt);
\vertex at (0.3,-0.25) (rb);

\vertex at (-1,0.2) (e1);
\vertex at (-1,-0.2) (e2);
\vertex at (-0.97,0.06) {.};
\vertex at (-0.97,-0.06) {.};
\vertex at (0.55,0.37) (e11);
\vertex at (0.35,0.5) (e12);
\vertex at (0.55,-0.37) (e13);
\vertex at (0.35,-0.5) (e14);
\vertex at (-0.5,-0.5) {\tiny $\O^{(n)}_{n-2}(a)$};

\diagram* {
	(l) -- [out=45,in=170] (rt) 
		-- [out=-30,in=90] (r)
		-- [out=-90,in=30] (rb)
		-- [out=190,in=-45] (l),
	(l) -- [] (e1),
	(l) -- [] (e2),
	(e11) --[] (rt) --[] (e12),
	(e13) --[] (rb) --[] (e14)
 };

\end{feynman}
\end{tikzpicture}
\end{gathered} \ .$$ 
Using textbook methods one easily extracts the following $1/\epsilon$ poles at one loop:%
\footnote{Note that here we do not rescale the operators by factors of $g$, in contrast to equation \eqref{bareoperators}.}
\begin{align*}
	 \Gamma_{m}[\O^{(n)}_{n\text{-}2}(a)]\Big|_{\eps^{\text{-}1}} &= 0\,, \quad \quad \text{for $m<n-2$ and $m>n$}\, ,\\
	 \Gamma_{n\text{-}2}[\O^{(n)}_{n\text{-}2}(a)]\Big|_{\eps^{\text{-}1}} 
	&= -g\left( \frac{3}{8}n^2 - \frac{11}{4}n +5 \right)\O^{(n)}_{n\text{-}2}(a)-g\left( \frac{n}{4} -1 \right) \O^{(n)}_{n\text{-}2;\partial^2}\\
	 \Gamma_n[\O^{(n)}_{n\text{-}2}(a)]\Big|_{\eps^{\text{-}1}} 
	 &= 
		g^2\left( \frac{5}{64}an^2(n-2)^2+\frac{1}{16}n^2(n-2)(n-4)
\right) \, \O^{(n)}_{n} \, .
\end{align*}
Here, we have explicitly isolated the physical operators and a redundant operator given by
$$\O^{(n)}_{n\text{-}2;\partial^2}\equiv \O^{(n)}_{n\text{-}2,1} + \O^{(n)}_{n\text{-}2,2} + \O^{(n)}_{n\text{-}2,3}$$
with Feynman rule
$$\frac{1}{2} \sum_{i=1}^{n\text{-}2} \left(p_i\right)^2\,. $$
The redundant operator must be replaced by $\O^{(n)}_{n}$ through a field redefinition that sets
	$$\E^{(n)}=  
	\O^{(n)}_{n\text{-}2;\partial^2} + \frac{-\frac{g}{2}\left[\left(\frac{n}{2}\right)!\right]^2}{  \left(\frac{n}{2}-1\right) ! \left(\frac{n}{2}-2\right) !}
	\, \O^{(n)}_{n} 
	= 0\,.$$
Therefore, the correlator $ \Gamma_{n\text{-}2}[\O^{(n)}_{n\text{-}2}(a)]$ requires the counterterm (including an overall minus to counter the divergence)
$$-\frac{1}{\eps}\frac{g^2}{64}n^2(n-2)(n-4) \, \O^{(n)}_{n} $$
in the bare physical Lagrangian, additionally to a counterterm proportional to $\O^{(n)}_{n\text{-}2}(a)$, which we are not interested in. Note that this is independent of $a$. This means that the mixing of $c^{(n)}_{n\text{-}2}$ into $c^{(n)}_{n}$ in the physical basis is governed by the renormalization constant
\begin{align*}
	Z^{(n)}_{n,n-2}/g^2
		&= -\frac{5}{64}n^2(n-2)\big(a(n-2)+(n-4)\big) \,, 
\end{align*}
which is zero for
$$a = \frac{a_1}{a_2}=\frac{a_1}{a_3}=- \frac{n-4}{n-2} \, .$$
Remarkably, this is exactly the condition that results in the conformal $(n-2)$-point operator, see \eqref{eq:confnn-2}.

\section{Orthonormal operators, and a symmetric and diagonal ADM} \label{s:ortho}

We will now give a proof that the one-loop intra-length anomalous dimension sub-matrix is symmetric for a basis of conformal primaries that are orthonormal with respect to integration over phase space. 
This result is known in the literature, see for example \cite{Craigie1985, Hogervorst:2015akt}, with the inner product defined through the two point correlation functions of operators.
It follows from unitarity that the phase space integral over a product of two operators is related to the imaginary part of the two point correlator,
\begin{align*}
	\mathrm{Im}\braket{\mathcal{O}_i, \, \mathcal{O}_j}_{P\neq0} 
	& =\vcentcolon \left[ 
\prod_{i=1}^{N} \int \hspace{-1.5mm} d^4p_i \ \delta^{+} \hspace{-1.3mm} \left(p_i^2\right) \, \right]
 \delta\hspace{-0.8mm}\left(P-\mathlarger{\Sigma}_i p_i\right)
	 \ \mathcal{O}_1(p_i) \cdot \mathcal{O}_2(\text{-}\,p_i) 
\\ &\overset{N{=}4}{=\joinrel=}
  \begin{gathered}\begin{tikzpicture}\begin{feynman}[small]
  \tikzfeynmanset{every vertex={dot,black,minimum size=1.5mm}}
  \vertex (left) at (-0.7,0);
  \vertex (right) at (0.7,0);
    \tikzfeynmanset{every vertex={dot,minimum size=0mm}}
  \vertex at (-0.9,-0.25) {\tiny $\mathcal{O}_i$};
  \vertex at (0.9,-0.25) {\tiny $\mathcal{O}_j$};
  \vertex (Pleft) at (-1.7,0) {\textcolor{darkgreen}{P}};
  \vertex (Pright) at (1.7,0) {\textcolor{darkgreen}{P}};
  \vertex (cuttop) at (0.1,0.8);
  \vertex (cutbottom) at (0.1,-0.8);
  \diagram* {
   (left) -- [out=70, in=110, fermion] (right),
   (left) -- [out=-70, in=-110, fermion] (right),
   (left) -- [out=20,in=160, fermion] (right),
   (left) -- [out=-20,in=-160,fermion] (right),
   (Pleft) -- [darkgreen, fermion, line width=0.5mm] (left),
   (right) -- [darkgreen, fermion, line width=0.5mm] (Pright),
   (cuttop) -- [dashed] (cutbottom),
  };
  \end{feynman}\end{tikzpicture}\end{gathered}
\end{align*}
where $\braket{\mathcal{O}_i, \, \mathcal{O}_j}$ is the leading order contribution to (or free theory limit of) the correlator
$\Gamma^{P\neq0}_{0}[\O_i,\O_j]$ with no external particles but a nonzero momentum flow $P$. 
It is straightforward to evaluate such phase space integrals {\it e.g.}~via an explicit parameterization of phase space that manifests the momentum conservation and on-shell conditions, for instance~\cite{Cox:2018wce,Larkoski:2020thc}.

The same phase space integral arises in the calculation of anomalous dimensions through the form factor method \cite{Caron-Huot:2016cwu}. The relevant master equation for mixing between operators of the same number of fields, without infrared divergences and at one loop, is
\begin{equation*}
  \braket{p_1,...,p_{N} | \mathcal{M} \otimes \O_i|0}
  =
  \text{-}\,\pi\sum_{j} \gamma_{ji} \braket{p_1,...,p_{N}|\O_j|0}\,.
\end{equation*}
Here, the $\otimes$ refers to a two-particle cut between a tree level renormalizable amplitude $\mathcal{M}$ and the tree level form factor with the insertion of $\O_i$. The set of operators that should be considered, with on-shell constraints but without momentum conservation, is described in section \ref{sec:operatorbasis}.
Diagrammatically, the master equation can be represented by (for $N=4$ here)
\begin{equation}\label{eq:nis4cut}
    \begin{gathered}\begin{tikzpicture}\begin{feynman}[small]
  \tikzfeynmanset{every vertex={dot,black,minimum size=2mm}}
  \vertex (left) at (0,0);
    \tikzfeynmanset{every vertex={dot,minimum size=0mm}}
  \vertex at (-0.2,-0.4) {$\O_i$};
  \vertex (Pleft) at (-1.7,0) {\textcolor{darkgreen}{P}};
  \vertex (v1) at (2.5,0);
  \vertex (outtop) at (3.3,0.3) {$p_2$};
  \vertex (outdown) at (3.3,-0.3) {$p_3$};
 \vertex (top) at (2.2,1) {$p_1$};
\vertex (bottom) at (2.2,-1) {$p_4$};
\vertex (cuttop) at (1.55,0.6);
\vertex (cutdown) at (1.55,-0.6);
  \diagram* {
   (left) -- [out=25, in=155, fermion] (v1),
   (left) -- [out=-25, in=-155, fermion] (v1),
   (left) -- [fermion, out=50,in=180] (top),
   (left) -- [fermion, out=-50,in=180] (bottom),
   (v1) --[fermion, out=25,in=-170] (outtop),
   (v1) --[fermion,out=-25,in=170] (outdown),
   (Pleft) --[line width=0.65mm, fermion, darkgreen] (left),
   (cuttop) --[dashed, red, line width=0.6mm] (cutdown),
  };
  \end{feynman}\end{tikzpicture}\end{gathered}
\ +perms \ = 
\text{-}\,\pi\sum_{j} \gamma_{ji} \ \cdot \ 
  \begin{gathered}\begin{tikzpicture}\begin{feynman}[small]
  \tikzfeynmanset{every vertex={dot,black,minimum size=2mm}}
  \vertex (left) at (0,0);
    \tikzfeynmanset{every vertex={dot,minimum size=0mm}}
  \vertex at (-0.2,-0.4) {$\O_j$};
  \vertex (Pleft) at (-1.3,0);
  \vertex at (-1,0.2) {\textcolor{darkgreen}{P}};
 \vertex (top) at (2.2,1) {$p_1$};
 \vertex (top2) at (2.2,0.35) {$p_2$};
\vertex (down2) at (2.2,-0.35) {$p_3$};
\vertex (bottom) at (2.2,-1) {$p_4$};
  \diagram* {
   (left) -- [out=25, in=180, fermion] (top2),
   (left) -- [out=-25, in=-180, fermion] (down2),
   (left) -- [fermion, out=50,in=180] (top),
   (left) -- [fermion, out=-50,in=180] (bottom),
   (Pleft) --[line width=0.65mm, fermion, darkgreen] (left),
  };
  \end{feynman}\end{tikzpicture}\end{gathered}
  \ \ .
\end{equation}

\vspace{2mm}

\noindent
Now, let us assume that the set of $N$-point operators at the considered mass dimension forms an orthonormal basis:
  $$
  \mathrm{Im}\braket{\mathcal{O}_i, \, \mathcal{O}_j}_{P\neq0} = (P^2)^x \, \delta_{ij} \, ,
$$
for some power $x$ depending on the considered mass dimension. 
One can take the additional cut with an operator insertion $\O_k$ on the both sides of
\eqref{eq:nis4cut}, to find
\begin{align*}
  \begin{gathered}\begin{tikzpicture}\begin{feynman}[small]
  \tikzfeynmanset{every vertex={dot,black,minimum size=2mm}}
  \vertex (left) at (0,0);
   \vertex (vright) at (4,0);
    \tikzfeynmanset{every vertex={dot,minimum size=0mm}}
  \vertex at (-0.2,-0.4) {$\O_i$};
    \vertex at (4.2,-0.4) {$\O_k$};
  \vertex at (-0.3,-0.6);
  \vertex (Pleft) at (-1.4,0) {\textcolor{darkgreen}{P}};
  \vertex (Pright) at (5.4,0) {\textcolor{darkgreen}{P}};
  \vertex (v1) at (2,0);
\vertex (cuttop) at (1.15,0.5);
\vertex (cutdown) at (1.15,-0.5);
\vertex (cutrighttop) at (3.1,0.5);
\vertex (cutrightdown) at (3.1,-0.5);
\vertex (cuttoptop) at (2.1,1.2);
\vertex (cuttopdown) at (2.1,0.55);
\vertex (cutdowntop) at (2.1,-1.2);
\vertex (cutdowndown) at (2.1,-0.55);
  \diagram* {
   (left) -- [out=25, in=155, fermion] (v1),
   (left) -- [out=-25, in=-155, fermion] (v1),
   (v1) -- [out=25, in=155, fermion] (vright),
   (v1) -- [out=-25, in=-155, fermion] (vright),
   (left) -- [fermion, out=45, in=135] (vright),
   (left) -- [fermion, out=-45, in=-135] (vright),
   (Pleft) --[line width=0.65mm, fermion, darkgreen] (left),
   (vright) --[line width=0.65mm, fermion, darkgreen] (Pright),
   (cuttop) --[dashed, red, line width=0.6mm] (cutdown),
   (cutrighttop) --[dashed, red, line width=0.6mm] (cutrightdown),
   (cuttoptop) --[dashed, red, line width=0.6mm] (cuttopdown),
   (cutdowntop) --[dashed, red, line width=0.6mm] (cutdowndown),
  };
  \end{feynman}\end{tikzpicture}\end{gathered}
&= 
\text{-}\,\pi\sum_{j} \gamma_{ji} \
  \begin{gathered}\begin{tikzpicture}\begin{feynman}[small]
  \tikzfeynmanset{every vertex={dot,black,minimum size=1.5mm}}
  \vertex (left) at (-0.7,0);
  \vertex (right) at (0.7,0);
    \tikzfeynmanset{every vertex={dot,minimum size=0mm}}
  \vertex at (-0.9,-0.25) {\tiny $\mathcal{O}_j$};
  \vertex at (0.9,-0.25) {\tiny $\mathcal{O}_k$};
  \vertex (Pleft) at (-1.7,0) {\textcolor{darkgreen}{P}};
  \vertex (Pright) at (1.7,0) {\textcolor{darkgreen}{P}};
  \vertex (cuttop) at (0.1,0.8);
  \vertex (cutbottom) at (0.1,-0.8);
  \diagram* {
   (left) -- [out=70, in=110, fermion] (right),
   (left) -- [out=-70, in=-110, fermion] (right),
   (left) -- [out=20,in=160, fermion] (right),
   (left) -- [out=-20,in=-160,fermion] (right),
   (Pleft) -- [darkgreen, fermion, line width=0.5mm] (left),
   (right) -- [darkgreen, fermion, line width=0.5mm] (Pright),
   (cuttop) -- [dashed] (cutbottom),
  };
  \end{feynman}\end{tikzpicture}\end{gathered}
\\&= 
\text{-}\,\pi\,(P^2)^x\,\gamma_{ki}\,.
\end{align*}
It then follows from the symmetry of the left hand side under interchanging $\O_i$ and $\O_k$ that the anomalous dimension sub-matrix $\gamma_{ki}$ is symmetric. For this to work, we note that the phase space integral of the left diagram is the same when all arrows are inverted, since $\O(p_i) = \O(\text{-}\,p_i)$ for all considered operators.

Up to this point, we have worked in a framework with a nonzero momentum flow $P$, which requires a larger set of operators than the $S$-matrix. 
To isolate information about the anomalous dimensions of physical operators, the non-physical operators must be chosen to be pure descendants, \textit{i.e.}~total derivatives. 
From the fact that conformal primaries are orthogonal to descendants (see \textit{e.g.}~\cite{Braun:2003rp}), the orthonormality assumption requires that the physical basis in this argument necessarily consists of conformal primaries. 
This leads to the conclusion that \textit{for physical anomalous dimensions, the intra-length AD sub-matrix is symmetric when the conformal primary part of the operators is orthonormal.}
The specification of the primary part in this result is relevant since the non-primary part of an operator does not affect intra-length mixing (see Sec.\ \ref{s:structure}), while it may affect the normalization.

\subsubsection*{A symmetric and a diagonal ADM at one loop}
To exemplify the above result, let us consider the ADM of the complex scalar at mass dimension 8, which is presented in Section~\ref{sec:results}. There are two 4-point operators in a physical basis (where all operators are of on-shell type), and there is a freedom in choosing the particular linear combination given by eqs.~\eqref{eq:choice8a},\,\eqref{eq:choice8b}, leading to the transformation matrix in eq.~\eqref{eq:choiceBmat}.

We wish to explore a choice of 4-point operators that form an orthonormal basis. One can check that the operators 
\begin{align}\label{eq:angledbasis}
\mathcal{O}^{(8)c}_{4}\left(a, \, b \right) &=
 \mathcal{O}^{(8)c}_{4}\left(3\sin{\theta},
 \ 13\sin{\theta}+2\sqrt{5}\cos{\theta} \right) 
 \nn\\
%
%
\mathcal{O}^{(8)c}_{4}\left(a', \, b' \right) &=
\mathcal{O}^{(8)c}_{4}\left(3\cos{\theta}, \ 13\cos{\theta}-2\sqrt{5}\sin{\theta} \right)
\end{align}
have the same norm and are orthogonal for any value of the angle $\theta$.
This choice of basis results in the following anomalous dimension sub-matrix of the 4-point operators at one loop,
\begin{align}
\gamma'_{\text{sub}} &=
\begin{pmatrix}
\begin{matrix} a & a' \\ b & b' 
\end{matrix}
\end{pmatrix} ^{-1}
\begin{pNiceMatrix}[first-row]
\begin{matrix}
\scaleto{\textcolor{gray}{g\,\O^{(8)c}_4(1,0)}}{11pt} \\[2mm]
\end{matrix}& 
\begin{matrix}
\scaleto{\textcolor{gray}{g\,\O^{(8)c}_4(0,1)}}{11pt}\\[2mm]
\end{matrix}\\
\frac{11g}{3} & -\frac{4g}{3} \\[2mm] \frac{46g}{3} & -\frac{19g}{3} 
\end{pNiceMatrix} 
\begin{pmatrix}
\begin{matrix} a & a' \\ b & b' \end{matrix}\end{pmatrix} 
\nonumber\\[5mm] \label{eq:symdim8}
&= 
\begin{pNiceMatrix}[first-row]
\begin{matrix}
\textcolor{gray}{g\,\O^{(8)c}_4(a,b)} \\[2mm]
\end{matrix}& 
\begin{matrix}
\textcolor{gray}{g\,\O^{(8)c}_4(a',b')}\\[2mm]
\end{matrix}
\\
  -\frac{g}{9} (12 - 7 \cos{2\theta } + 8 \sqrt{5} \sin{2 \theta}) 
  & -\frac{g}{9}(8 \sqrt{5} \cos{2 \theta} + 7 \sin{2 \theta})\\ 
  -\frac{g}{9} (8 \sqrt{5} \cos{2 \theta} + 7 \sin{2 \theta})
   & -\frac{g}{9} (12 + 7 \cos{2 \theta} - 8 \sqrt{5} \sin{2 \theta})
\end{pNiceMatrix} \,,
\end{align}
which we indeed observe to be symmetric. 

We can further illustrate the choice of basis that diagonalizes the full one-loop ADM of the complex scalar at mass dimension 8. 
Note from \eqref{eq:symdim8}, a choice of $\theta$ that diagonalizes the 4-point sub-matrix is
\begin{align*}
  \theta =-\frac{1}{2}\arctan{\left(\frac{8\sqrt{5}}{7}\right)}\,.
\end{align*}
Instead of using this solution directly in \eqref{eq:angledbasis}, we can additionally rescale both operators separately to tidy up the expressions.\footnote{The effect of rescaling an operator (let's say the fourth) on the AD happens through the matrix $B=\text{diag}(1,1,1,x)$. This affects the normalization of the operator, and hence a general matrix would be made non-symmetric, but a diagonal (sub-)matrix remains diagonal.
}
For example, let us work with the operator basis%
\footnote{This choice of 4-point operators can be found as the eigenvectors of the 4-point sub-matrix in the original anomalous dimension matrix $\gamma_\textsc{c}^{(8)}$, which informs about the similarity transformation $B$ required for diagonalization.}
$$
\left\{ \mathcal{O}^{(8)c}_{8}, \ \mathcal{O}^{(8)c}_{6}, \ 
\mathcal{O}^{(8)c}_{4}\left(\frac{1}{46} (15 + \sqrt{41}), 1 \right), \ 
\mathcal{O}^{(8)c}_{4}\left({\frac{1}{46} (15 - \sqrt{41}), 1} \right)
\right\}
= \vcentcolon \left\{\mathcal{O}^{(8)c}_{8}, \ \mathcal{O}^{(8)c}_{6}, \ \mathcal{O}^{(8)c}_{4,\textsc{a}} , \ \mathcal{O}^{(8)c}_{4,\textsc{b}} \right\}
\,.
$$
which has the one-loop anomalous dimension matrix (omitting the factor of $g$)
{
\setlength\arraycolsep{5pt}
$$
\gamma = \begin{pNiceMatrix}[first-row]
\begin{matrix}
\scaleto{\textcolor{gray}{g^3\,\O^{(8)c}_8}}{14pt} \\[2mm]
\end{matrix}& 
\begin{matrix}
\scaleto{\textcolor{gray}{g^2\,\O^{(8)c}_6}}{14pt}\\[2mm]
\end{matrix}&
\begin{matrix}
\scaleto{\textcolor{gray}{g\,\mathcal{O}^{(8)c}_{4,\textsc{a}} }}{14pt} \\[2mm]
\end{matrix}& 
\begin{matrix}
\scaleto{\textcolor{gray}{g\,\mathcal{O}^{(8)c}_{4,\textsc{b}} }}{14pt}\\[2mm]
\end{matrix}\\
29& 0& \frac{108}{115} (1879 + 251 \sqrt{41})& \frac{108}{115} (1879 - 251 \sqrt{41})\\[2mm]
  0& 4& -\frac{1}{230} (2641 + 679 \sqrt{41})& -\frac{1}{230} (2641 - 679 \sqrt{41})\\[2mm] 
  0& 0& -\frac{1}{3} (4 - \sqrt{41})& 0\\[2mm]
  0& 0& 0& -\frac{1}{3} (4 + \sqrt{41})
\end{pNiceMatrix}
  \,.
$$
The} eigenvalues of this matrix can directly be read off from the diagonal entries. The corresponding eigenvectors give rise to the basis in which the AD is diagonal.
\begin{align}
  &\mathcal{O}^{(8)c}_{8}\nonumber\, , \ \mathcal{O}^{(8)c}_{6}\nonumber \, ,\\
&\left( 
  -\frac{324}{115} (22 + 3 \sqrt{41}) \mathcal{O}^{(8)c}_{8} + \frac{3}{9890} (14019 + 2701 \sqrt{41}) \mathcal{O}^{(8)c}_{6} + \mathcal{O}^{(8)c}_{4,\textsc{a}} 
\right) 
\,,\nonumber\\\label{eq:diagbasis}
&\left( 
  \frac{324}{115} (-22 + 3 \sqrt{41}) \mathcal{O}^{(8)c}_{8} - \frac{3}{9890} (-14019 + 2701 \sqrt{41}) \mathcal{O}^{(8)c}_{6} + \mathcal{O}^{(8)c}_{4,\textsc{b}} 
\right) \,.
\end{align}
We finally note that a field redefinition can be used to transform the obtained basis back into a basis where each operator has a fixed number of fields, via
\begin{align*}
  8! \, \mathcal{O}^{(8)c}_{8} & \hspace{2mm} = \hspace{2mm}
 \begin{gathered}
			\begin{tikzpicture}	
			\begin{feynman}[small, baseline=g1]
					\tikzfeynmanset{every vertex={dot,black,minimum size=1mm}}
				\vertex (g1);
				\vertex [below =0.3cm of g1] (g2) ;
				\vertex [below =0.3cm of g2] (g3);
				\vertex [below =0.3cm of g3] (g4);
					\tikzfeynmanset{every vertex={empty dot,black,minimum size=1mm}}
				\vertex [right =0.3cm of g1] (g6);
				\vertex [below =0.3cm of g6] (g7);
				\vertex [below =0.3cm of g7] (g8);
				\vertex [below =0.3cm of g8] (g9);
				\diagram* {
				};	
			\end{feynman}
			\end{tikzpicture}
			\end{gathered}
 \hspace{2mm} 
 \, \stackrel{\textsc{fr\hspace{0.5mm}}}{\longrightarrow} \,
 \hspace{2mm} \left(\frac{g^2}{4}\right)\left( \hspace{2mm}
\begin{gathered}
	\begin{tikzpicture}
	\begin{feynman}[small, baseline=g4]
	\tikzfeynmanset{every vertex={dot,black,minimum size=1mm}}
	\vertex (g1);
	\vertex [below =0.3cm of g1] (g3) ;	
	\tikzfeynmanset{every vertex={empty dot,black,minimum size=1mm}}
	\vertex [right =0.3cm of g1] (g2) ;
	\vertex [right =0.3cm of g3] (g4) ;
	\tikzfeynmanset{every vertex={dot,red,minimum size=0mm}}
	\vertex [above =2.5mm of g1] (loop1);
	\vertex [above =2.5mm of g2] (loop2);
	\diagram* {
			(g1) -- [out=45,in=0] (loop1) -- [out=180,in=135] (g1),
		(g2) -- [out=45,in=0] (loop2) -- [out=180,in=135] (g2)
	};
	\end{feynman}
	\end{tikzpicture}
\end{gathered}
\hspace{2mm} \right)
\\[2mm]
(-12)\, \mathcal{O}^{(8)c}_{6}& \hspace{2mm} = \hspace{2mm}
-2 \hspace{2mm} 
			\begin{gathered}
			\begin{tikzpicture}	
			\begin{feynman}[small, baseline=g1]
					\tikzfeynmanset{every vertex={dot,black,minimum size=1mm}}
				\vertex (g1);
				\vertex [below =0.3cm of g1] (g3);
				\vertex [below =0.3cm of g3] (l1);
					\tikzfeynmanset{every vertex={empty dot,black,minimum size=1mm}}
				\vertex [right =0.3cm of g1] (g2) ;
				\vertex [below =0.3cm of g2] (r1);
				\vertex [below =0.3cm of r1] (r2);
				\diagram* {
					(g1) -- [] (g2)
				};	
			\end{feynman}
			\end{tikzpicture}
			\end{gathered}
 \hspace{2mm} + \hspace{2mm}
			\begin{gathered}
			\begin{tikzpicture}	
			\begin{feynman}[small, baseline=g1]
					\tikzfeynmanset{every vertex={dot,black,minimum size=1mm}}
				\vertex (g1);
				\vertex [below =0.3cm of g1] (g3);
				\vertex [below =0.3cm of g3] (l1);
					\tikzfeynmanset{every vertex={empty dot,black,minimum size=1mm}}
				\vertex [right =0.3cm of g1] (g2) ;
				\vertex [below =0.3cm of g2] (r1);
				\vertex [below =0.3cm of r1] (r2);
				\diagram* {
					(g1) -- [] (g3)
				};	
			\end{feynman}
			\end{tikzpicture}
			\end{gathered}
 \hspace{2mm} + \hspace{2mm}
			\begin{gathered}
			\begin{tikzpicture}	
			\begin{feynman}[small, baseline=g1]
					\tikzfeynmanset{every vertex={dot,black,minimum size=1mm}}
				\vertex (g1);
				\vertex [below =0.3cm of g1] (g3);
				\vertex [below =0.3cm of g3] (l1);
					\tikzfeynmanset{every vertex={empty dot,black,minimum size=1mm}}
				\vertex [right =0.3cm of g1] (g2) ;
				\vertex [below =0.3cm of g2] (g4);
				\vertex [below =0.3cm of g4] (r2);
				\diagram* {
					(g2) -- [] (g4)
				};	
			\end{feynman}
			\end{tikzpicture}
			\end{gathered}
\hspace{2mm} \stackrel{\textsc{fr}}{\rightarrow} \hspace{2mm} \left(-\frac{g}{2}\right)\left( \hspace{2mm}
 - \hspace{2mm}
\begin{gathered}
	\begin{tikzpicture}
	\begin{feynman}[small, baseline=g4]
	\tikzfeynmanset{every vertex={dot,black,minimum size=1mm}}
	\vertex (g1);
	\vertex [below =0.3cm of g1] (g3) ;	
	\tikzfeynmanset{every vertex={empty dot,black,minimum size=1mm}}
	\vertex [right =0.3cm of g1] (g2) ;
	\vertex [right =0.3cm of g3] (g4) ;
	\tikzfeynmanset{every vertex={dot,red,minimum size=0mm}}
	\vertex [above =2.5mm of g1] (loop1);
	\vertex [above =2.5mm of g2] (loop2);
	\diagram* {
			(g1) -- [out=45,in=0] (loop1) -- [out=180,in=135] (g1),
		(g3) -- (g4)
	};
	\end{feynman}
	\end{tikzpicture}
\end{gathered}
 \hspace{2mm}- \hspace{2mm}
\begin{gathered}
	\begin{tikzpicture}
	\begin{feynman}[small, baseline=g4]
	\tikzfeynmanset{every vertex={dot,black,minimum size=1mm}}
	\vertex (g1);
	\vertex [below =0.3cm of g1] (g3) ;	
	\tikzfeynmanset{every vertex={empty dot,black,minimum size=1mm}}
	\vertex [right =0.3cm of g1] (g2) ;
	\vertex [right =0.3cm of g3] (g4) ;
	\tikzfeynmanset{every vertex={dot,red,minimum size=0mm}}
	\vertex [above =2.5mm of g1] (loop1);
	\vertex [above =2.5mm of g2] (loop2);
	\diagram* {
			(g2) -- [out=45,in=0] (loop2) -- [out=180,in=135] (g2),
		(g3) -- (g4)
	};
	\end{feynman}
	\end{tikzpicture}
\end{gathered}
 \hspace{2mm}+ \hspace{2mm}
\begin{gathered}
	\begin{tikzpicture}
	\begin{feynman}[small, baseline=g4]
	\tikzfeynmanset{every vertex={dot,black,minimum size=1mm}}
	\vertex (g1);
	\vertex [below =0.3cm of g1] (g3) ;	
	\tikzfeynmanset{every vertex={empty dot,black,minimum size=1mm}}
	\vertex [right =0.3cm of g1] (g2) ;
	\vertex [right =0.3cm of g3] (g4) ;
	\tikzfeynmanset{every vertex={dot,red,minimum size=0mm}}
	\vertex [above =2.5mm of g1] (loop1);
	\vertex [above =2.5mm of g2] (loop2);
	\diagram* {
			(g2) -- [out=45,in=0] (loop2) -- [out=180,in=135] (g2),
		(g1) -- (g3)
	};
	\end{feynman}
	\end{tikzpicture}
\end{gathered}
 \hspace{2mm}+ \hspace{2mm}
\begin{gathered}
	\begin{tikzpicture}
	\begin{feynman}[small, baseline=g4]
	\tikzfeynmanset{every vertex={dot,black,minimum size=1mm}}
	\vertex (g1);
	\vertex [below =0.3cm of g1] (g3) ;	
	\tikzfeynmanset{every vertex={empty dot,black,minimum size=1mm}}
	\vertex [right =0.3cm of g1] (g2) ;
	\vertex [right =0.3cm of g3] (g4) ;
	\tikzfeynmanset{every vertex={dot,red,minimum size=0mm}}
	\vertex [above =2.5mm of g1] (loop1);
	\vertex [above =2.5mm of g2] (loop2);
	\diagram* {
			(g1) -- [out=45,in=0] (loop1) -- [out=180,in=135] (g1),
		(g2) -- (g4)
	};
	\end{feynman}
	\end{tikzpicture}
\end{gathered}
\hspace{2mm} \right)
\end{align*}
(note this is not a unique procedure due to the freedom of performing integration by parts both before and after the field redefinitions).
This exemplifies the claim that to diagonalize the ADM one needs to use non-primary operators in the choice of basis. The form of the operators in \eqref{eq:diagbasis} is quite obtuse, and we have not established any rule that characterizes the basis which fully diagonalizes the ADM at one loop.

\bibliographystyle{JHEP}
\bibliography{refs}
 
\end{document}